\documentclass{article}

\usepackage{mathtools}
\usepackage[T1]{fontenc} 
\usepackage{fourier} 
\usepackage[english]{babel} 
\usepackage{sectsty} 
\usepackage{enumerate} 
\usepackage{titlesec} 
\usepackage{amsfonts}
\usepackage{mdframed}
\usepackage{cite}
\usepackage{amssymb}
\usepackage{amsthm}
\usepackage{slashed}
\usepackage{gensymb}
\usepackage[bottom]{footmisc} 
\usepackage{mathrsfs}
\usepackage{subcaption}
\usepackage{float}
\usepackage{physics} 
\usepackage{listings}
\usepackage{graphicx}
\usepackage{fancyhdr}
\usepackage[margin=1in]{geometry}
\usepackage{setspace}
\usepackage{multicol}
\usepackage{mdframed}
\usepackage{xcolor}
\usepackage{eucal}
\usepackage{tikz}
	\usetikzlibrary[decorations.pathmorphing, positioning]
\usepackage{tikz-cd}
\usepackage[compat=1.1.0]{tikz-feynman}
\usepackage{mathdots}
\usepackage[sort&compress,numbers]{natbib} 

\newcommand{\ii}{\textrm{i}} 
\newcommand{\ee}{\textrm{e}}

\newcommand{\sfrac}[2]{{\textstyle \frac{#1}{#2}}} 

\newenvironment{eqn}
	{
		\equation 
		\aligned
	}	
	{
		\endaligned
		\endequation
	}

\DeclareDocumentCommand\gradientvec{ g o d() }{ 
	\IfNoValueTF{#1}{
		\IfNoValueTF{#2}{
			\IfNoValueTF{#3}{\vec{\vnabla}}{\fbraces{\lparen}{\rparen}{\vec{\vnabla}}{#3}}
			}
			{\fbraces{\lbrack}{\rbrack}{\vec{\vnabla}}{#2} \IfNoValueTF{#3}{}{(#3)}}
		}
		{\vec{\vnabla} #1 \IfNoValueTF{#2}{}{[#2]} \IfNoValueTF{#3}{}{(#3)}}
	}

\DeclareDocumentCommand\divergencevec{ g o d() }{ 
	\IfNoValueTF{#1}{
		\IfNoValueTF{#2}{
			\IfNoValueTF{#3}{\vec{\vnabla} \vdot}{\vec{\vnabla} \vdot \quantity(#3)}
			}
			{\vec{\vnabla} \vdot \quantity[#2] \IfNoValueTF{#3}{}{(#3)}}
		}
		{\vec{\vnabla} \vdot #1 \IfNoValueTF{#2}{}{[#2]} \IfNoValueTF{#3}{}{(#3)}}
	}

\DeclareDocumentCommand\curlvec{ g o d() }{ 
	\IfNoValueTF{#1}{
		\IfNoValueTF{#2}{
			\IfNoValueTF{#3}{\vec{\vnabla} \cross}{\vec{\vnabla} \cross \quantity(#3)}
			}
			{\vec{\vnabla} \cross \quantity[#2] \IfNoValueTF{#3}{}{(#3)}}
		}
		{\vec{\vnabla} \cross #1 \IfNoValueTF{#2}{}{[#2]} \IfNoValueTF{#3}{}{(#3)}}
	}

\DeclareDocumentCommand\laplacianvec{ g o d() }{ 
	\IfNoValueTF{#1}{
		\IfNoValueTF{#2}{
			\IfNoValueTF{#3}{\vec{\nabla}^2}{\fbraces{\lparen}{\rparen}{\vec{\nabla}^2}{#3}}
			}
			{\fbraces{\lbrack}{\rbrack}{\vec{\nabla}^2}{#2} \IfNoValueTF{#3}{}{(#3)}}
		}
		{\vec{\nabla}^2 #1 \IfNoValueTF{#2}{}{[#2]} \IfNoValueTF{#3}{}{(#3)}}
	}

\DeclareMathAlphabet{\mathpzc}{OT1}{pzc}{m}{it}
\renewcommand{\mathcal}[1]{\CMcal{#1}}

\usepackage{multirow}
\usepackage{authblk}
\usepackage{xr-hyper} 
\usepackage{graphicx}
\usepackage{dcolumn}
\usepackage{bm}
\usepackage{amsmath,amssymb,amsfonts}
\usepackage{mathrsfs}
\usepackage{bbm}
\usepackage{slashed}
\usepackage{verbatim}
\usepackage[T1]{fontenc}
\usepackage[utf8]{inputenc}
\usepackage{hyperref}
\usepackage{ifthen}
\usepackage{forloop}
\usepackage[english]{babel}
\usepackage{mathbbol}
\usepackage[usenames,dvipsnames]{xcolor}
\usepackage[normalem]{ulem}
\usepackage{tabularx, booktabs}
\usepackage{subcaption}

\newcommand{\bea}{\begin{eqnarray}}
\newcommand{\eea}{\end{eqnarray}}
\newcommand{\be}{\begin{equation}}

\definecolor{darkgreen}{rgb}{0,0.4,0}

\graphicspath{{figures/}}

\title{Euclidean Correlation Functions in Quantum Gravity}

\author[1]{Jack Laiho \thanks{jwlaiho@syr.edu}}
\author[2]{Kenny Ratliff \thanks{kratliff@fastmail.com}}

\affil[1]{Department of Physics, Syracuse University, Syracuse, NY 13244}

\affil[2]{Department of Physics, Syracuse University, Syracuse, NY 13244 and Department of Physics, Kenyon College, Gambier, OH 43022}

\begin{document}
\maketitle

\abstract{We calculate Euclidean correlation functions through next-to-leading order in the low energy effective theory of gravity.  We focus on correlation functions of curvature and volume operators, calculating these functions through one-loop order. We show that quantum fluctuations of the background spacetime must be taken into account in order to obtain gauge invariant expressions, and we point out a subtlety associated with the analytic continuation of the conformal mode.  Our final expressions for the correlation functions involve only Newton's constant and the source-sink separation, and they are a universal prediction of the low energy effective theory.  Thus, they serve as a useful point of comparison for nonperturbative lattice formulations of gravity.}

\normalsize
\section{\label{sec:intro}Introduction}


The formulation of quantum gravity remains an important outstanding problem in theoretical physics.  Although a theory of quantum gravity valid at all scales is not yet established, it is well-known how to formulate the low energy effective theory of gravity, valid at long distances and expressed in terms of propagating gravitons \cite{Donoghue:1994dn}.  The effective theory is not renormalizable, so that at each order in the perturbative expansion, new low energy constants must be introduced, representing unknown short-distance physics.  This leads to a loss of predictive power of the perturbation expansion.  Even so, the leading non-analytic behavior coming from radiative corrections can be calculated, giving rise to universal predictions that are expressible in terms of a small number of couplings, often just the Newton constant $G$.  Graviton-graviton scattering \cite{Dunbar:1994bn, Donoghue:1999qh} and the leading quantum corrections to the Newtonian potential \cite{Bjerrum-Bohr:2002gqz} are examples of this.


Attempts to formulate quantum gravity on the lattice aim to go beyond the effective theory and to capture non-perturbative effects \cite{Hamber:1985gx, Agishtein:1991cv, Ambjorn:1991pq, Catterall:1994pg, Ambjorn:2005qt, Ambjorn:2012jv, Laiho:2016nlp, Dai:2024vjc}.  Such approaches typically attempt to demonstrate the existence of a non-trivial fixed point that would allow the definition of an ultra-violet (UV)-complete theory, thus realizing Weinberg's asymptotic safety scenario for gravity \cite{Weinberg:1980gg}.  Many challenges to this program exist, though it is fair to say that asymptotic safety has not been excluded as a possibility, and promising evidence has been presented \cite{Wetterich:1992yh, Reuter:1996cp, Percacci:2007sz, Reuter:2012id, Eichhorn:2018yfc, Bonanno:2020bil, Ambjorn:2005db, Ambjorn:2005qt, Ambjorn:2007jv, Laiho:2016nlp, Ambjorn:2024qoe}.  Whether or not the lattice realizes this scenario, 
it should still be possible for lattice methods to recover, at least in principle, the effective theory at long distances.  Indeed, results from lattice simulations show the emergence of de Sitter space in four-dimensions \cite{Ambjorn:2004qm, Ambjorn:2005qt, Dai:2024vjc}, with some simulations showing that the properties of the (quenched) matter sector behave as expected \cite{Catterall:2018dns, Dai:2021fqb} if they are to make contact with the correct low energy theory.  Semiclassical fluctuations about de Sitter space can be used to determine $G$ in lattice units, thus determining the lattice spacing in terms of the Planck length \cite{Ambjorn:2007jv, Ambjorn:2008wc, Bassler:2021pzt}.  We are motivated by the possibility of performing further checks of the lattice approach where a comparison to the low energy effective theory may be possible.


It is natural to consider two-point Euclidean correlation functions on the lattice, where operators are constructed from geometric quantities.  Correlations between local curvature fluctuations and between local volume fluctuations were looked at in lattice calculations decades ago \cite{Hamber:1993rb, deBakker:1995he, Ambjorn:1998vd}, as well as more recently \cite{vanderDuin:2024pxb, Maas:2025rug}.  In pure gravity in four dimensions, where the degrees of freedom are massless gravitons, these correlation functions have a power-law fall-off.  The leading order curvature correlator (about flat space) was first calculated in Ref.~\cite{Modanese:1992ir}, where it was shown to be (derivatives of) a delta function in the source-sink separation, thus vanishing everywhere away from the origin.  Such a contact term would not be visible in the Euclidean correlation functions accessible to numerical lattice calculations.  The first non-vanishing contribution to the curvature correlator thus starts at next-to-leading order and requires a one-loop calculation.  The effective theory calculation of the volume correlator has not been considered in the literature, to the best of our knowledge.  In this work we fill this gap and present the calculation of both the curvature and volume correlation functions through next-to-leading (one-loop) order, expressed in terms of the Euclidean source-sink separation.  These expressions feature Newton's constant as the only input parameter, so they represent unambiguous predictions of the low energy effective theory.


There are several challenging aspects to the correlation function calculation presented here.  There is the usual issue in perturbative gravity calculations that the Feynman rules are complicated.  The three-graviton vertex has, after accounting for all of the momentum permutations, on the order of one hundred terms.  Individual diagrams can have tens of thousands of terms at intermediate stages of the calculation, requiring computer algebra packages for their evaluation.  Other difficulties are more subtle.  Foremost of these is the presence of coordinate corrections, which appear in addition to the standard field theory contributions, and are needed to account for the fact that the source and sink locations of the correlation functions live on a spacetime that is itself undergoing quantum fluctuations \cite{Brunetti:2016hgw, Frob:2017lnt}.  Since the source-sink separation is most conveniently expressed in terms of distances on the flat background spacetime about which we expand, additional corrections will appear when spacetime fluctuations are taken into account.  These corrections were shown to be required to produce gauge-invariant results (in a generalized de Donder gauge) for the two-point function of a scalar field including gravitational corrections \cite{Frob:2017lnt}.  We verify that the coordinate corrections are necessary in order to obtain similarly gauge-invariant results for the curvature and volume correlation functions considered in this work.  The gauge-invariance of our final results is a highly non-trivial check of the calculations.  There is also a subtlety arising from the analytic continuation to Euclidean signature of the conformal mode \cite{Gibbons:1978ac} that requires some care.     


This paper is organized as follows.  In Section~\ref{sec:observ} we give a detailed discussion of the construction of diffeomorphism-invariant observables, and we use the formalism to compute the coordinate corrections to the scalar curvature and volume operators.  In Section~\ref{sec:Feyn} we review the Feynman rules for the effective theory of gravity to the order that we require.  Section~\ref{sec:corr} presents the calculation of the curvature and volume correlation functions that are the focus of this work.  We conclude in Section~\ref{sec:conclude}.

\section{\label{sec:observ}Diffeomorphism-invariant observables}

The definition of gauge-invariant observables in general relativity is itself a subtle one; we review here a solution to this problem, following the recent development of {\em relational observables} \cite{C_Rovelli_1991, Dittrich_2006, Giddings:2005id, Tambornino:2011vg, Khavkine:2015fwa, Marolf:2015jha, Brunetti:2013maa, Frob:2017lnt, Frob:2021mpb}. The essential idea of relational observables is to define a shared ``master'' coordinate system $\mathsf{X}$ to which all other coordinate systems $\mathsf{x}$ refer their observations. These master coordinates are defined to be harmonic with respect to the full metric, $\grad^2 \mathsf{X} = 0$, and are in this sense therefore the ``straightest possible'' coordinates, even when quantum fluctuations of spacetime are taken into account. Two-point functions of relational observables are then functions of the ``master coordinate distance'' between the points at which the relational observables are measured.  In this section we begin with some conventions.  We then introduce the coordinate scalars needed to define our ``master'' coordinate system, and finally we show how to construct relational geometric observables. 
We use this formalism to obtain the explicit perturbative expansions of the relational observables constructed from the volume factor $\sqrt{-g}$ and from the Ricci scalar curvature $R$, which to the best of our knowledge have not previously been obtained in the literature.

\subsection{Conventions}

Throughout this work we denote the $d$-dimensional spacetime manifold by $M$ and a generic coordinate system on $M$ as
$\mathsf{x} : M \to \mathbb{R}^d$. Note that the sans serif symbol $\mathsf{x}$ represents the {\em map} which takes a point $p \in M$ to its coordinates. The italic symbol $x$ rather denotes an actual {\em value} of the coordinates:
		\begin{eqn}
			\mathsf{x} : p \in M \mapsto \mathsf{x}(p) = x \in \mathbb{R}^d.
		\end{eqn}
This may seem unnecessarily pedantic, but we find this distinction to be useful in our discussion of relational observables. Adopting the usual conventions, we denote the coordinate frame by $\partial_\mu$ and the coordinate coframe by $\dd{\mathsf{x}^\mu}$. When we need a second coordinate system we denote it with tildes, so that e.g. the coordinate system $\tilde{\mathsf{x}} : M \to \mathbb{R}^d$ has coordinate frame $\tilde{\partial}_\mu$. We denote a generic diffeomorphism of $M$ by $F : M \to M$ and assume that some such diffeomorphism relates $\mathsf{x}$ and $\tilde{\mathsf{x}}$ as $\tilde{\mathsf{x}} = \mathsf{x} \circ F^{-1}$. We also typically abbreviate ``diffeomorphism'' as ``diff''.

\subsection{The coordinate scalars}
\label{sec:CoordinateScalars}

In this subsection and the next we describe a recently-developed program \cite{C_Rovelli_1991, Dittrich_2006, Giddings:2005id, Tambornino:2011vg, Khavkine:2015fwa, Marolf:2015jha, Brunetti:2013maa, Frob:2017lnt, Frob:2021mpb}, called the {\em relational approach}, by which to construct, given any local observable, a corresponding diff-invariant but nonlocal observable.  To understand the relational approach we begin with the following example.  Suppose that two observers wish to measure some observable, which for simplicity we will take to be a scalar field $\phi : M \to \mathbb{R}$. Then each would set up their coordinate systems and make their measurements and record where and when they did so. Thus each would measure some value $\phi \circ \mathsf{x}^{-1}(x)$ and $\phi \circ {\tilde{\mathsf{x}}}^{-1}(\tilde{x})$, where $x$ and $\tilde{x}$ are the coordinate {\em values} at which each observer made their measurement and $\mathsf{x}$ and $\tilde{\mathsf{x}}$ are the coordinate {\em systems} each has set up. If these observers wished they could then sit down later and talk about it, and they could (in principle) figure out whether their different sets of numbers $x$ and $\tilde{x}$ corresponded to the same point in space. However this does not change the fact that in order for the observable to be diff-invariant it must appear the same to them in their own frames, without them sitting down and figuring out how to translate from one of their systems to the other, and if $\mathsf{x} \neq \tilde{\mathsf{x}}$ and $x = \tilde{x}$ (i.e. if the observers are distinct and make their measurements at the same coordinate values) then their measurements cannot in general be the same in both frames.

The above also points to a resolution to the problem: if it were somehow possible to ``signpost'' each point in spacetime, so that observers in different coordinate systems could still agree on the point at which to make their measurements, then it would be possible to make diff-invariant observations - each observer would simply mark down the signpost at which the measurement was made, instead of the coordinates in their own system.\footnote{This is all framed from the passive perspective. From the active perspective we would want our signposts to get pushed around by diffeomorphisms in the same manner as the fields of the theory.} To put this more quantitatively, we would hope to construct a ``master'' coordinate system, and provide to every observer the means to obtain these master coordinates given only information in their frame. In this subsection we review the construction of such a coordinate system \cite{Frob:2017lnt, Frob:2021mpb}. 

\subsubsection{The coordinate scalars as a function of the background coordinates}
\label{sec:CoordScalarsOfBackgroundCoords}

Our first order of business is to construct the master coordinates. We denote these by $\mathsf{X} : M \to \mathbb{R}^d$, with the italic symbol $X$ referring to an arbitrary value, i.e. $\mathsf{X}(p) = X$ for $p \in M$. Since $\mathsf{X}$ are not ``coordinates'' in the usual sense, with instead each component $\mathsf{X}^\mu$ transforming under diffeomorphisms in precisely the same way as an arbitrary scalar field, we refer to these as the {\em coordinate scalars}.

The construction of the coordinate scalars depends on the setting in which they are constructed. In our case we are interested in perturbations about flat space, which implies the following.
\begin{itemize}

	\item We assume some pre-existing coordinate system $\mathsf{x} = (\mathsf{t}, \vb*{\mathsf{x}}) : M \to \mathbb{R}^d$, an arbitrary value of which is denoted $\mathsf{x}(p) = x = (t, \vb*{x}) \in \mathbb{R}^d$. We call these the {\em background coordinates}.
	
	\item We assume the existence of a metric $g_{\mu\nu}$ on $M$, whose perturbation about flat space is denoted in the usual way,
		\begin{eqn}
			g_{\mu\nu}
				&= \eta_{\mu\nu} + \kappa h_{\mu\nu},
		\end{eqn}
	with $\eta_{\mu\nu}$ the flat metric and $\kappa$ the gravitational coupling, given in terms of Newton's constant by $\kappa = \sqrt{32 \pi G}$ in four dimensions.  The background metric is itself not a well-defined geometric object on $M$ in either the active or passive pictures: in the active picture the background metric is unaffected by diffeomorphisms while all other fields (including the full metric) are pulled around, while in the passive picture the background metric is defined to have the same components in any coordinate system.

\end{itemize}

\noindent
Following \cite{Frob:2017lnt}, we begin from the observation that the background coordinates are harmonic with respect to the background metric, $\bar{\grad}^2 \mathsf{x}^\mu = 0$ (where $\bar{\grad}$ denotes the gradient with respect to the background metric). Since we are perturbing about flat space we then define the coordinate scalars (a) to be harmonic with respect to the perturbed (full) metric, and (b) to reduce to the background coordinates when the metric perturbation vanishes. In other words we define the coordinate scalars to satisfy
		\begin{eqn}
		\label{eq:CoordinateScalarsAreHarmonic}
			\grad^2 \mathsf{X}^\mu
				&= 0,
		\end{eqn}
	(where $\grad$ denotes the gradient operator with respect to the {\em full} metric $g_{\mu\nu}$) and construct them perturbatively as
		\begin{eqn}
		\label{eq:CoordinateScalarExpansion}
			\mathsf{X}
				&= \sum_{a=0}^\infty \kappa^a \mathsf{X}_a,
		&
			\mathsf{X}_0(p)
				&= \mathsf{x}(p).
		\end{eqn}
	Note that the coordinate scalars depend nontrivially, by construction, on the background coordinates from which we build them: if we first proceed through the next paragraphs and then perform a passive coordinate transformation $\mathsf{x} \mapsto \tilde{\mathsf{x}} = \mathsf{x} \circ F^{-1}$, the explicitly constructed coordinate scalars will obey $\mathsf{X}_0 = \mathsf{x}$ and {\em not}  $\mathsf{X}_0 = \tilde{\mathsf{x}}$. Equivalently, if we perform an active transformation then the coordinate scalars will themselves change as any other scalar field, $\mathsf{X}^\mu \mapsto \tilde{\mathsf{X}}^\mu = F^* \mathsf{X}^\mu$, meaning that $\tilde{\mathsf{X}}_0(p) = \mathsf{X}_0 \circ F(p) \neq \mathsf{x}(p)$. (Even though the components of $\mathsf{X}$ carry an index which looks superficially like a vector index they are all individually scalars, not the components of a vector.)

We can reexpress the equation (\ref{eq:CoordinateScalarsAreHarmonic}) in a more perturbatively useful form by recalling \cite{Lee2012} that we can write the full Laplacian $\grad^2$ in terms of the full metric $g_{\mu\nu}$, its determinant $g$, and the coordinate frame $\partial_\mu$ of the $\mathsf{x}$ coordinates as
		\begin{eqn}
		\label{eq:LaplaceOperator}
			\grad^2 \mathsf{X}^\mu
				&= \frac{1}{\sqrt{-g}} \partial_\alpha 
					\big( \sqrt{-g} g^{\alpha \beta} \partial_\beta \mathsf{X}^\mu \big).
		\end{eqn}
	We may then expand this expression in $\kappa$ and solve for the $\mathsf{X}_a$'s order by order, which proceeds as follows. Expanding the various pieces gives
	\begin{eqn}
		\label{eq:metric-inverse-expansion}
			g^{\mu \nu}
				&= \eta^{\mu \nu}
					- \kappa h^{\mu \nu}
					+ \kappa^2 h^{\mu \alpha} h {}_\alpha {}^\nu
					- \kappa^3 h^{\mu \alpha} h^{\nu \beta} h_{\alpha \beta}
					+ \kappa^4 h^{\mu \alpha} h^{\nu \beta} h {}_\alpha {}^\gamma
						h_{\beta \gamma}
					+ O(\kappa^5)
		\end{eqn}
	for the inverse metric and
		\begin{eqn}
		\label{eq:volume-expansion}
			\sqrt{-g}
				&= 1
					+ \sfrac{1}{2} \kappa h
					+ \kappa^2 \bigg( \sfrac{1}{8} h^2 
						- \sfrac{1}{4} h_{\mu \nu} h^{\mu \nu} \bigg)
					+ \kappa^3 \bigg( 
						\sfrac{1}{6} h^{\mu \nu} h {}_\mu {}^\alpha h_{\alpha \nu}
						- \sfrac{1}{8} h h_{\mu \nu} h^{\mu \nu}
						+ \sfrac{1}{48} h^3 \bigg)
		\\
				&\hspace{1cm}
					+ \kappa^4 \bigg(
							- \sfrac{1}{8} h^{\mu \nu} h {}_\mu {}^\alpha h {}_\nu {}^\beta
								h_{\alpha \beta}
							+ \sfrac{1}{12} h h^{\mu \nu} h {}_\mu {}^\alpha h_{\alpha \nu}
							+ \sfrac{1}{32} (h_{\mu \nu} h^{\mu \nu})^2
							- \sfrac{1}{32} h^2 h_{\mu \nu} h^{\mu \nu}
							+ \sfrac{1}{384} h^4
						\bigg)
					+ O(\kappa^5)
		\end{eqn}
	for the volume factor.
	We review the derivations of these expansions in Sec. \ref{sec:inverse-metric-expansion} and \ref{sec:volume-factor-expansion}. We also make use of the {\sc Mathematica} \cite{Mathematica} package {\sc xAct} \cite{Martin-Garcia:2007bqa, Martin-Garcia:2008yei, Martin-Garcia:2008ysv, Brizuela:2008ra, Gomez-Lobo:2011kaw, Pitrou:2013hga, Nutma:2013zea, Levi:2017kzq} to confirm our result here, as well as for the lengthier perturbative expansions to come. The $O(\kappa^0)$ term in $\grad^2 \mathsf{X}^\mu$ is then
		\begin{eqn}
			\grad^2 \mathsf{X}^\mu
				&= \partial_\alpha 
					\big( \eta^{\alpha \beta} \partial_\beta \mathsf{X}_0^\mu \big)
					+ O(\kappa)
				= \bar{\grad}^2 \mathsf{X}_0^\mu + O(\kappa).
		\end{eqn}
	Since we impose $\grad^2 \mathsf{X}^\mu = 0$, this term must vanish, although we in fact already knew this, since we also impose that $\mathsf{X}_0 = \mathsf{x}$.

Things are less trivial at $O(\kappa)$. When we expand the right hand side of eq. (\ref{eq:LaplaceOperator}) to this order we find one term featuring $\mathsf{X}_1^\mu$, and the rest of the terms feature $\mathsf{X}_0^\mu = \mathsf{x}^\mu$ and various factors of $h_{\mu\nu}$. The $\mathsf{X}_1^\mu$ term reduces to $\bar{\grad}^2 \mathsf{X}_1^\mu$, and (since $\partial_\mu \mathsf{X}_0^\nu = \partial_\mu \mathsf{x}^\nu = \delta^\nu_\mu$) the rest form a linear polynomial in $h_{\mu\nu}$, which we may denote $J_1^\mu$. Setting $\grad^2 \mathsf{X}^\mu = 0$ at this order then implies that $\mathsf{X}_1^\mu$ satisfies an equation of the form
		\begin{eqn}
		\label{eq:X1-condition}
			\bar{\grad}^2 \mathsf{X}_1^\mu
				&= J_1^\mu,
		\end{eqn}
	where using the expansions (\ref{eq:metric-inverse-expansion}) and (\ref{eq:volume-expansion}) and turning the crank yields
		\begin{eqn}
		\label{eq:J1-def}
			J_1^\mu
				&= \partial_\alpha h^{\alpha \mu} - \frac{1}{2} \partial^\mu h.
		\end{eqn}
	The objects in Eqs.~(\ref{eq:X1-condition}) and (\ref{eq:J1-def}) are functions of spacetime, e.g. $\mathsf{X}_1^\mu : M \to \mathbb{R}$. To obtain an explicit expression for $\mathsf{X}_1$ we need to rewrite these as functions of the background coordinates $\mathsf{x} : M \to \mathbb{R}^d$. We therefore denote the coordinate representations of these objects with hats, e.g. $\hat{\mathsf{X}} = \mathsf{X} \circ \mathsf{x}^{-1}$, so that $\hat{\mathsf{X}}(x)$ tells us the value of the coordinate scalars $\mathsf{X}$ at the spacetime point $p \in M$ whose background coordinates are $x \in \mathbb{R}^d$. The right hand side of Eq.~(\ref{eq:X1-condition}) expands to the standard coordinate representation of the flat-space Laplacian,
		\begin{eqn}
			\big( \bar{\grad}^2 \mathsf{X}_1^\mu \big) \circ \mathsf{x}^{-1}(x)
				&= \eta^{\alpha \beta} \pdv[2]{\hat{\mathsf{X}}_1^\mu}{x^\alpha}{x^\beta}
					(x)
				= \bigg( \Box \hat{\mathsf{X}}_1^\mu \bigg) (x),
		\end{eqn}
	denoting by $\Box$ the explicit coordinate representation $\eta^{\alpha \beta} \pdv*[2]{}{x^\alpha}{x^\beta}$ of the flat-space Laplacian $\bar{\grad}^2$. Eq. (\ref{eq:X1-condition}) then becomes
		\begin{eqn}
			\Box \hat{\mathsf{X}}_1^\mu
				&= \hat{J}_1^\mu,
		\end{eqn}
	meaning that, given a Green function $G(x,x')$ of $\Box$, we obtain the explicit solution
		\begin{eqn}
		\label{eq:X1-result}
			\hat{\mathsf{X}}_1^\mu(x)
				&= \int \dd[d]{x'} G(x,x') \hat{J}_1^\mu(x').
		\end{eqn}

	Similar logic applies at $O(\kappa^2)$. There is one term featuring $\mathsf{X}_2^\mu$ and with every other factor evaluated at $O(\kappa^0)$, which term reduces to $\bar{\grad}^2 \mathsf{X}_2^\mu$. There is then a collection of terms featuring $\mathsf{X}_1^\mu$, in each of which one of the other factors is evaluated at $O(\kappa)$ and the rest are at $O(\kappa^0)$. This collection takes the form of a differential operator linear in $h_{\mu\nu}$ acting on $\mathsf{X}_1^\mu$, which we may denote $K_1 \mathsf{X}_1^\mu$. Finally, there is another collection of terms featuring $\mathsf{X}_0^\mu$, and this collection reduces to a quadratic polynomial in $h_{\mu \nu}$, which we denote $J_2^\mu$, following Ref.~\cite{Frob:2017lnt}. Setting $\grad^2 \mathsf{X}^\mu = 0$ at this order therefore implies that $\mathsf{X}_2^\mu$ satisfies an equation of the form
		\begin{eqn}
		\label{eq:X2-condition}
			\bar{\grad}^2 \mathsf{X}_2^\mu
				&= J_2^\mu + K_1 \mathsf{X}_1^\mu,
		\end{eqn}
	and again turning the crank yields
		\begin{eqn}
		\label{eq:J2-K1-def}
			K_1
				&= h^{\alpha \beta} \partial_\alpha \partial_\beta
					+ J_1^\alpha \partial_\alpha,
		&
			J_2^\mu
				&= \frac{1}{2} \bigg( h_{\alpha \beta} \partial^\mu h^{\alpha \beta}
					+ h^{\alpha \mu} \partial_\alpha h \bigg)
					- \partial_\alpha \bigg( h^{\alpha \beta} h {}_\beta {}^\mu \bigg).
		\end{eqn}
	By the same logic as for $\mathsf{X}_1$ we then obtain an explicit solution for $\mathsf{X}_2^\mu$ as
		\begin{eqn}
		\label{eq:X2-result}
			\hat{\mathsf{X}}_2^\mu(x)
				&= \int \dd[d]{x'} G(x,x') 
					\bigg( \hat{J}_2^\mu(x') + \hat{K}_1 \hat{\mathsf{X}}_1^\mu (x') \bigg),
		\end{eqn}
	again with $G(x,x')$ a Green function of $\Box$ and with hats denoting the coordinate representations in $\mathsf{x}$.
	
The expressions (\ref{eq:X1-result}) and (\ref{eq:X2-result}) make manifest the tradeoff we are making in this construction. As we see below, the $\mathsf{X}$'s do allow us to define invariant forms of tensor components of arbitrary rank. However these ``invariantized'' tensor components will be written in terms of the $\mathsf{X}$'s, which contain explicit integrations over all of spacetime, and hence the gauge-invariant observables so defined are nonlocal. This is to be expected for observables in quantum gravity \cite{Giddings:2005id}.

	We can make systematic the above construction to all orders as follows. We define
		\begin{eqn}
			{\cal D}^\mu 
				&\equiv\frac{1}{\sqrt{-g}} \partial_\alpha \big( \sqrt{-g} g^{\alpha \mu} \big)
				\equiv - \sum_{n=0}^\infty \kappa^n J_n^\mu.
		\end{eqn}
	Since $g^{\alpha \mu} = g^{\alpha \beta} \partial_\beta x^\mu$ we can interpret ${\cal D}^\mu$ as the Laplacian of the background coordinate component $x^\mu$. As we show below the $J_n^\mu$'s defined here include precisely the $J_1^\mu$ and $J_2^\mu$ we have already met.  We also define, given any $\kappa$-independent function $f \in \text{C}^\infty(M)$,
		\begin{eqn}
			\grad^2 f
				&\equiv - \sum_{n=0}^\infty \kappa^2 K_n f.
		\end{eqn}
	We show that the differential operators $K_n$ include the same $K_1$ obtained above.

We can relate the $K$'s and $J$'s, so defined, by expanding the Laplacian of our $\kappa$-independent $f$:
		\begin{eqn}
			\grad^2 f
				&= \frac{1}{\sqrt{-g}} \partial_\alpha \big( \sqrt{-g} g^{\alpha \mu} \partial_\mu f \big)
				= {\cal D}^\mu \partial_\mu f + g^{\alpha \mu} \partial_\alpha \partial_\mu f.
		\end{eqn}
	Defining the perturbative expansion of the full inverse metric by
		\begin{eqn}
			g^{\mu \nu}
				&= \sum_{n=0}^\infty \kappa^n \tilde{g}_n^{\mu \nu}
		\end{eqn}
	we then have \footnote{Note that $\tilde{g}^{\mu \nu}$ here is unrelated to $\tilde{x}$.}
		\begin{eqn}
			\grad^2 f
				&= \sum_{n=0}^\infty \kappa^n \bigg( - J_n^\mu \partial_\mu
						+ \tilde{g}_n^{\alpha \mu} \partial_\alpha \partial_\mu
					\bigg) f,
		\end{eqn}
	from which we can conclude that
		\begin{eqn}
			K_n
				&= J_n^\mu \partial_\mu
					- \tilde{g}_n^{\alpha \mu} \partial_\alpha \partial_\mu.
		\end{eqn}

So far we have only considered the action of the Laplacian on the $\kappa$-independent function $f$. However, what we are actually interested in is the action of the Laplacian on the coordinate scalars, which are not $\kappa$-independent, but are infinite series in $\kappa$. This action can still be represented in terms of the $K$'s, and therefore in terms of the $J$'s, by expanding both the Laplacian and the coordinate scalar itself:
		\begin{eqn}
			\grad^2 \mathsf{X}^\mu
				&= - \sum_{r=0}^\infty \kappa^r K_r \mathsf{X}^\mu
				= - \sum_{r=0}^\infty \sum_{s=0}^\infty \kappa^{r+s} K_r \mathsf{X}_s^\mu
				= - \sum_{n=0}^\infty \kappa^n \sum_{r=0}^n K_r \mathsf{X}_{n-r}^\mu,
		\end{eqn}
	or to $O(\kappa^2)$,
		\begin{eqn}
			\grad^2 \mathsf{X}^\mu
				&= - K_0 \mathsf{X}_0^\mu
					- \kappa \bigg( K_0 \mathsf{X}_1^\mu + K_1 \mathsf{X}_0^\mu \bigg)
					- \kappa^2 \bigg( K_0 \mathsf{X}_2^\mu + K_1 \mathsf{X}_1^\mu 
						+ K_2 \mathsf{X}_0^\mu \bigg)
					+ O(\kappa^3).
		\end{eqn}

To find the explicit forms of the $K$'s we need the expansions of $g^{\mu \nu}$ and $\sqrt{-g}$, which are given by Eqs.~(\ref{eq:metric-inverse-expansion}) and (\ref{eq:volume-expansion}). At zeroth order we therefore have
		\begin{eqn}
			{\cal D}^\mu
				&= \partial_\alpha \eta^{\alpha \mu} + O(\kappa)
				= 0 + O(\kappa)
		\implies
			J_0^\mu
				= 0,
		\end{eqn}
	and hence
		\begin{eqn}
			K_0
				= - \tilde{g}_0^{\mu \nu} \partial_\mu \partial_\nu
				= - \bar{\grad}^2,
		\end{eqn}
	as it must. Thus at $O(\kappa^0)$ and using our assumption $\mathsf{X}_0^\mu = x^\mu$ we find that the condition $\grad^2 \mathsf{X}^\mu = 0$ reduces to the harmonic gauge condition on the background coordinates, $\bar{\grad}^2 x^\mu = 0$, as expected, and at arbitrary $O(\kappa^n)$ the same condition yields a differential equation of the form
		\begin{eqn}
		\label{eq:general-condition-coordinate-scalar-expansion}
			\bar{\grad}^2 \mathsf{X}_n^\mu
				= \sum_{r=1}^n K_r \mathsf{X}_{n-r}^\mu.
		\end{eqn}

At first order we have
		\begin{eqn}
			{\cal D}^\mu
				&= \partial_\alpha
					\bigg( \big( 1 + \sfrac{1}{2} \kappa h \big) 
						\big( \eta^{\alpha \mu} - \kappa h^{\alpha \mu} \big)
					\bigg)
					+ O(\kappa^2)
				= \kappa \bigg( 
					\sfrac{1}{2} \partial^\mu h
					- \partial_\alpha h^{\alpha \mu}
					\bigg)
					+ O(\kappa^2)
		\implies
			J_1^\mu
				= \partial_\alpha h^{\alpha \mu}
					- \sfrac{1}{2} \partial^\mu h,
		\end{eqn}
	in terms of which
		\begin{eqn}
			K_1
				&= J_1^\mu \partial_\mu + h^{\mu \nu} \partial_\mu \partial_\nu,
		\end{eqn}
	both of which agree with the prior results (\ref{eq:J1-def}) and (\ref{eq:J2-K1-def}). From eq. (\ref{eq:general-condition-coordinate-scalar-expansion}) we then find the equation for $\mathsf{X}_1^\mu$,
		\begin{eqn}
			\bar{\grad}^2 \mathsf{X}_1^\mu
				&= K_1 \mathsf{X}_0^\mu
				= J_1^\alpha \partial_\alpha x^\mu
					+ h^{\alpha \beta} \partial_\alpha \partial_\beta x^\mu
				= J_1^\mu,
		\end{eqn}
	in agreement with (\ref{eq:X1-condition}). Proceeding similarly at second order we have
		\begin{eqn}
			{\cal D}^\mu
				&= \big( 1 - \sfrac{1}{2} \kappa h \big) \partial_\alpha \Bigg\{
					\bigg( 1 + \sfrac{1}{2} \kappa h 
						+ \kappa^2 \big( \sfrac{1}{8} h^2 
						- \sfrac{1}{4} h_{\rho \sigma} h^{\rho \sigma} \big)
					\bigg)
					\bigg( \eta^{\alpha \mu} - \kappa h^{\alpha \mu} 
						+ \kappa^2 h^{\mu \nu} h {}_\nu {}^\alpha \bigg) \Bigg\}
		\\
				&= O(\kappa) + \kappa^2 \Bigg\{
						- \sfrac{1}{2} h_{\alpha \beta} \partial^\mu h^{\alpha \beta} 
						- \sfrac{1}{2} h^{\alpha \mu} \partial_\alpha h
						+ \partial_\alpha \big( h^{\mu \nu} h {}_\nu {}^\alpha \big)
					\Bigg\}
					+ O(\kappa^3),
		\end{eqn}
	from which we can read off
		\begin{eqn}
			J_2^\mu
				= \sfrac{1}{2} \big( h_{\alpha \beta} \partial^\mu h^{\alpha \beta}
					+ h^{\alpha \mu} \partial_\alpha h \big)
					- \partial_\alpha \big( h {}^{\alpha \beta} h {}_\beta {}^\mu \big),
		\end{eqn}
		in agreement with eq. (\ref{eq:J2-K1-def}). In terms of $J_2$ we then have
		\begin{eqn}
			K_2
				&= J_2^\mu \partial_\mu - h^{\mu \alpha} h {}_\alpha {}^\nu \partial_\mu \partial_\nu,
		\end{eqn}
	and the general condition (\ref{eq:general-condition-coordinate-scalar-expansion}) at this order gives us the differential equation (\ref{eq:X2-condition}) for $\mathsf{X}_2$,
		\begin{eqn}
			\bar{\grad}^2 \mathsf{X}_2^\mu
				&= K_1 \mathsf{X}_1^\mu
					+ K_2 \mathsf{X}_0^\mu
				= K_1 \mathsf{X}_1^\mu
					+ J_2^\mu,
		\end{eqn}
	again using the fact that $\mathsf{X}_0^\mu = x^\mu$.	

\subsubsection{Toy: inverting a perturbatively-constructed function}
\label{sec:toy-inverting-r-function}

In sec. \ref{sec:CoordScalarsOfBackgroundCoords} we obtained a perturbative expression for the coordinate scalars $\mathsf{X}$ as a function of the background coordinates $\mathsf{x}$,
		\begin{eqn}
			\hat{\mathsf{X}}(x)
				&= x + \kappa \hat{\mathsf{X}}_1(x) + \kappa^2 \hat{\mathsf{X}}_2(x)
					+ O(\kappa^3),
		\end{eqn}
	where $\hat{\mathsf{X}} \equiv \mathsf{X} \circ \mathsf{x}^{-1}$ is the background-coordinate representation of $\mathsf{X} : M \to \mathbb{R}^d$ and the $\hat{\mathsf{X}}_a$'s are given in Eqs.~(\ref{eq:X1-result}) and (\ref{eq:X2-result}). However, our goal is to express the tensor fields of a theory, which we know as functions of the background coordinates, in terms of the coordinate scalars, and thus our goal is to invert the relationship $\hat{\mathsf{X}}(x)$ to obtain the background coordinates as a function of the coordinate scalars. In fact, we may obtain this inverse in terms of the same $\hat{\mathsf{X}}_a$'s as above. To make this procedure clear we, in this section, demonstrate the analogous logic as applied to a simple function $\mathbb{R} \to \mathbb{R}$.

Suppose therefore that we have some $f : \mathbb{R} \to \mathbb{R}$, analogous to $\hat{\mathsf{X}}(x)$, which is known to us as a Taylor expansion in some parameter $\kappa$ and which at $O(\kappa^0)$ is the identity map:
		\begin{eqn}
		\label{eq:toy:f-expansion}
			f(x)
				= \sum_a \kappa^a f_a(x)
				&= x + \kappa f_1(x) + \kappa^2 f_2(x) + O(\kappa^3).
		\end{eqn}
	Our goal is to obtain an expression for the inverse of $f$, which we denote $g = f^{-1} : \mathbb{R} \to \mathbb{R}$, in terms of the $f_a$'s. We begin from the fact that $f \circ g$ is the identity map by definition and then use our defining expansion of $f$, evaluated at $g(y)$ (writing an arbitrary element of the domain of $f$ as $x$ and an arbitrary element of its range as $y$):
		\begin{eqn}
			y
				&= (f \circ g)(y)
				= g(y) + \kappa f_1 \big( g(y) \big) + \kappa^2 f_2 \big( g(y) \big) 
					+ O(\kappa^3),
		\end{eqn}
	or
		\begin{eqn}
		\label{eq:toy:g-first-result}
			g(y)
				&= y - \kappa f_1 \big( g(y) \big) - \kappa^2 f_2 \big( g(y) \big)
					+ O(\kappa^3).
		\end{eqn}
	We can systematically eliminate the explicit dependence on the unknown $g$ on the right hand side as follows. The above tells us that at $O(\kappa^0)$ the function $g$ is just the identity map:
		\begin{eqn}
		\label{eq:toy:g-zero-order}
			g(y)
				&= y + O(\kappa).
		\end{eqn}
	The full function $g$ contains $O(\kappa^n)$ terms for, in principle, arbitrarily large $n \geq 0$, so the superficially $O(\kappa)$ term in Eq.~(\ref{eq:toy:g-first-result}), $- \kappa f_1 \big( g(y) \big)$, in fact contributes at all orders $n \geq 1$. However by using eq. (\ref{eq:toy:g-zero-order}) in the argument we may explicitly isolate the $O(\kappa)$ contribution:
		\begin{eqn}
			f_1 \big( g(y) \big)
				&= f_1(y) + O(\kappa),
		\end{eqn}
	which yields an explicit expression for $g(y)$ up to $O(\kappa)$ in terms of the (assumed known) $f_a$'s:
		\begin{eqn}
		\label{eq:toy:g-first-order}
			g(y)
				&= y - \kappa f_1 (y) + O(\kappa^2).
		\end{eqn}
	Now that we know $g(y)$ to $O(\kappa)$ we may isolate the $O(\kappa^2)$ contribution to $g(y)$ in a similar manner from the $O(\kappa)$ contribution to $f_1 \big( g(y) \big)$ and the $O(\kappa^0)$ contribution to $f_2 \big(g(y) \big)$. For the former we find
		\begin{eqn}
			f_1 \big(g(y) \big)
				&= f_1 \big( y - \kappa f_1(y) \big) + O(\kappa^2)
				= f_1(y) - \kappa f_1(y) f_1'(y) + O(\kappa^2),
		\end{eqn}
	and for the latter
		\begin{eqn}
			f_2 \big( g(y) \big)
				&= f_2 (y) + O(\kappa),
		\end{eqn}
	which yields to $O(\kappa^2)$
		\begin{eqn}
		\label{eq:toy:g-2nd-order}
			g(y)
				&= y - \kappa f_1(y)
					+ \kappa^2 \bigg( f_1(y) f_1'(y) - f_2(y) \bigg)
					+ O(\kappa^3).
		\end{eqn}
	This procedure may in principle be continued to arbitrary order in $\kappa$ to obtain an expression for $g(y)$ in terms of the expansion functions $f_a$, although for our purposes $O(\kappa^2)$ is sufficient. Once the calculation has been performed to some $O(\kappa^n)$ an explicit calculation will verify that $g = f^{-1}$ to that same order. For example combining eqs. (\ref{eq:toy:f-expansion}) and (\ref{eq:toy:g-2nd-order}) yields the expected results
		\begin{eqn}
			(f \circ g)(y)
				&= y + O(\kappa^3),
		&
			(g \circ f)(x)
				&= x + O(\kappa^3).
		\end{eqn}
Note that in eq. (\ref{eq:toy:g-2nd-order}) the functions $f_a(y)$ are the exact same as the functions $f_a(x)$ that appear in eq. (\ref{eq:toy:f-expansion}) -- if $f_2(x) = x^2$ in the latter, then $f_2(y) = y^2$ in the former. This observation turns out to be useful in the next subsection.

\subsubsection{The background scalars as a function of the background coordinates}
\label{sec:BackgroundCoordsOfCoordScalars}

We return to the task at hand. We have an expression for the coordinate scalars $\mathsf{X} : M \to \mathbb{R}^d$ as a function of the background coordinates $\mathsf{x} : M \to \mathbb{R}^d$,
		\begin{eqn}
		\label{eq:CoordsOfBackgroundRestatement}
			\hat{\mathsf{X}}(x)
				&= x + \kappa \hat{\mathsf{X}}_1 (x) + \kappa^2 \hat{\mathsf{X}}_2 (x)
					+ O(\kappa^3),
		\end{eqn}
	where $\hat{\mathsf{X}} = \mathsf{X} \circ \mathsf{x}^{-1}$; the italic $x$ is an arbitrary value of the background coordinates; and the functions $\hat{\mathsf{X}}_a(x)$ are given in Eqs.~(\ref{eq:X1-result}) and (\ref{eq:X2-result}). Our goal is to obtain an expression $\hat{\mathsf{x}} = \mathsf{x} \circ \mathsf{X}^{-1}$ for the background coordinates as a function of the coordinate scalars, i.e. to invert $\hat{\mathsf{X}}(x)$ for $\hat{\mathsf{x}}(X)$, where the italic $X$ is an arbitrary value of the coordinate scalars. This is hardly any more complicated than the toy calculation of the previous section! Since we are working entirely in terms of the coordinate representations $\hat{\mathsf{X}} = \mathsf{X} \circ \mathsf{x}^{-1} : x \in \mathbb{R}^d \mapsto X \in \mathbb{R}^d$ and $\hat{\mathsf{x}} = \mathsf{x} \circ \mathsf{X}^{-1} : X \in \mathbb{R}^d \mapsto x \in \mathbb{R}^d$, the problem is simply the $d$-dimensional generalization of the previous and is independent of the geometrical origins of these functions.

We therefore follow the exact same steps as in Sec.~\ref{sec:toy-inverting-r-function}, which we outline here in brief. Starting from Eq.~(\ref{eq:CoordsOfBackgroundRestatement}) we use the fact that $X = \big( \hat{\mathsf{X}} \circ \hat{\mathsf{x}} \big)(X)$ to obtain
		\begin{eqn}
			\hat{\mathsf{x}}^\mu(X)
				&= X^\mu - \kappa \hat{\mathsf{X}}_1^\mu \big( \hat{\mathsf{x}}(X) \big)
					- \kappa^2 \hat{\mathsf{X}}_2^\mu \big( \hat{\mathsf{x}}(X) \big)
					+ O(\kappa^3),
		\end{eqn}
	analogous to Eq.~(\ref{eq:toy:g-first-result}). From this we have 
		\begin{eqn}
			\hat{\mathsf{x}}^\mu(X) 
				= X^\mu + O(\kappa),
		\end{eqn}
	analogous to Eq.~(\ref{eq:toy:g-zero-order}), using which in the $O(\kappa)$ term yields
		\begin{eqn}
			\hat{\mathsf{x}}^\mu(X)
				&= X^\mu - \kappa \hat{\mathsf{X}}_1^\mu(X)
					+ O(\kappa^2),
		\end{eqn}
	analogous to Eq.~(\ref{eq:toy:g-first-order}), and using which in turn in the $O(\kappa)$ and $O(\kappa^2)$ terms yields
		\begin{eqn}
		\label{eq:BackgroundCoordsOfScalars-2nd-order}
			\hat{\mathsf{x}}^\mu(X)
				&= X^\mu - \kappa \hat{\mathsf{X}}_1^\mu (X)
					+ \kappa^2 \Bigg( 
						\hat{\mathsf{X}}_1^\alpha (X)
							\pdv{\hat{\mathsf{X}}_1^\mu}{x^\alpha} (X)
						- \hat{\mathsf{X}}_2^\mu(X) \Bigg)
					+ O(\kappa^3),
		\end{eqn}
	analogous to Eq.~(\ref{eq:toy:g-2nd-order}).

It is here that the point raised at the end of the previous subsection becomes important. Recall that in the toy model we have emphasized that the $f_a(y)$'s that appear in the expansion of $g(y)$ have the same functional dependence on $y$ as the $f_a(x)$'s have on $x$ in the expansion of $f(x)$. In the exact same way, the $\hat{\mathsf{X}}_a^\mu(X)$'s that appear in the above expansion of $\hat{\mathsf{x}}^\mu(X)$ have the same functional dependence on the coordinate scalar value $X$ as the $\hat{\mathsf{X}}_a^\mu(x)$'s have on the background coordinate value $x$ in the expansion of $\hat{\mathsf{X}}^\mu(x)$. In other words there is {\em nothing} implicit in eq. (\ref{eq:BackgroundCoordsOfScalars-2nd-order}): $\hat{\mathsf{X}}_1^\mu(X)$ (for instance) means the function $\hat{\mathsf{X}}_1^\mu : \mathbb{R}^d \to \mathbb{R}$ evaluated at $X \in \mathbb{R}^d$, and nothing more. In particular, even though $X$ represents an arbitrary value of the {\em coordinate scalars,} we are feeding it directly into $\hat{\mathsf{X}}_1^\mu = \mathsf{X}_1^\mu \circ \mathsf{x}^{-1}$ in the slot where we would expect to put a value of the {\em background coordinates.} While this may not feel right, it is in fact critical to the usefulness of this whole construction -- we have explicit expressions for $\hat{\mathsf{X}}_1^\mu$ and $\hat{\mathsf{X}}_2^\mu$ in eqs. (\ref{eq:X1-result}) and (\ref{eq:X2-result}) as functions of the background coordinates, and Eq.~(\ref{eq:BackgroundCoordsOfScalars-2nd-order}) tells us how to use these results, with the desired value of the coordinate scalars playing the role of the background coordinates, to obtain (a second-order approximation of) the value of the background coordinates that corresponds to that value of the coordinate scalars.

\subsubsection{Derivatives of and with respect to the background coordinates and the coordinate scalars}

We have constructed two coordinate systems on spacetime: the background coordinates $\mathsf{x} : M \to \mathbb{R}^d$, $p \mapsto \mathsf{x}(p) = x$, and the coordinate scalars $\mathsf{X} : M \to \mathbb{R}^d$, $p \mapsto \mathsf{X}(p) = X$. In this section we carefully discuss the basis frames of each of these coordinate systems.

It is important to keep in mind for this discussion that we are engaging in a slight abuse of notation here: namely, in this work the lowercase italic symbol $x$ refers both to a generic value of the background coordinates and to the canonical coordinates on $\mathbb{R}^d$ themselves. This is directly relevant in the construction of the basis frames as follows. 
The basis frame of any coordinate system $\mathsf{x} : M \to \mathbb{R}^d$ is given by the pushforward by $\mathsf{x}^{-1}$ of the canonical coordinate frame on $\mathbb{R}^d$:
		\begin{eqn}
		\label{eq:background-basis-frame}
			\partial_\mu
				&= \big( \mathsf{x}^{-1} \big)_* \pdv{x^\mu}
		\implies
			\big( \partial_\mu f \big)_p
				= \pdv{\big( f \circ \mathsf{x}^{-1} \big)}{x^\mu} 
					\big( \mathsf{x}(p) \big).
		\end{eqn}
	Changing the coordinate system whose frame you are interested in does {\em not} change the basis frame on $\mathbb{R}^d$ which you push forward -- it only changes the map $\mathsf{x}^{-1}$ by which you push it forward. Thus the basis frame of the coordinate scalars is the pushforward of the {\em same} coordinate frame $\pdv*{x^\mu} \in X(\mathbb{R}^d)$, just by $\mathsf{X}^{-1}$ this time:\footnote{We use $D_\mu$ for the basis from of the coordinate scalars in keeping with the general theme of ``lowercase for background, uppercase for scalars''. Note that $D_\mu$ does {\em not} in this work refer to the gauge covariant derivative of some Yang-Mills theory.}
		\begin{eqn}
		\label{eq:scalars-basis-frame}
			D_\mu
				&= \big( \mathsf{X}^{-1} \big)_* \pdv{x^\mu}
		\implies
			\big( D_\mu f \big)_p
				= \pdv{\big( f \circ \mathsf{X}^{-1} \big)}{x^\mu} \big( \mathsf{X}(p) \big).
		\end{eqn}
	We make this point to emphasize that the denominator in Eq.~(\ref{eq:scalars-basis-frame}) should {\em not} be a capital $X^\mu$ -- we are differentiating the coordinate scalar representation $f \circ \mathsf{X}^{-1} : \mathbb{R}^d \to \mathbb{R}$ with respect to the {\em same} coordinates on $\mathbb{R}^d$ as those with respect to which we differentiate the background coordinate representation $f \circ \mathsf{x}^{-1} : \mathbb{R}^d \to \mathbb{R}$ in Eq.~(\ref{eq:background-basis-frame}). The only differences are the coordinate representations $f \circ \mathsf{x}^{-1}$ and $f \circ \mathsf{X}^{-1}$ themselves, and the coordinate values $\mathsf{x}(p)$ and $\mathsf{X}(p)$ at which we evaluate the derivatives. 

This is directly relevant to explicit calculations in that, if we did write $\pdv*{X^\mu}$ instead of $\pdv*{x^\mu}$, that would then mistakenly suggest that we need an extra factor of $\pdv*{\hat{\mathsf{X}}^\mu}{x^\nu}$ to relate $D_\mu$ and $\partial_\mu$, and including this extra factor would lead to errors in our calculations. This is especially important when we construct the relational Christoffel symbols -- including an extra $\pdv*{\hat{\mathsf{X}}^\mu}{x^\nu}$ next to the partial derivatives in that construction would then lead to an incorrect invariantized Ricci scalar.

%
%
%
%
%

\subsection{Relational observables}
\label{sec:RelationalObservables}

We now come to the crux of this section: the construction, given any tensor field $C \in \Gamma^k_\ell M$, of a set of corresponding diffeomorphism-invariant observables.

\subsubsection{Defining relational observables}

The {\em relational observable} $\mathcal{C} {}^\mu {}_\nu$ corresponding to any component $C {}^\mu {}_\nu$ of $C$ is defined \cite{Frob:2017lnt, Frob:2021mpb} to be that component in the coordinate system defined by the coordinate scalars:
		\begin{eqn}
			C
				&\equiv {\cal C} {}^\mu {}_\nu D_\mu \otimes \dd \mathsf{X}^\nu.
		\end{eqn}
	If the tensor field has a name then the corresponding set of relational observables is its {\em invariantized} form (e.g. the invariantized metric in Sec.~\ref{sec:InvariantizedMetric}).
	
	In terms of the components $C {}^\mu {}_\nu$ of $C$ in the background coordinates the invariantized form is found by transforming from $\mathsf{x}$ to $\mathsf{X}$ as one would transform between any coordinate systems, namely
		\begin{eqn}
		\label{eq:relational-observable-def}
			\mathcal{C} {}^\mu {}_\nu
				&= \big( \partial_\alpha \mathsf{X}^\mu \big) 
					\big( D_\nu \mathsf{x}^\beta \big)
					C {}^\alpha {}_\beta.
		\end{eqn}
	Note that the above is just the standard rule for the passive transformation of the components of a tensor field, with the coordinate scalars $\mathsf{X}^\mu$ and the corresponding frame $D_\mu$ playing the role of the ``new'' coordinates and frame $\tilde{\mathsf{x}}$ and $\tilde{\partial}_\mu$.

Evaluating Eq.~(\ref{eq:relational-observable-def}) at a point $p \in M$ yields
		\begin{eqn}
		\label{eq:relational-observable-def-pointwise}
			\mathcal{C} {}^\mu {}_\nu(p)
				&= \big( \partial_\alpha \mathsf{X}^\nu \big)_p
					\big( D_\nu \mathsf{x}^\beta \big)_p
						C {}^\alpha {}_\beta (p).
		\end{eqn}
	We rewrite the above more explicitly in terms of functions of the coordinates, starting with the transformation matrices. 
    For the first matrix we find
		\begin{eqn}
		\label{eq:forward-deriv-pointwise}
			\big( \partial_\mu \mathsf{X}^\nu \big)_p
				&= \big( \mathsf{x}^{-1} \big)_* 
					\Bigg( \pdv{x^\mu} \Bigg|_{\mathsf{x}(p)} \Bigg) \mathsf{X}^\nu
				= \pdv{\big( \mathsf{X}^\nu \circ \mathsf{x}^{-1} \big)}{x^\mu} 
					\big( \mathsf{x}(p) \big)
				= \pdv{\hat{\mathsf{X}}^\nu}{x^\mu} \big( \mathsf{x}(p) \big),
		\end{eqn}
	and similarly for the second
		\begin{eqn}
		\label{eq:backward-deriv-pointwise}
			\big( D_\mu \mathsf{x}^\nu \big)_p
				&= \big( \mathsf{X}^{-1} \big)_*
					\Bigg( \pdv{x^\mu} \Bigg|_{\mathsf{X}(p)} \Bigg) \mathsf{x}^\nu
				= \pdv{\hat{\mathsf{x}}^\nu}{x^\mu} \big( \mathsf{X}(p) \big).
		\end{eqn}
	In Sec.~\ref{sec:pert-exp-trans-mats} we expand these transformation matrices in terms of the $\hat{\mathsf{X}}^\mu_a$'s.

In the literature these matrices are often written more concisely as $\pdv*{X^\nu}{x^\mu}$ and $\pdv*{x^\nu}{X^\mu}$ \cite{Frob:2021ore}. However, we emphasize once again that we are not making a mistake by leaving the denominator in the latter lowercase -- in both matrices we differentiate the transition map and its inverse using the {\em same} basis frame on $\mathbb{R}^d$, but we evaluate the matrices at the different coordinate values $\mathsf{x}(p)$ and $\mathsf{X}(p)$. It is this latter difference which is more concisely indicated by the differing denominators in the literature. We make the distinction here to make it clear that there is no extra factor of $\pdv*{X^\mu}{x^\nu}$ needed to relate the derivatives in the two matrices.

We return to the question of writing Eq.~(\ref{eq:relational-observable-def-pointwise}) in terms of functions of the coordinates. Since the invariantized $\mathcal{C} {}^\mu {}_\nu$ is a component of the tensor $C$ in the $\mathsf{X}$ coordinate system, its natural coordinate representation is as a function of the background coordinates:
		\begin{eqn}
			\hat{\mathcal{C}} {}^\mu {}_\nu
				\equiv \mathcal{C} {}^\mu {}_\nu \circ \mathsf{X}^{-1}.
		\end{eqn}
	We should therefore compose both sides of eq. (\ref{eq:relational-observable-def-pointwise}) with $\mathsf{X}^{-1}$:
		\begin{eqn}
			\hat{\mathcal{C}} {}^\mu {}_\nu(X)
				&= \big( \partial_\alpha \mathsf{X}^\mu \big)_{\mathsf{X}^{-1}(X)}
					\big( D_\nu \mathsf{x}^\beta \big)_{\mathsf{X}^{-1}(X)}
					C {}^\alpha {}_\beta \circ \mathsf{X}^{-1}(X).
		\end{eqn}
	From Eqs.~(\ref{eq:forward-deriv-pointwise}) and (\ref{eq:backward-deriv-pointwise}) we can simplify the derivative matrices. For the first we find
		\begin{eqn}
		\label{eq:forward-deriv-coordinate}
			\big( \partial_\mu \mathsf{X}^\nu \big)_{\mathsf{X}^{-1}(X)}
				&= \pdv{\hat{\mathsf{X}}^\nu}{x^\mu} \big( \hat{\mathsf{x}}(X) \big),
		\end{eqn}
	i.e. the $\mu^\text{th}$ derivative of $\hat{\mathsf{X}}^\nu$, evaluated at the background coordinate value of the point with coordinate scalar value $X$. For the second we find
		\begin{eqn}
		\label{eq:backward-deriv-coordinate}
			\big( D_\mu \mathsf{x}^\nu \big)_{\mathsf{X}^{-1}(X)}
				&= \pdv{\hat{\mathsf{x}}^\nu}{x^\mu} (X).
		\end{eqn}
	i.e. the $\mu^\text{th}$ derivative of $\hat{\mathsf{x}}^\nu$, evaluated directly at the coordinate scalar value $X$. Again, there is no mistake in the denominator being lowercase.  Finally, we rewrite $C {}^\alpha {}_\beta \circ \mathsf{X}^{-1}(X)$ in terms of the natural coordinate representation\footnote{For the interested reader we note that it may be straightforwardly verified that this definition of $\hat{C} {}^\mu {}_\nu$ is equivalent to defining $\hat{C} = \big( \mathsf{x}^{-1} \big)^* C \in \Gamma^k_\ell \mathbb{R}^d$ and taking the components of the result in the canonical basis frame and coframe on $\mathbb{R}^d$. (An analogous statement holds for $\hat{\mathcal{C}}$ and $\mathsf{X}$.)} $\hat{C} {}^\mu {}_\nu \equiv C {}^\mu {}_\nu \circ \mathsf{x}^{-1}$ as
		\begin{eqn}
			C {}^\alpha {}_\beta \circ \mathsf{X}^{-1}(X)
				&= \hat{C} {}^\alpha {}_\beta 
					\big( \hat{\mathsf{x}}(X) \big).
		\end{eqn}
	Thus, in terms of the natural coordinate representations, Eq.~(\ref{eq:relational-observable-def-pointwise}) becomes
		\begin{eqn}
			\hat{\mathcal{C}} {}^\mu {}_\nu (X)
				&= \pdv{\hat{\mathsf{X}}^\mu}{x^\alpha} \big( \hat{\mathsf{x}}(X) \big)
					\pdv{\hat{\mathsf{x}}^\beta}{x^\nu} \big( X \big)
					\hat{C} {}^\alpha {}_\beta
						\big( \hat{\mathsf{x}}(X) \big).
		\end{eqn}

%
%
%
%
%

\subsubsection{Perturbative expansion of the transformation matrices}
\label{sec:pert-exp-trans-mats}

To obtain an explicit expression for a relational observable we need the derivative matrices that transform tensor components from the background coordinates $\mathsf{x}$ to the coordinate scalars $\mathsf{X}$.

We begin with the ``forward'' derivative $\partial_\mu \mathsf{X}^\nu$, whose coordinate representation we know from Eq.~(\ref{eq:forward-deriv-coordinate}). To explicitly write it in terms of the $\hat{\mathsf{X}}_a^\mu$'s we start by differentiating the expansion of $\hat{\mathsf{X}}(x)$ in $\kappa$,
		\begin{eqn}
		\label{eq:forward-deriv-1}
			\pdv{\hat{\mathsf{X}}^\nu}{x^\mu} (x)
				&= \delta^\nu_\mu
					+ \kappa \pdv{\hat{\mathsf{X}}_1^\nu}{x^\mu} (x)
					+ \kappa^2 \pdv{\hat{\mathsf{X}}_2^\nu}{x^\mu} (x)
					+ O(\kappa^3).
		\end{eqn}
	Now evaluate the above at $x = \hat{\mathsf{x}}(X)$, using the expansion (\ref{eq:BackgroundCoordsOfScalars-2nd-order}). In fact since the $O(\kappa^0)$ term in eq. (\ref{eq:forward-deriv-1}) is independent of $x$ we only need $\hat{\mathsf{x}}(X)$ to $O(\kappa)$,
		\begin{eqn}
			\hat{\mathsf{x}}(X)
				&= X - \kappa \hat{\mathsf{X}}_1(X)
					+ O(\kappa),
		\end{eqn}
	from which we find
		\begin{eqn}
		\label{eq:forward-deriv}
			\pdv{\hat{\mathsf{X}}^\nu}{x^\mu} \big( \hat{\mathsf{x}}(X) \big)
				&= \delta^\nu_\mu
					+ \kappa \pdv{\hat{\mathsf{X}}_1^\nu}{x^\mu} (X)
					+ \kappa^2
						\Bigg(
							\pdv{\hat{\mathsf{X}}_2^\nu}{x^\mu} (X)
							- \hat{\mathsf{X}}_1^\alpha (X)
								\pdv[2]{\hat{\mathsf{X}}_1^\nu}{x^\alpha}{x^\mu} (X) 
						\Bigg)
					+ O(\kappa^3).
		\end{eqn}
	Note that while on the left hand side of Eq.~(\ref{eq:forward-deriv}) the coordinate scalar value $X$ is converted to a background coordinate value by $\hat{\mathsf{x}}$, there is no such $\hat{\mathsf{x}}$ implicit on the right hand side. For example $\hat{\mathsf{X}}_1^\nu : \mathbb{R}^d \to \mathbb{R}$ is a function of the background coordinates $x \in \mathbb{R}^d$, which we differentiate with respect to the $\mu^\text{th}$ canonical coordinate $x^\mu$ on $\mathbb{R}^d$ to obtain $\pdv*{\hat{\mathsf{X}}_1^\nu}{x^\mu} : \mathbb{R}^d \to \mathbb{R}$, and we then plug the coordinate scalar value $X \in \mathbb{R}^d$ directly into this function.

For the ``backward'' derivative $D_\mu \mathsf{x}^\nu$ we similarly use the coordinate representation (\ref{eq:backward-deriv-coordinate}) and the expansion (\ref{eq:BackgroundCoordsOfScalars-2nd-order}) of $\hat{\mathsf{x}}(X)$ in terms of the $\hat{\mathsf{X}}_a^\mu$'s, from which we obtain
		\begin{eqn}
		\label{eq:backward-deriv}
			\pdv{\hat{\mathsf{x}}^\nu}{x^\mu} (X)
				&= \delta^\nu_\mu
					- \kappa \pdv{\hat{\mathsf{X}}^\nu_1}{x^\mu} (X)
					+ \kappa^2
					\Bigg(
						\hat{\mathsf{X}}_1^\alpha (X)
							\pdv[2]{\hat{\mathsf{X}}_1^\nu}{x^\alpha}{x^\mu} (X)
						+ \pdv{\hat{\mathsf{X}}_1^\alpha}{x^\mu} (X) 
							\pdv{\hat{\mathsf{X}}_1^\nu}{x^\alpha} (X)
						- \pdv{\hat{\mathsf{X}}_2^\nu}{x^\mu} (X)
					\Bigg)
					+ O(\kappa^3).
		\end{eqn}
	Note that it may be straightforwardly checked that the above results satisfy the condition
		\begin{eqn}
			\pdv{x^\nu}{x^\mu} (X)
				&= \pdv{\big(\hat{\mathsf{X}} \circ \hat{\mathsf{x}} \big)^\nu}{x^\mu} (X)
				= \pdv{\hat{\mathsf{X}}^\nu}{x^\alpha} \big(\hat{\mathsf{x}}(X) \big)
					\pdv{\hat{\mathsf{x}}^\alpha}{x^\mu} (X)
		\end{eqn}
	to $O(\kappa^2)$, as they must.

\subsubsection{Invariantized scalars}
\label{sec:InvariantizedScalars}

The simplest example of a relational observable is the invariantized form $\Phi = \phi \circ \mathsf{X}^{-1} : \mathbb{R}^d \to \mathbb{R}$ of a real scalar field $\phi : M \to \mathbb{R}$. Our goal is to obtain an explicit expression for $\Phi$ entirely in terms of quantities that are known in the background coordinate system, namely:
\begin{itemize}

	\item the coordinate representation of the scalar field, $\hat{\phi} = \phi \circ \mathsf{x}^{-1} : \mathbb{R}^d \to \mathbb{R}$.
	
	\item the coordinate representation of the perturbative expansion of the coordinate scalars, i.e. the $\hat{\mathsf{X}}_a^\mu$'s. 

\end{itemize}

\noindent
Before proceeding we note that for the scalar field we have only three distinct quantities -- the original scalar $\phi : M \to \mathbb{R}$, the background coordinate representation $\hat{\phi} = \phi \circ \mathsf{x}^{-1}$, and the invariantized scalar $\Phi = \phi \circ \mathsf{X}^{-1}$, which we are here conflating with its own coordinate representation. This is in contrast with a tensor field of nontrivial rank, for which there are four distinct quantities -- the original tensor components $C {}^\mu {}_\nu : M \to \mathbb{R}$, the background coordinate representation $\hat{C} {}^\mu {}_\nu = C {}^\mu {}_\nu \circ \mathsf{x}^{-1}$ of those components, the invariantized components $\mathcal{C} {}^\mu {}_\nu : M \to \mathbb{R}$, and the coordinate representation $\hat{\mathcal{C}} {}^\mu {}_\nu = \mathcal{C} {}^\mu {}_\nu \circ \mathsf{X}^{-1}$ of those invariantized components. For the scalar we may conflate the latter two simply because a scalar field does not have different components in different coordinate systems, so the only new quantity introduced by the relational program is the coordinate representation $\Phi = \phi \circ \mathsf{X}^{-1}$ of the scalar field with respect to the coordinate scalars.

We use the fact that $\mathsf{X}^{-1} = \mathsf{x}^{-1} \circ \hat{\mathsf{x}}$ to write
		\begin{eqn}
			\Phi
				&= \hat{\phi} \circ \hat{\mathsf{x}},
		\end{eqn}
	and expand $\Phi(X)$ using the expansion (\ref{eq:BackgroundCoordsOfScalars-2nd-order}) of $\hat{\mathsf{x}}(X)$ in terms of the $\hat{\mathsf{X}}_a^\mu$'s:
		\begin{eqn}
		\label{eq:invariantized-scalar-expansion}
			\Phi(X)
				&= \hat{\phi} \Bigg(
					X - \kappa \hat{\mathsf{X}}_1 (X) + 
						\kappa^2 \Bigg[
							\hat{\mathsf{X}}_1^\alpha(X) 
								\pdv{\hat{\mathsf{X}}_1}{x^\alpha} (X)
							- \hat{\mathsf{X}}_2(X)
						\Bigg]
					 \Bigg)
					+ O(\kappa^3)
		\\
				&= \hat{\phi}
					- \kappa \hat{\mathsf{X}}_1^\alpha \pdv{\hat{\phi}}{x^\alpha}
					+ \kappa^2 \Bigg(
							\frac{1}{2} \hat{\mathsf{X}}_1^\alpha
								\hat{\mathsf{X}}_1^\beta
								\pdv[2]{\hat{\phi}}{x^\alpha}{x^\beta}
							+ \hat{\mathsf{X}}_1^\alpha
								\pdv{\hat{\mathsf{X}}_1^\beta}{x^\alpha}
								\pdv{\hat{\phi}}{x^\beta}
							- \hat{\mathsf{X}}_2^\alpha \pdv{\hat{\phi}}{x^\alpha}
						\Bigg)
					+ O(\kappa^3),
		\end{eqn}
	where every quantity in the last line is evaluated at the coordinate scalar value $X$.  This agrees with the expression found in \cite{Frob:2017lnt}.

Note that the above applies to {\em any} scalar field, including one that is built out of a tensor or tensors of nontrivial rank. In particular, the invariantized Ricci scalar $\mathcal{R}(X)$ is obtained from the coordinate representation $\hat{R} = R \circ \mathsf{x}^{-1}$ of the Ricci scalar $R$ in the exact same way,
		\begin{eqn}
		\label{eq:invariant-ricci-as-scalar-field}
			\mathcal{R}(X)
				&= \hat{R} - \kappa \hat{\mathsf{X}}_1^\alpha 
					\pdv{\hat{R}}{x^\alpha} 
					+ \kappa^2 \Bigg(
							\frac{1}{2} \hat{\mathsf{X}}_1^\alpha
								\hat{\mathsf{X}}_1^\beta
								\pdv[2]{\hat{R}}{x^\alpha}{x^\beta}
							+ \hat{\mathsf{X}}_1^\alpha
								\pdv{\hat{\mathsf{X}}_1^\beta}{x^\alpha}
								\pdv{\hat{R}}{x^\beta}
							- \hat{\mathsf{X}}_2^\alpha \pdv{\hat{R}}{x^\alpha}
						\Bigg)
					+ O(\kappa^3),
		\end{eqn}
	every quantity on the right hand side again being evaluated at the coordinate scalar value $X$. In Secs. \ref{sec:InvariantizedMetric} and \ref{sec:InvariantizedRicci} we verify this result in the context of perturbation theory by properly constructing the invariantized metric and the resulting Christoffel symbols.

\subsubsection{The invariantized metric}
\label{sec:InvariantizedMetric}

In this section we obtain the explicit expansions of the invariantized metric and its inverse, whose coordinate representations are given by
		\begin{eqn}
		\label{eq:invariant-met-metinv-def}
			\hat{\mathcal{G}}_{\mu \nu} (X)
				&= \pdv{\hat{\mathsf{x}}^\alpha}{x^\mu} (X)
					\pdv{\hat{\mathsf{x}}^\beta}{x^\nu} (X)
					\hat{g}_{\alpha \beta} \big( \hat{\mathsf{x}} (X) \big),
		&
			\hat{\mathcal{G}}^{\mu \nu}(X)
				&= \pdv{\hat{\mathsf{X}}^\mu}{x^\alpha} \big( \hat{\mathsf{x}}(X) \big)
					\pdv{\hat{\mathsf{X}}^\nu}{x^\beta} \big( \hat{\mathsf{x}}(X) \big)
					\hat{g}^{\alpha \beta} \big( \hat{\mathsf{x}}(X) \big),
		\end{eqn}
	in which $\hat{g}_{\mu \nu} = g_{\mu \nu} \circ \mathsf{x}^{-1}$ is the background coordinate representation of the components of $g_{\mu\nu}$ and analogously for $\hat{g}^{\mu \nu}$ and $g_{\mu\nu}^{-1}$.

These calculations are somewhat more complicated than the analogous calculation for the invariantized scalar field $\Phi$. Recall that in the scalar case we needed only to evaluate the background coordinate representation $\hat{\phi}$ at the background coordinate value $\hat{\mathsf{x}}(X)$ of the spacetime point whose coordinate scalar value is $X$ and use our known expansion of $\hat{\mathsf{x}}(X)$ to obtain an expansion of $\Phi$ in terms of quantities that are known in the background coordinates. We still need to do that when we invariantize the metric and its inverse -- that is how we handle the $\hat{g}_{\alpha \beta} \big( \hat{\mathsf{x}} (X) \big)$ and $\hat{g}^{\alpha \beta} \big( \hat{x} (X) \big)$ factors -- but we also need to multiply that result by expansions of the transformation matrices, which are given by Eqs.~(\ref{eq:forward-deriv}) and (\ref{eq:backward-deriv}).

This process is simplified by the fact that we are interested in obtaining expressions for $\hat{\mathcal{G}}_{\mu \nu}$ and $\hat{\mathcal{G}}^{\mu \nu}$ not in terms of the full metric $g_{\mu \nu}$ but the metric perturbation $h_{\mu \nu}$, in terms of which the metric and its inverse are
		\begin{eqn}
		\label{eq:metric-expansion}
			\hat{g}_{\alpha \beta}
				&= \eta_{\alpha \beta} + \kappa \hat{h}_{\alpha \beta},
		&
			\hat{g}^{\alpha \beta}
				&= \eta^{\alpha \beta} - \kappa \hat{h}^{\alpha \beta}
					+ \kappa^2 \hat{h}^{\alpha \sigma} \hat{h} {}_\sigma {}^\beta
					+ \mathcal{O}(\kappa^3).
		\end{eqn}
	Thus, expanding $\hat{g}_{\alpha \beta}\big( \hat{\mathsf{x}}(X) \big)$ and $\hat{g}^{\alpha \beta}\big( \hat{\mathsf{x}}(X) \big)$ in $\kappa$ consists of two steps: first, apply the expansion of the argument, which proceeds identically to the steps which led to the invariantized scalar field (\ref{eq:invariantized-scalar-expansion}) and hence yields identical results but with $\hat{g}_{\alpha \beta}$ and $\hat{g}^{\alpha \beta}$ in place of $\hat{\phi}$; and second, apply the expansions (\ref{eq:metric-expansion}). This latter step simplifies things a great deal, since (a) all partial derivatives of $\eta_{\alpha \beta}$ vanish and (b) we need only keep the terms up to $O(\kappa)$ in eq. (\ref{eq:metric-expansion}) when calculating the $O(\kappa)$ terms in eq. (\ref{eq:invariantized-scalar-expansion}), and even better we need only keep the $O(\kappa^0)$ terms in the former -- whose derivatives, again, vanish -- when calculating the $O(\kappa^2)$ terms in the latter, meaning that all the terms in brackets in eq. (\ref{eq:invariantized-scalar-expansion}) actually vanish. We are left with the results
		\begin{eqn}
		\label{eq:expanded-metric-inv}
			\hat{g}_{\alpha \beta} \big( \hat{\mathsf{x}}(X) \big)
				&= \eta_{\alpha \beta}
					+ \kappa \hat{h}_{\alpha \beta}
					- \kappa^2 \hat{\mathsf{X}}_1^\sigma 
						\pdv{\hat{h}_{\alpha \beta}}{x^\sigma}
					+ O(\kappa^3),
		\\
			\hat{g}^{\alpha \beta} \big( \hat{\mathsf{x}}(X) \big)
				&= \eta^{\alpha \beta}
					- \kappa \hat{h}^{\alpha \beta}
					+ \kappa^2 \Bigg( \hat{h}^{\alpha \sigma} \hat{h} {}_\sigma {}^\beta
						+ \hat{\mathsf{X}}_1^\sigma \pdv{\hat{h}^{\alpha \beta}}{x^\sigma}
						\Bigg)
					+ O(\kappa^3),
		\end{eqn}
	again with all quantities on the right hand sides evaluated directly at $X$. As a check it's straightforward to verify that the above satisfy $\hat{g}_{\alpha \sigma} \hat{g}^{\sigma \beta} = \delta_\alpha^\beta + O(\kappa^3)$.

We are not yet done, however.  It remains to plug these results into the definitions (\ref{eq:invariant-met-metinv-def}) of the invariantized metric and inverse metric and apply the expansions (\ref{eq:forward-deriv}) and (\ref{eq:backward-deriv}) of the derivative matrices.  In the interest of clarity and brevity we omit the intermediate steps and organize the results by defining
		\begin{eqn}
		\label{eq:invariant-metric-expansion-def}
			\hat{\mathcal{G}}_{\mu \nu}
				&\equiv \sum_a \kappa^a \hat{\mathcal{G}}^a_{\mu \nu},
		&
			\hat{\mathcal{G}}^{\mu \nu}
				&\equiv \sum_a \kappa^a \hat{\mathcal{G}}_a^{\mu \nu},
		\end{eqn}
	in terms of which we find at zeroth order
		\begin{eqn}
			\hat{\mathcal{G}}^0_{\mu \nu}
				&= \eta_{\mu \nu},
		&
			\hat{\mathcal{G}}_0^{\mu \nu}
				&= \eta^{\mu \nu};
		\end{eqn}
	at first order
		\begin{eqn}
			\hat{\mathcal{G}}^1_{\mu \nu}
				&= \hat{h}_{\mu \nu}
					- \pdv{\hat{\mathsf{X}}_{1\nu}}{x^\mu}
					- \pdv{\hat{\mathsf{X}}_{1\mu}}{x^\nu},
		&
			\hat{\mathcal{G}}^{\mu \nu}_1
				&= - \Bigg( \hat{h}^{\mu \nu}
					- \pdv{\hat{\mathsf{X}}_1^\nu}{x_\mu}
					- \pdv{\hat{\mathsf{X}}_1^\mu}{x_\nu} \Bigg);
		\end{eqn}
	and at second order
		\begin{eqn}
		\label{eq:invariant-metric-second-order}
			\hat{\mathcal{G}}^2_{\mu \nu}
				&= 
					\Bigg\{ 
						\pdv{\hat{\mathsf{X}}_1^\sigma}{x^\mu} 
							\pdv{\hat{\mathsf{X}}_{1 \nu}}{x^\sigma}
						+ \pdv{\hat{\mathsf{X}}_1^\sigma}{x^\nu}
							\pdv{\hat{\mathsf{X}}_{1 \mu}}{x^\sigma}
						+ \hat{\mathsf{X}}_1^\sigma
							\pdv[2]{\hat{\mathsf{X}}_{1 \nu}}{x^\sigma}{x^\mu}
						+ \hat{\mathsf{X}}_1^\sigma
							\pdv[2]{\hat{\mathsf{X}}_{1 \mu}}{x^\sigma}{x^\nu}
						+ \pdv{\hat{\mathsf{X}}_1^\sigma}{x^\mu} 
							\pdv{\hat{\mathsf{X}}_{1 \sigma}}{x^\nu} 
					\Bigg\}
		\\
				&\hspace{5cm}
					- \Bigg\{ 
						\pdv{\hat{\mathsf{X}}_{2 \nu}}{x^\mu}
						+ \pdv{\hat{\mathsf{X}}_{2 \mu}}{x^\nu}
					\Bigg\}
					- \Bigg\{
						\hat{\mathsf{X}}_1^\sigma \pdv{\hat{h}_{\mu \nu}}{x^\sigma}
						+ \hat{h}_{\mu \sigma} \pdv{\hat{\mathsf{X}}_1^\sigma}{x^\nu}
						+ \hat{h}_{\sigma \nu} \pdv{\hat{\mathsf{X}}_1^\sigma}{x^\mu}
					\Bigg\},
		\\
			\hat{\mathcal{G}}_2^{\mu \nu}
				&= 
					\Bigg\{
						\pdv{\hat{\mathsf{X}}_1^\mu}{x^\sigma}
							\pdv{\hat{\mathsf{X}}_1^\nu}{x_\sigma}
						- \hat{\mathsf{X}}_1^\sigma
							\pdv[2]{\hat{\mathsf{X}}_1^\nu}{x^\sigma}{x_\mu}
						- \hat{\mathsf{X}}_1^\sigma
							\pdv[2]{\hat{\mathsf{X}}_1^\mu}{x^\sigma}{x_\nu}
					\Bigg\}
					+ \Bigg\{
						\pdv{\hat{\mathsf{X}}_2^\nu}{x_\mu}
						+ \pdv{\hat{\mathsf{X}}_2^\mu}{x_\nu}
					\Bigg\}
		\\
				&\hspace{5cm}
					+ \Bigg\{
						\hat{h}^{\mu \sigma} \hat{h} {}_\sigma {}^\nu
						+ \hat{\mathsf{X}}_1^\sigma \pdv{\hat{h}^{\mu \nu}}{x^\sigma}
						- \hat{h}^{\mu \sigma} \pdv{\hat{\mathsf{X}}_1^\nu}{x^\sigma}
						- \hat{h}^{\sigma \nu} \pdv{\hat{\mathsf{X}}_1^\mu}{x^\sigma}
					\Bigg\},
		\end{eqn}
	where we use brackets to separate distinct classes of terms (those quadratic in $\hat{\mathsf{X}}_1$, those linear in $\hat{\mathsf{X}}_2$, and those containing at least one factor of $\hat{h}$).
	
	It may be verified that the above do indeed satisfy $\hat{\mathcal{G}}_{\mu \alpha} \hat{\mathcal{G}}^{\alpha \nu} = \delta_\mu^\nu$, as they must. Additionally, given the expansion coefficients $\hat{\mathcal{G}}^a_{\mu \nu}$ for the invariantized metric and the fact that at zeroth order both $\hat{\mathcal{G}}_{\mu \nu}$ and $\hat{\mathcal{G}}^{\mu \nu}$ are flat, it is straightforward to show that the expansion coefficients $\hat{\mathcal{G}}_a^{\mu \nu}$ for the invariantized inverse metric are given by
			\begin{eqn}
				\hat{\mathcal{G}}_1^{\mu \nu}
					&= - \hat{\mathcal{G}}^{1 \mu \nu},
			&
				\hat{\mathcal{G}}_2^{\mu \nu}
					&= \hat{\mathcal{G}}^{1 \mu \alpha} \hat{\mathcal{G}}^1_\alpha {}^\nu
						- \hat{\mathcal{G}}^{2 \mu \nu},
			\end{eqn}
		and it may be shown that these relations are satisfied by the above.

\subsubsection{The invariantized metric perturbation and volume factor}

From the invariantized metric we can immediately define the invariantized metric perturbation \cite{Frob:2021mpb} to be the metric perturbation in the coordinate system $\mathsf{X}$:
		\begin{eqn}
			\mathcal{G}_{\mu \nu}
				&= \eta_{\mu \nu} + \kappa \mathcal{H}_{\mu \nu},
		&
			&\text{i.e.}
		&
			\mathcal{H}_{\mu \nu}
				&= \frac{1}{\kappa} \big( \mathcal{G}_{\mu \nu} - \eta_{\mu \nu} \big).
		\end{eqn}
	In terms of the metric expansion coefficients defined above the coordinate representation of the invariantized metric perturbation is therefore
		\begin{eqn}\label{eq:invariantized-metpert}
			\hat{\mathcal{H}}_{\mu \nu}
				&= \hat{\mathcal{G}}^1_{\mu \nu}
					+ \kappa \hat{\mathcal{G}}^2_{\mu \nu}
					+ O(\kappa^2)
		\\
				&= \Bigg\{ \hat{h}_{\mu \nu}
						- \pdv{\hat{\mathsf{X}}_{1\nu}}{x^\mu}
						- \pdv{\hat{\mathsf{X}}_{1\mu}}{x^\nu} \Bigg\}
					+ \kappa 
					\Bigg\{ 
						\pdv{\hat{\mathsf{X}}_1^\sigma}{x^\mu} 
							\pdv{\hat{\mathsf{X}}_{1 \nu}}{x^\sigma}
						+ \pdv{\hat{\mathsf{X}}_1^\sigma}{x^\nu}
							\pdv{\hat{\mathsf{X}}_{1 \mu}}{x^\sigma}
						+ \hat{\mathsf{X}}_1^\sigma
							\pdv[2]{\hat{\mathsf{X}}_{1 \nu}}{x^\sigma}{x^\mu}
						+ \hat{\mathsf{X}}_1^\sigma
							\pdv[2]{\hat{\mathsf{X}}_{1 \mu}}{x^\sigma}{x^\nu}
						+ \pdv{\hat{\mathsf{X}}_1^\sigma}{x^\mu} 
							\pdv{\hat{\mathsf{X}}_{1 \sigma}}{x^\nu} 
		\\
				&\hspace{6cm}
					- 
						\pdv{\hat{\mathsf{X}}_{2 \nu}}{x^\mu}
						- \pdv{\hat{\mathsf{X}}_{2 \mu}}{x^\nu}
					- 
						\hat{\mathsf{X}}_1^\sigma \pdv{\hat{h}_{\mu \nu}}{x^\sigma}
						- \hat{h}_{\mu \sigma} \pdv{\hat{\mathsf{X}}_1^\sigma}{x^\nu}
						- \hat{h}_{\sigma \nu} \pdv{\hat{\mathsf{X}}_1^\sigma}{x^\mu}
					\Bigg\}
				+ O(\kappa^2).
		\end{eqn}
	In keeping with the convention that $h^{\mu \nu} = \eta^{\mu \alpha} \eta^{\nu \beta} h_{\alpha \beta} \neq (g^{\mu \nu} - \eta^{\mu \nu})/\kappa$, we can also define $\mathcal{H}^{\mu \nu} = \eta^{\mu \alpha} \eta^{\nu \beta} \mathcal{H}_{\alpha \beta}$, although we do not need to make use of this here.

The invariantized metric perturbation was used in \cite{Frob:2021mpb} to compute gauge-invariant corrections to the Newtonian potential. We use it here to obtain the invariantized volume factor $\sqrt{ - \det \mathcal{G}}$, which we do as follows. In any coordinate system the volume factor $\sqrt{- \det g}$ can be expanded in the metric perturbation $g_{\mu \nu} = \eta_{\mu \nu} + \kappa h_{\mu \nu}$:
		\begin{eqn}\label{eq:standard-volume-expansion-restated}
			\sqrt{ - \det g}
				&= 1 + \sfrac{1}{2} \kappa h
					+ \kappa^2 \bigg( 
						\sfrac{1}{8} h^2 
						- \sfrac{1}{4} h_{\mu \nu} h^{\mu \nu} \bigg)
					+ O(\kappa^3).
		\end{eqn}
 Eq. (\ref{eq:standard-volume-expansion-restated}) provides the expansion of the volume factor $\sqrt{- \det g}$ in terms of the metric perturbation $h_{\mu \nu}$, evaluated in any coordinate system. It therefore follows that the invariantized volume factor is given by the exact same equation, evaluated in the $\mathsf{X}$-coordinate system:
		\begin{eqn}
			\sqrt{- \det \mathcal{G}}
				&= 1 + \sfrac{1}{2} \kappa \mathcal{H}
					+ \kappa^2 \bigg( 
						\sfrac{1}{8} \mathcal{H}^2 
						- \sfrac{1}{4} \mathcal{H}_{\mu \nu} \mathcal{H}^{\mu\nu} \bigg)
					+ O(\kappa^3),
		\end{eqn}
	where $\mathcal{H} = \mathcal{H} {}^\mu {}_\mu$. Using eq. (\ref{eq:invariantized-metpert}) we can write this in terms of $h_{\mu \nu}$ and the $\mathsf{X}$'s as
		\begin{eqn}
		\label{eq:invar-volume-expansion}
			\sqrt{ - \det \mathcal{G}}
				&= 1 + \kappa \bigg( \sfrac{1}{2} h - \partial_\mu \mathsf{X}_1^\mu \bigg)
					+ \kappa^2 \bigg(
						\sfrac{1}{2} \partial_\mu \mathsf{X}_1^\mu \partial_\nu \mathsf{X}_1^\nu
						+ \mathsf{X}_1^\mu \partial_\mu \partial_\nu \mathsf{X}_1^\nu
						+ \sfrac{1}{2} \partial_\mu \mathsf{X}_{1\nu} \partial^\nu \mathsf{X}_1^\mu
						- \partial_\mu \mathsf{X}_2^\mu
		\\
				&\hspace{7cm}
						- \sfrac{1}{2} \mathsf{X}_1^\mu \partial_\mu h
						- \sfrac{1}{2} h \partial_\mu \mathsf{X}_1^\mu
						+ \sfrac{1}{8} h^2
						- \sfrac{1}{4} h_{\mu \nu} h^{\mu \nu}
					\bigg)
					+ O(\kappa^3).
		\end{eqn}

\subsubsection{The invariantized Christoffel symbols and Ricci scalar}
\label{sec:InvariantizedRicci}

Finally let us return to the claim at the end of Sec.~\ref{sec:InvariantizedScalars} that the invariantized Ricci scalar may be obtained from the invariantized metric.

{\sc The Ricci scalar in perturbation theory.} We begin with an expansion of the standard (non-invariantized) Ricci scalar from the expansion of the metric about flat space, $g_{\mu \nu} = \eta_{\mu \nu} + \kappa h_{\mu \nu}$. The Christoffel symbols are
			\begin{eqn}
				\Gamma^\rho_{\mu \nu}
					&= \frac{1}{2} g^{\rho \alpha}
						\big( \partial_\mu g_{\alpha \nu} + \partial_\nu g_{\alpha \mu}
							- \partial_\alpha g_{\mu \nu} \big),
			\end{eqn}
		where $\partial_\mu$ is the frame of the coordinate system in which the metric components are $g_{\mu \nu}$, and the Riemann tensor is
			\begin{eqn}
				R {}^\mu {}_{\nu \rho \sigma}
					&= \partial_\rho \Gamma^\mu_{\sigma \nu}
						- \partial_\sigma \Gamma^\mu_{\sigma \mu}
						+ \Gamma^\mu_{\rho \alpha} \Gamma^\alpha_{\sigma \nu}
						- \Gamma^\mu_{\sigma \alpha} \Gamma^\alpha_{\rho \nu},
			\end{eqn}
		from which the Ricci tensor and Ricci scalar are obtained as
			\begin{eqn}
				R_{\mu \nu}
					&= R {}^\alpha {}_{\mu \alpha \nu},
			&
				R
					&= g^{\mu \nu} R_{\mu \nu}
					= g^{\mu \nu} R {}^\alpha {}_{\mu \alpha \nu}.
			\end{eqn}
		Since every term in the Christoffel symbol contains at least one partial derivative on the metric, and the partial derivatives of $\eta_{\mu \nu}$ vanish, it follows that the Christoffel symbols begin at $O(\kappa)$. It then follows that the Ricci scalar also begins at this order, so that we may write
			\begin{eqn}
				R
					&= \kappa R_1 + \kappa^2 R_2 + O(\kappa^3).
			\end{eqn}
		An explicit calculation yields for the expansion terms
			\begin{eqn}
				R_1
					&= \partial_\mu \partial_\nu h^{\mu \nu}
						- \partial^2 h,
			\\
				R_2
					&= h^{\mu \nu} \partial_\mu \partial_\nu h
						- \frac{1}{4} \partial_\mu h \partial^\mu h
						- \partial_\mu h^{\mu \nu} \partial_\rho h {}_\nu {}^\rho
						+ \partial^\mu h \partial_\nu h {}_\mu {}^\nu
						- 2 h^{\mu \nu} \partial_\nu \partial_\rho h {}_\mu {}^\rho
			\\
					&\hspace{5cm}
						+ h^{\mu \nu} \partial^2 h_{\mu \nu}
						- \frac{1}{2} \partial_\nu h_{\mu \rho} \partial^\rho h^{\mu \nu}
						+ \frac{3}{4} \partial_\rho h_{\mu \nu} \partial^\rho h^{\mu \nu}.
			\end{eqn}

{\sc The invariantized Ricci scalar, obtained as a scalar field.} Using the expansion in Eq.~(\ref{eq:invariant-ricci-as-scalar-field}) yields an expression for the invariantized Ricci scalar in terms of the expansions of both the metric and the coordinate scalars:
			\begin{eqn}
			\label{eq:invar-R-expanded-from-scalar}
				\mathcal{R}
					&= \kappa \hat{R}_1
						+ \kappa^2
							\Bigg(
								\hat{R}_2
									- \hat{\mathsf{X}}_1^\alpha \pdv{\hat{R}_1}{x^\alpha}
							\Bigg)
						+ O(\kappa^3),
			\end{eqn}
		where $\mathcal{R} = R \circ \mathsf{X}^{-1}$ and $\hat{R}_a = R_a \circ \mathsf{x}^{-1}$.
		
		This result does not in itself rely on the fact that $R$ is defined in terms of any higher-rank tensor field -- given any scalar field $\phi$ expressed as an expansion in $\kappa$ and whose $O(\kappa^0)$ contribution vanishes, the invariantized $\phi$ would have this exact same form. In what follows we show that this form may also be obtained by correctly constructing the invariantized curvature tensors from the invariantized metric.

	{\sc A tempting but incorrect derivation from the invariantized metric.}
	Before proceeding to the correct derivation we demonstrate the problem with the formulation that is most tempting in the standard, more concise notation. It is most common to conflate the basis frame $\partial_\mu$ on $M$ corresponding to a coordinate system with the partial derivative $\pdv*{x^\mu}$ with respect to those coordinates, and to conflate a tensor field component $g_{\mu \nu}$ with its coordinate representation $\hat{g}_{\mu \nu}$. In this notation one might then think to construct the Christoffel symbols in the ${\mathsf{X}}$-coordinate system as
		\begin{eqn}
		\label{eq:wrong-invar-christoffel}
			\Gamma^\rho_{\mu \nu}
				&= \frac{1}{2} \mathcal{G}^{\rho \alpha}
					\Bigg( \pdv{\mathcal{G}_{\alpha \nu}}{X^\mu} 
						+ \pdv{\mathcal{G}_{\alpha \mu}}{X^\nu}
						- \pdv{\mathcal{G}_{\mu \nu}}{X^\alpha} \Bigg),
		\end{eqn}
	the Riemann tensor as
		\begin{eqn}
			\mathcal{R} {}^\mu {}_{\nu \rho \sigma}
				&= \pdv{\Gamma^\mu_{\sigma \nu}}{X^\rho}
					- \pdv{\Gamma^\mu_{\rho \nu}}{X^\sigma}
					+ \Gamma^\mu_{\rho \alpha} \Gamma^\alpha_{\sigma \nu}
					- \Gamma^\mu_{\sigma \alpha} \Gamma^\alpha_{\rho \nu},
		\end{eqn}
	and the Ricci tensor and scalar as
		\begin{eqn}
			\mathcal{R}_{\mu \nu}
				&= \mathcal{R} {}^\rho {}_{\mu \rho \nu},
		&
			\mathcal{R}
				&= \mathcal{G}^{\mu \nu} \mathcal{R}_{\mu \nu}.
		\end{eqn}
	If one wished to then expand the Ricci scalar in $\kappa$ one would then rightly use the known expansion of $\mathcal{G}_{\mu \nu}$ and $\mathcal{G}^{\mu \nu}$. 
	
	The problem with this notation is that one would also think that, in order to reduce the expression to one involving only functions we know in the background coordinates -- namely, partial derivatives of background coordinate functions with respect to the background coordinates -- one must also convert the $\pdv*{X^\mu}$'s to $\pdv*{x^\mu}$'s via the chain rule:
		\begin{eqn}
			\Gamma^\rho_{\mu \nu}
				&= \frac{1}{2} \mathcal{G}^{\rho \alpha}
					\Bigg( 
						\pdv{x^\beta}{X^\mu}
							\pdv{\mathcal{G}_{\alpha \nu}}{x^\beta} 
						+ \pdv{x^\beta}{X^\nu}
							\pdv{\mathcal{G}_{\alpha \mu}}{x^\beta}
						- \pdv{x^\beta}{X^\alpha}
							\pdv{\mathcal{G}_{\mu \nu}}{X^\beta}
					\Bigg),
		\\
			\mathcal{R} {}^\mu {}_{\nu \rho \sigma}
				&= 
					\pdv{x^\alpha}{X^\rho}
						\pdv{\Gamma^\mu_{\sigma \nu}}{x^\alpha}
					- \pdv{x^\alpha}{X^\sigma}
						\pdv{\Gamma^\mu_{\rho \nu}}{x^\alpha}
					+ \Gamma^\mu_{\rho \alpha} \Gamma^\alpha_{\sigma \nu}
					- \Gamma^\mu_{\sigma \alpha} \Gamma^\alpha_{\rho \nu}.
		\end{eqn}
	That this construction is incorrect may be seen directly by following it through and observing that the result disagrees with the result (\ref{eq:invar-R-expanded-from-scalar}) obtained from treating $R$ like any other scalar field. This is by itself fatal: all we are really doing in obtaining the invariantized scalar field is transforming from an arbitrary coordinate system $\mathsf{x}$ to the specified coordinate system $\mathsf{X}$, meaning that if the relationship between $R$ and $\mathcal{R}$ differs from the relationship between a generic scalar field $\phi$ and its invariantized form $\Phi$ then the Ricci scalar does not transform like a scalar at all.
	
	The problem with the derivative prescription above may also be seen by considering the actual meaning of a partial derivative with respect to coordinates on a manifold. This is most apparent by comparing to the more careful development below, but we may also understand it as follows. Suppose we have a function $f : M \to \mathbb{R}^d$ and some coordinate system $\mathsf{x} : M \to \mathbb{R}^d$ with frame $\partial_\mu$. If we wish to take the $\mu^\text{th}$ partial derivative of $f$ with respect to this coordinate system then we ``think of $f$ as a function of the coordinates'', i.e. construct its coordinate representation $f \circ \mathsf{x}^{-1}$, and then take the $\mu^\text{th}$ derivative of that function. If we also have another coordinate system $\tilde{\mathsf{x}}$ with frame $\tilde{\partial}_\mu$ and we want to take the $\mu^\text{th}$ partial derivative of $f$ with respect to these other coordinates then we do the same thing: we construct the coordinate representation $f \circ \tilde{\mathsf{x}}^{-1}$ and take its $\mu^\text{th}$ derivative. 
	
	The key point here is that, once we have the coordinate representations $f_\mathsf{x} \equiv f \circ \mathsf{x}^{-1}$ and $f_{\tilde{\mathsf{x}}} \equiv f \circ \tilde{\mathsf{x}}^{-1}$, we do {\em the exact same thing} to each -- we are differentiating these different coordinate representations with respect to {\em the same} coordinates on $\mathbb{R}^d$, and therefore we do not a priori need any extra chain-rule factor to relate the two derivatives. More explicitly, evaluating $\partial_\mu f$ and $\tilde{\partial}_\mu f$ at $p \in M$ such that $\mathsf{x}(p) = x$ and $\tilde{\mathsf{x}}(p) = \tilde{x}$, we have
			\begin{eqn}
				\big( \partial_\mu f \big)_p
					&= \pdv{f_\mathsf{x}}{x^\mu} (x),
			&
				\big( \tilde{\partial}_\mu f \big)_p
					&= \pdv{f_{\tilde{\mathsf{x}}}}{x^\mu} (\tilde{x}).
			\end{eqn}
		Of course we can then relate the two derivatives by the chain rule if we wish by writing $f_{\tilde{\mathsf{x}}}(\tilde{x}) = f_\mathsf{x} \big( \hat{\mathsf{x}}(\tilde{x}) \big)$ with $\hat{\mathsf{x}} = \mathsf{x} \circ \tilde{\mathsf{x}}^{-1}$, so that
			\begin{eqn}
				\big( \tilde{\partial}_\mu f \big)_p
					&= \pdv{f_{\tilde{\mathsf{x}}}}{x^\mu} (\tilde{x})
					= \pdv{\hat{\mathsf{x}}^\alpha}{x^\mu} (\tilde{x})
						\pdv{f_\mathsf{x}}{x^\alpha} \big( \hat{\mathsf{x}}(\tilde{x}) \big)
					= \pdv{\hat{\mathsf{x}}^\alpha}{x^\mu} (\tilde{x})
						\big( \partial_\mu f \big)_p,
			\end{eqn}
		but the expression containing the partial derivative matrix does not also contain the new coordinate representation of the function $f$.
		
		In short, the problem with the intuitive construction (\ref{eq:wrong-invar-christoffel}) is that, implicitly, we are simultaneously including the partial derivative matrix {\em and} differentiating the {\em new} coordinate representation, when really we should be doing one or the other. Thus the correct invariantized Christoffel symbols are
			\begin{eqn}
			\label{eq:correct-christoffel-1}
			\Gamma^\rho_{\mu \nu}
				&= \frac{1}{2} \mathcal{G}^{\rho \alpha}
					\Bigg( \pdv{\mathcal{G}_{\alpha \nu}}{x^\mu} 
						+ \pdv{\mathcal{G}_{\alpha \mu}}{x^\nu}
						- \pdv{\mathcal{G}_{\mu \nu}}{x^\alpha} \Bigg),
			\end{eqn}
		in terms of which the correct invariantized Riemann tensor is
			\begin{eqn}
			\label{eq:correct-riemann-1}
				\mathcal{R} {}^\mu {}_\nu \rho \sigma
					&= \pdv{\Gamma^\mu_{\sigma \nu}}{x^\rho}
						- \pdv{\Gamma^\mu_{\rho \nu}}{x^\sigma}
						+ \Gamma^\mu_{\rho \alpha} \Gamma^\alpha_{\sigma \nu}
						- \Gamma^\mu_{\sigma \alpha} \Gamma^\alpha_{\rho \nu}.
			\end{eqn}
		To more rigorously justify the above results we obtain the above from a more careful construction in which spacetime- and coordinate-dependent objects are not conflated.

{\sc The correct derivation from the invariantized metric.}
	In a general coordinate system $\mathsf{x} : M \to \mathbb{R}^d$ with coordinate frame $\partial_\mu = \big( \mathsf{x}^{-1} \big)_* \big( \pdv*{x^\mu} \big)$ and in which the metric has components $g_{\mu \nu}$ the Christoffel symbols are defined by
			\begin{eqn}
				\Gamma^\rho_{\mu \nu}
					&= \frac{1}{2} g^{\rho \alpha}
						\big( \partial_\mu g_{\alpha \nu} + \partial_\nu g_{\alpha \mu}
							- \partial_\alpha g_{\mu \nu} \big).
			\end{eqn}
		We want to write down the Christoffel symbols in the coordinate system $\mathsf{X} : M \to \mathbb{R}^d$. Thus, not conflating anything and being careful to write $D_\mu = \big( \mathsf{X}^{-1} \big)_* \big( \pdv*{x^\mu} \big)$ for the frame of this coordinate system and $g_{\mu \nu}$ for the metric components, the Christoffel symbols are
			\begin{eqn}
				\Gamma^\rho_{\mu \nu}
					&= \frac{1}{2} \mathcal{G}^{\rho \alpha}
						\big( D_\mu \mathcal{G}_{\alpha \nu} + D_\nu \mathcal{G}_{\alpha \mu}
							- D_\alpha \mathcal{G}_{\mu \nu} \big).
			\end{eqn}
		Note that in the above each metric component $\mathcal{G}_{\mu \nu}$ is a real-valued function of spacetime and hence {\em distinct} from its coordinate representation $\hat{\mathcal{G}}_{\mu \nu} = \mathcal{G}_{\mu \nu} \circ \mathsf{X}^{-1}  
        $, which is a real-valued function of the coordinate scalars. 
		
		To write the Christoffel symbols in this coordinate system as a function of the coordinates let us evaluate at a point $p$. For a representative derivative term we find
			\begin{eqn}
				\big(D_\alpha \mathcal{G}_{\mu \nu} \big)_p
					&= \pdv{\big( \mathcal{G}_{\mu \nu} 
						\circ \mathsf{X}^{-1} \big)}{x^\alpha} \big( \mathsf{X}(p) \big)
					= \pdv{\hat{\mathcal{G}}_{\mu \nu}}{x^\alpha} \big( \mathsf{X}(p) \big),
			\end{eqn}
		meaning that the coordinate representation $\hat{\Gamma}^\rho_{\mu \nu} = \Gamma^\rho_{\mu \nu} \circ \mathsf{X}^{-1}$ of the Christoffel symbols is
			\begin{eqn}
			\label{eq:correct-christoffel}
				\hat{\Gamma}^\rho_{\mu \nu}
					&= \frac{1}{2} \hat{\mathcal{G}}^{\rho \alpha}
						\Bigg( \pdv{\hat{\mathcal{G}}_{\alpha \nu}}{x^\mu}
							+ \pdv{\hat{\mathcal{G}}_{\alpha \mu}}{x^\nu}
							- \pdv{\hat{\mathcal{G}}_{\mu \nu}}{x^\alpha} \Bigg),
			\end{eqn}
		in agreement with eq. (\ref{eq:correct-christoffel-1}). Similarly the invariantized Riemann tensor is
		\begin{eqn}
			\mathcal{R} {}^\mu {}_{\nu \rho \sigma}
				&= D_\rho \Gamma^\mu_{\sigma \nu}
					- D_\sigma \Gamma^\mu_{\rho \nu}
					+ \Gamma^\mu_{\rho \alpha} \Gamma^\alpha_{\sigma \nu}
					- \Gamma^\mu_{\sigma \alpha} \Gamma^\alpha_{\rho \nu},
		\end{eqn}
	which yields the coordinate representation
		\begin{eqn}
		\label{eq:correct-riemann}
			\hat{\mathcal{R}} {}^\mu {}_{\nu \rho \sigma}
				&= \pdv{\hat{\Gamma}^\mu_{\sigma \nu}}{x^\rho}
					- \pdv{\hat{\Gamma}^\mu_{\rho \nu}}{x^\sigma}
					+ \hat{\Gamma}^\mu_{\rho \alpha} \hat{\Gamma}^\alpha_{\sigma \nu}
					- \hat{\Gamma}^\mu_{\sigma \alpha} \hat{\Gamma}^\alpha_{\rho \nu},
		\end{eqn}
	in agreement with eq. (\ref{eq:correct-riemann-1}).
	
The invariantized Ricci scalar is, finally, given by
		\begin{eqn}
			\mathcal{R}(X)
				&= \hat{\mathcal{G}}^{\mu \nu}(X)
					\hat{\mathcal{R}}^\rho {}_{\mu \rho \nu} (X).
		\end{eqn}
	To turn this into an expression for $\mathcal{R}$ in terms of quantities known in the background coordinate system one would (i) use Eqs.~(\ref{eq:correct-christoffel}) and (\ref{eq:correct-riemann}) to write the Riemann tensor in terms of the invariantized metric, yielding an expression for $\mathcal{R}$ entirely in terms of $\mathcal{G}_{\mu \nu}$ and $\mathcal{G}^{\mu \nu}$; and then (ii) use Eqs.~(\ref{eq:invariant-metric-expansion-def}) through (\ref{eq:invariant-metric-second-order}) to expand this result in terms of $\hat{h}_{\mu \nu}$ and the $\hat{\mathsf{X}}_a^\mu$'s, which are themselves given in terms of $\hat{h}_{\mu \nu}$ by Eqs.~(\ref{eq:J1-def}) through (\ref{eq:X2-result}). Doing so confirms that this construction agrees with the result (\ref{eq:invar-R-expanded-from-scalar}) obtained by treating the Ricci scalar like any other scalar field.

\subsection{Summarizing and cleaning up the notation}

Throughout this section we have made a careful distinction between objects defined on spacetime and their coordinate representations to clarify certain subtle points in the construction of relational observables. For the remainder of this work we not need be quite so explicit, and we bring our notation more in line with convention \cite{Frob:2017lnt, Frob:2021ore} as follows.
\begin{itemize}

	\item We conflate functions of spacetime and their coordinate representations, meaning that we drop all the hats, writing, for example, $h_{\mu \nu}(x)$ and $\mathsf{X}(x)$.
	
	\item A partial derivative, e.g. $\partial_\mu$, may denote either the coordinate frame (which acts on functions of spacetime) or the actual partial derivative (which acts on functions of the coordinates).

\end{itemize}

\noindent
Additionally, in the remainder of this paper we make reference to the conformal mode of the metric tensor, which we denote $\Phi$. This should not be confused with the generic invariantized scalar field operator, which does not appear in what follows.

Finally, in the interest of clarity, we summarize the main results of this section in this more conventional notation. The coordinate scalars as a function of the background coordinates are
		\begin{eqn}
			\mathsf{X}(x)
				&= x + \kappa \mathsf{X}_1(x) + \kappa^2 \mathsf{X}_2(x) 
					+ O(\kappa^3).
		\end{eqn}
	The expansion terms are
		\begin{eqn}
			\mathsf{X}_1(x)
				&= \int \dd[d]{x'} G(x,x') J_1(x'),
		\end{eqn}
	where $G(x,x')$ is a Green function of $\partial^2$ and
		\begin{eqn}
			J_1^\mu
				&= \partial_\alpha h^{\alpha \mu}
					- \sfrac{1}{2} \partial^\mu h;
		\end{eqn}
	we also have
		\begin{eqn}
			\mathsf{X}_2(x)
				&= \int \textrm{d}^d{x'} G(x,x')
					\bigg( J_2(x') + K_1 \mathsf{X}_1(x') \bigg),
		\end{eqn}
	where
		\begin{eqn}
            \label{eq:K1andJ2}
			K_1
				&= h^{\alpha \beta} \partial_\alpha \partial_\beta
					+ J_1^\alpha \partial_\alpha,
		&
			J_2^\mu
				&= \sfrac{1}{2} \bigg( h_{\alpha \beta} \partial^\mu h^{\alpha \beta}
						+ h^{\alpha \mu} \partial_\alpha h \bigg)
					- \partial_\alpha \bigg( h^{\alpha \beta} h {}_\beta {}^\mu \bigg).
		\end{eqn}
	Given a scalar field $\phi$, its invariantized form is
		\begin{eqn}
			\hat{\phi}
				&= \phi \circ \mathsf{X}^{-1}
				= \phi - \kappa \mathsf{X}_1^\alpha \partial_\alpha \phi
					+ \kappa^2 \bigg( \sfrac{1}{2} 
						\mathsf{X}_1^\alpha \mathsf{X}_1^\beta 
							\partial_\alpha \partial_\beta \phi
						+ \mathsf{X}_1^\alpha \partial_\alpha \mathsf{X}_1^\beta
							\partial_\beta \phi
						- \mathsf{X}_2^\alpha \partial_\alpha \phi \bigg)
					+ O(\kappa^3),
		\end{eqn}
	and considering in particular the scalar curvature yields
		\begin{eqn}
            \label{eq:RwCoordinateCorrections}
			\mathcal{R}
				&= R \circ \mathsf{X}^{-1}
				= \kappa R_1
					+ \kappa^2 \bigg( R_2 - \mathsf{X}_1^\alpha \partial_\alpha R_1 \bigg)
                    + \kappa^3 \bigg( R_3 - \mathsf{X}_1^\alpha \partial_\alpha R_2 + \sfrac{1}{2} 
						\mathsf{X}_1^\alpha \mathsf{X}_1^\beta 
							\partial_\alpha \partial_\beta R_1
						+ \mathsf{X}_1^\alpha \partial_\alpha \mathsf{X}_1^\beta
							\partial_\beta R_1
						- \mathsf{X}_2^\alpha \partial_\alpha R_1 \bigg)
					+ O(\kappa^4),
		\end{eqn}
	where
		\begin{eqn} \label{eq:R_1,2}
				R_1
					&= \partial_\mu \partial_\nu h^{\mu \nu}
						- \partial^2 h,
			\\
				R_2
					&= h^{\mu \nu} \partial_\mu \partial_\nu h
						- \frac{1}{4} \partial_\mu h \partial^\mu h
						- \partial_\mu h^{\mu \nu} \partial_\rho h {}_\nu {}^\rho
						+ \partial^\mu h \partial_\nu h {}_\mu {}^\nu
						- 2 h^{\mu \nu} \partial_\nu \partial_\rho h {}_\mu {}^\rho
			\\
					&\hspace{1cm}
						+ h^{\mu \nu} \partial^2 h_{\mu \nu}
						- \frac{1}{2} \partial_\nu h_{\mu \rho} \partial^\rho h^{\mu \nu}
						+ \frac{3}{4} \partial_\rho h_{\mu \nu} \partial^\rho h^{\mu \nu},
			\\
			R_3
				&= - \sfrac{3}{4} h^{\mu \nu} \partial_\mu h^{\alpha \beta} 
						\partial_\nu h_{\alpha \beta}
					+ \sfrac{1}{4} h^{\mu \nu} \partial_\mu h \partial_\nu h
					- h^{\mu \nu} \partial_\mu h \partial_\alpha h {}_\nu {}^\alpha
					- h^{\mu \nu} \partial_\mu h {}_\nu {}^\alpha \partial_\alpha h
					- h^{\mu \nu} h {}_\mu {}^\alpha \partial_\nu \partial_\alpha h
		\\
				&\hspace{1cm}
					+ \sfrac{1}{2} h^{\mu \nu} \partial_\alpha h_{\mu \nu} \partial^\alpha h
					+ h^{\mu \nu} \partial_\alpha h {}_\mu {}^\alpha 
						\partial_\beta h {}_\nu {}^\beta
					+ 2 h^{\mu \nu} \partial_\nu h {}_\mu {}^\alpha 
						\partial_\beta h {}_\alpha {}^\beta
					- h^{\mu \nu} \partial_\alpha h_{\mu \nu} \partial_\beta h^{\alpha \beta}
					+ h^{\mu \nu} h^{\alpha \beta} \partial_\mu \partial_\alpha h_{\nu \beta}
		\\
				&\hspace{1cm}
					- h^{\mu \nu} h^{\alpha \beta} \partial_\alpha \partial_\beta h_{\mu \nu}
					+ 2 h^{\mu \nu} h {}_\mu {}^\alpha \partial_\alpha \partial_\beta 
						h {}_\nu {}^\beta
					- h^{\mu \nu} h {}_\mu {}^\alpha \partial^2 h_{\nu \alpha}
					 + h^{\mu \nu} \partial_\nu h {}_\alpha {}^\beta 
					 	\partial_\beta h {}_\mu {}^\alpha
					+ \sfrac{1}{2} h^{\mu \nu} \partial_\alpha h {}_\mu {}^\beta
						 \partial_\beta h {}_\nu {}^\alpha
		\\
				&\hspace{1cm}
					- \sfrac{3}{2} h^{\mu \nu} \partial_\alpha h_{\mu \beta}
						\partial^\alpha h {}_\nu {}^\beta.
		\end{eqn}
The invariantized volume factor is
\begin{eqn}
		\label{eq:invar-volume-expansion}
			\sqrt{ - \det \mathcal{G}}
				&= 1 + \kappa \bigg( \sfrac{1}{2} h - \partial_\mu \mathsf{X}_1^\mu \bigg)
					+ \kappa^2 \bigg(
						\sfrac{1}{2} \partial_\mu \mathsf{X}_1^\mu \partial_\nu \mathsf{X}_1^\nu
						+ \mathsf{X}_1^\mu \partial_\mu \partial_\nu \mathsf{X}_1^\nu
						+ \sfrac{1}{2} \partial_\mu \mathsf{X}_{1\nu} \partial^\nu \mathsf{X}_1^\mu
						- \partial_\mu \mathsf{X}_2^\mu
		\\
				&\hspace{7cm}
						- \sfrac{1}{2} \mathsf{X}_1^\mu \partial_\mu h
						- \sfrac{1}{2} h \partial_\mu \mathsf{X}_1^\mu
						+ \sfrac{1}{8} h^2
						- \sfrac{1}{4} h_{\mu \nu} h^{\mu \nu}
					\bigg)
					+ O(\kappa^3).
		\end{eqn}

\section{\label{sec:Feyn}Feynman rules}

In this section we review the formulation of the low-energy effective theory of gravity and provide its Feynman rules.  As an effective theory, an infinite number of terms are required to define the gravitational action, including new low energy constants appearing at each order of perturbation theory.  Predictions at low order are still possible, however, since the radiative corrections in some cases are independent of the higher order couplings of the theory.  This turns out to be the case for the correlation functions calculated here, where only Newton's constant appears in the final results (when we ignore contact terms that vanish everywhere but the origin).  

In this work, we only need to consider the effective action through ${\cal O}(R^2)$, which we write as
		\begin{eqn}
			S
				&= S_\text{EH} + S_\text{gf} + S_\text{gh} + S_{R^2}.
		\end{eqn}
The Einstein-Hilbert action is
		\begin{eqn}
			S_\text{EH}
				&= - \frac{2}{\kappa^2} \int \textrm{d}^dx \sqrt{-g} R,
		\end{eqn}
where (in four dimensions) the coupling $\kappa$ is related to Newton's constant by $\kappa^2=32\pi G$.  As in Sect.~\ref{sec:observ}, we expand the metric $g_{\mu \nu}=\eta_{\mu\nu} + \kappa h_{\mu\nu}$ about flat space, with $\eta_{\mu\nu}$ the Minkowski metric.

The choice of gauge and the ghost action are determined by the gauge-fixing function $F$.  We work in a generalized harmonic (de Donder) gauge, with gauge-fixing function
		\begin{eqn}
			F_\mu
				&= \partial^\nu h_{\mu \nu}
					- \frac{1 + \beta}{d} \partial_\mu h,
		\end{eqn}
where $\beta$ is an arbitrary parameter, in terms of which the linearized harmonic gauge can be obtained by setting $\beta=\frac{d}{2}-1$.  The gauge-fixing action is then
    \begin{eqn}
			S_\text{gf}
				&= \frac{1}{2 \alpha}\int \textrm{d}^dx 
					\Bigg( \partial_\nu h^{\mu \nu} - \frac{1 + \beta}{d} \partial^\mu h \Bigg)
					\Bigg( \partial^\lambda h_{\mu \lambda} - \frac{1 + \beta}{d} \partial_\mu h \Bigg),
		\end{eqn}
with $\alpha$ another arbitrary gauge parameter.  Coordinate transformations $x^\mu \rightarrow x'^\mu= x^\mu - \xi^\mu(x)$ lead to a transformation of the metric
\bea  h_{\mu\nu} \rightarrow h'_{\mu\nu}=h_{\mu\nu} + \partial_\mu \xi_\nu + \partial_\nu \xi_\mu +\kappa\big(\xi^\rho\partial_\rho h_{\mu\nu} + h_{\rho\nu}\partial_\mu \xi^\rho +h_{\rho\mu}\partial_\nu \xi^\rho \big).
\eea
The Fadeev-Popov gauge-fixing procedure then leads to ghost degrees of freedom in the action, which can be obtained by a functional derivative of the gauge fixing function $F$ with respect to $\xi$.  For our choice of gauge, the ghost action is
    \begin{eqn} \label{eq:ghost_lagrangian}
			S_\text{gh}
				&= \int \textrm{d}^dx \Bigg\{ \bar{c}^\mu \partial^2 c_\mu
					+ \Bigg( 1 - \frac{2(1 + \beta)}{d} \Bigg) 
						\bar{c}^\mu \partial_\mu \partial^\nu c_\nu + \kappa \Bigg[
						\Bigg( 1 - \frac{2 (1 + \beta)}{d} \Bigg) 
							\bar{c}^\mu h_{\nu \rho} \partial^\rho \partial_\mu c^\nu
						- \frac{1 + \beta}{d} 
							\big( \bar{c}^\mu \partial_\mu c^\nu \partial_\nu h
								+ \bar{c}^\mu c^\nu \partial_\mu \partial_\nu h \big)
                    \\
				&\hspace{2cm}
						- \frac{2(1 + \beta)}{d} 
							\bar{c}^\mu \partial_\mu h_{\nu \rho} \partial^\rho c^\nu		
						+ \bar{c}^\mu \partial_\mu c^\nu \partial_\rho {h_\nu}^\rho
						+ \bar{c}^\mu c^\nu \partial_\nu \partial_\rho {h_\nu}^\rho
						+ \bar{c}^\mu h_{\mu \nu} \partial^2 c^\nu
						+ \bar{c}^\mu \partial_\nu h_{\mu \rho} \partial^\rho c^\nu
                    \\
				&\hspace{2cm}
						+ \bar{c}^\mu \partial_\rho h_{\mu \nu} \partial^\rho c^\nu
					\Bigg] \Bigg\},
		\end{eqn}
where the first two terms are ghost kinetic terms, and the remaining ones lead to a ghost anti-ghost graviton vertex.

In the basis we choose, the $O(R^2)$ terms in the action are
\bea\label{eq:R2Lagrangian}  S_{R^2} = \int \textrm{d}^d x \sqrt{-g}\left(c_1 R^2 +c_2 C^2 +c_3 G \right),
\eea
where $C^2 = R^{\mu \nu \alpha \beta}R_{\mu \nu \alpha \beta} - 2R^{\mu \nu}R_{\mu \nu} + \frac{1}{3} R^2$ is the square of the Weyl tensor, and $G=R^{\mu \nu \alpha \beta}R_{\mu \nu \alpha \beta} - 4R^{\mu \nu}R_{\mu \nu} + R^2$ is the Gauss-Bonnet term.  This is a convenient basis for the $O(R^2)$ operators, since the insertion of the Weyl squared tensor in tree-level diagrams vanishes for the correlation functions we consider.  The Gauss-Bonnet term $G$ is a total derivative in four dimensions, and its integral is a topological invariant.  Although this term contributes only an overall constant factor to the path integral in four dimensions, it can have non-zero contributions in dimensional regularization.  Its coefficient $c_3$ is proportional to a $1/\varepsilon$ pole, with $\varepsilon= \frac{1}{2}(4-d)$, as can be shown by a calculation that nearly coincides with that of the conformal anomaly \cite{Capper:1974ic, Capper:1975ig, Gibbons:1978ac, Goroff:1985th}.  Despite the possibility of this evanescent operator giving a non-zero contribution in the effective theory of gravity \cite{Bern:2015xsa}, we find that it does not contribute through next-to-leading order in our correlation functions, as its insertion in the relevant tree-level diagrams leads to them vanishing identically.

It is well known that the first two terms in Eq.~(\ref{eq:R2Lagrangian}) can be eliminated by a field redefinition of the metric, leading to the fact that pure gravity is renormalizable at one-loop order in perturbation theory \cite{tHooft:1974toh, AccettulliHuber:2019jqo}.  As a consequence, the contributions of these operators vanish for all on-shell graviton scattering S-matrix elements at tree level \cite{AccettulliHuber:2019jqo}.  Although the $\sqrt{-g}C^2$ and $\sqrt{-g}G$ operator insertions vanish at next-to-leading order in our correlation functions, the $\sqrt{-g}R^2$ operator does not.  It gives a gauge invariant, non-zero contribution through an insertion in a tree-level diagram, as it must, in order to provide a counterterm to cancel the divergences coming from the one-loop insertions of the Einstein-Hilbert action.  This does not present a contradiction with the usual picture, since we are considering Euclidean correlation functions, not on-shell S-matrix elements.  The field redefinition of the metric that removes the higher curvature operators from the action reintroduces their coefficients in the external operators of our correlation functions, which also involve the metric.  An explicit calculation shows that with or without the field redefinition that removes the ${\cal O}(R^2)$ terms from the action, we get the same non-vanishing dependence on the coefficient $c_1$ of the $\sqrt{-g}R^2$ term appearing at next-to-leading order.  A consequence of this result is that our correlation function calculation fixes the divergent part of the counterterm coefficient of the $\sqrt{-g}R^2$ operator in the effective theory of pure gravity in an unambiguous, gauge-invariant way (up to a choice of basis of ${\cal O}(R^2)$ operators).

\subsection{Expansion pieces}
\label{sec:expansion-pieces}

This subsection presents in some detail the expansions of the main geometric ingredients needed to construct the Feynman rules for gravity.  

\subsubsection{The inverse metric}
\label{sec:inverse-metric-expansion}

We begin with the perturbative expansion of the inverse of the metric.  We first define its expansion coefficients as
		\begin{eqn}
			g^{\mu \nu}
				&= \sum_n \kappa^n \tilde{g}_n^{\mu \nu}.
		\end{eqn}
	The $\tilde{g}_n^{\mu \nu}$'s can be obtained order-by-order by imposing the definition $g^{\mu \rho} g_{\rho \nu} = \delta^\mu_\nu$ with $g_{\mu \nu} = \eta_{\mu \nu} + \kappa h_{\mu \nu}$. At zeroth order we immediately find $\tilde{g}_0^{\mu \nu} = \eta^{\mu \nu}$. At first order we then have
		\begin{eqn}
			\delta^\mu_\nu
				&= \big( \eta^{\mu \rho} + \kappa \tilde{g}_1^{\mu \rho} \big)
					\big( \eta_{\rho \nu} + \kappa h_{\rho \nu} \big)
					+ O(\kappa^2)
				= \delta^\mu_\nu
					+ \kappa \big( \tilde{g}_1 {}^\mu {}_\nu + h {}^\mu {}_\nu \big)
					+ O(\kappa^2),
		\end{eqn}
	from which we find $\tilde{g}_1^{\mu \nu} = - h^{\mu \nu}$. At second order we then find
		\begin{eqn}
			\delta^\mu_\nu
				&= \big( \eta^{\mu \rho} - \kappa h^{\mu \rho} + \kappa^2 \tilde{g}_2^{\mu \rho} \big)
					\big( \eta_{\rho \nu} + \kappa h_{\rho \nu} \big)
					+ O(\kappa^3)
				= \delta^\mu_\nu
					+ \kappa^2 \big( 
						\tilde{g}_2 {}^\mu {}_\nu - h^{\mu \rho} h_{\rho \nu}
					\big)
					+ O(\kappa^3),
		\end{eqn}
	which yields $\tilde{g}_2^{\mu \nu} = h^{\mu \rho} h {}_\rho {}^\nu$, and proceeding in this way we find
		\begin{eqn}
			g^{\mu \nu}
				&= \eta^{\mu \nu}
					- \kappa h^{\mu \nu}
					+ \kappa^2 h^{\mu \alpha} h {}_\alpha {}^\nu
					- \kappa^3 h^{\mu \alpha} h^{\nu \beta} h_{\alpha \beta}
					+ \kappa^4 h^{\mu \alpha} h^{\nu \beta} h {}_\alpha {}^\gamma
						h_{\beta \gamma}
					+ O(\kappa^5).
		\end{eqn}

\subsubsection{The volume factor}
\label{sec:volume-factor-expansion}

To expand the volume factor appearing in the Einstein-Hilbert action, we begin by recalling the matrix identities
		\begin{eqn}
			\ln \big( \det A \big)
				&= \tr \big( \ln A \big),
		&
			\det(AB)
				&= \det A \det B,
		\end{eqn}
	and the Taylor expansions
		\begin{eqn}
			\ln(1 + x)
				&= x - \sfrac{1}{2} x^2 + O(x^3),
		&
			\sqrt{1 + x}
				&= 1 + \sfrac{1}{2} x - \sfrac{1}{8} x^2 + O(x^3).
		\end{eqn}
        With the standard notation $-g \equiv -\det(g_{\mu\nu})$, we have
		\begin{eqn}
			-g 
				&= - \det \big( \eta_{\mu\nu} + \kappa h_{\mu\nu} \big)
				= \det \big( \delta^\mu_\nu + \kappa \eta^{\mu\alpha} h_{\alpha\nu} \big)
				= \exp \Bigg[ \tr \big[ \ln \big( 1 + \kappa \eta^{\mu\alpha} h_{\alpha\nu} \big) \big]\Bigg],
		\end{eqn}
        where $1$ here refers to the identity matrix and we have used the fact that $\det (\eta_{\mu\nu}) = -1$.  As an intermediate step we expand the logarithm and take its trace, yielding
		\begin{eqn}
			\tr \big[ \ln \big( 1 + \kappa \eta^{\mu\alpha} h_{\alpha\nu} \big) \big]
				&= \tr \bigg[ \kappa \eta^{\mu\alpha} h_{\alpha\nu} 
					- \sfrac{1}{2} \kappa^2 \big( \eta^{\mu\alpha} h_{\alpha\nu} \big)^2
					+ O\big(\kappa^3\big) \bigg]
				= \kappa h - \sfrac{1}{2} \kappa^2 h^{\mu\nu}h_{\mu\nu} + O(\kappa^3).
		\end{eqn}
	Exponentiating this, taking the square root, and further expanding, we obtain
		\begin{eqn}\label{eq:standard-volume-expansion}
			\sqrt{ - g}
				&= 1 + \sfrac{1}{2} \kappa h
					+ \kappa^2 \bigg( 
						\sfrac{1}{8} h^2 
						- \sfrac{1}{4} h_{\mu \nu} h^{\mu \nu} \bigg)
					+ O(\kappa^3).
		\end{eqn}

One may proceed in the manner above to arbitrarily high order. Doing so introduces nothing new conceptually but the algebra becomes tedious and potentially error prone.  To perform these expansions to the order needed we make use of the computer program {\sc xAct}. The result through ${\cal O}(\kappa^4)$ is 
		\begin{eqn}
			\sqrt{-g}
				&= 1
					+ \sfrac{1}{2} \kappa h
					+ \kappa^2 \bigg( \sfrac{1}{8} h^2 
						- \sfrac{1}{4} h_{\mu \nu} h^{\mu \nu} \bigg)
					+ \kappa^3 \bigg( 
						\sfrac{1}{6} h^{\mu \nu} h {}_\mu {}^\alpha h_{\alpha \nu}
						- \sfrac{1}{8} h h_{\mu \nu} h^{\mu \nu}
						+ \sfrac{1}{48} h^3 \bigg)
		\\
				&\hspace{1cm}
					+ \kappa^4 \bigg(
							- \sfrac{1}{8} h^{\mu \nu} h {}_\mu {}^\alpha h {}_\nu {}^\beta
								h_{\alpha \beta}
							+ \sfrac{1}{12} h h^{\mu \nu} h {}_\mu {}^\alpha h_{\alpha \nu}
							+ \sfrac{1}{32} (h_{\mu \nu} h^{\mu \nu})^2
							- \sfrac{1}{32} h^2 h_{\mu \nu} h^{\mu \nu}
							+ \sfrac{1}{384} h^4
						\bigg)
					+ O(\kappa^5).
		\end{eqn}
	For later use we denote the terms in the expansion as
		\begin{eqn}
			\sqrt{-g}
				&= \sum_n \kappa^n \gamma_n.
		\end{eqn}

\subsubsection{The scalar curvature}
\label{sec:scalar-curvature-expansion}

The expansion of the scalar curvature to the needed order is quite lengthy. We define its expansion coefficients as
		\begin{eqn}
			R
				&= \sum_n \kappa^n R_n.
		\end{eqn}
	At zeroth order it vanishes
		\begin{eqn}
			R_0
				&= 0.
		\end{eqn}
    The expansions for $R_1$, $R_2$, and $R_3$ are given in Eq.~(\ref{eq:R_1,2}).

\subsection{The propagators}

\subsubsection{The ghost propagator}

The kinetic terms in the ghost Lagrangian are
		\begin{eqn}
			L_\text{gh,kin}
				&= \bar{c}^\mu \partial^2 c_\mu
					+ \Bigg( 1 - \frac{2(1 + \beta)}{d} \Bigg) 
						\bar{c}^\mu \partial_\mu \partial^\nu c_\nu,
		\end{eqn}
	which yield the vector propagator
		\begin{eqn}
			S_{\mu \nu}(p)
				&= \frac{\ii}{p^2}
					\Bigg\{
						\Bigg( 1 - \frac{d}{2 (d-1-\beta)} \Bigg) p_\mu p_\nu - g_{\mu \nu}
					\Bigg\}.
		\end{eqn}
	We represent the ghost propagator diagrammatically as
		\begin{eqn}
    	\begin{tikzpicture}[baseline=(v1.base)]
    		\begin{feynman}
    			\vertex (v1) {$\mu$};
				\vertex at ($(v1) + (2cm, 0)$) (v2) {$\nu$};
    			\diagram*{
    				(v1) --[fermion, dashed, momentum = $p$] (v2) 
    			};
    		\end{feynman}
    	\end{tikzpicture}
			&=  S_{\mu \nu}(p).
		\end{eqn}

\subsubsection{The graviton propagator}

The graviton's kinetic terms come from the second-order expansion of the Einstein-Hilbert action,
		\begin{eqn}
			L_\text{EH,kin}
				&= - \partial h^\mu \partial h_\mu
					- \frac{1}{2} h^{\mu \nu} \partial^2 h_{\mu \nu}
					+ \partial h^\nu \partial_\nu h
					+ \frac{1}{2} h \partial^2 h,
		\end{eqn}
	along with the gauge-fixing action
		\begin{eqn}
			L_\text{gf}
				&= \frac{1}{2 \alpha}
					\Bigg( \partial h^\mu - \frac{1 + \beta}{d} \partial^\mu h \Bigg)
					\Bigg( \partial h_\mu - \frac{1 + \beta}{d} \partial_\mu h \Bigg).
		\end{eqn}
	The resulting propagator is conveniently given as follows. 
    We define the symmetric tensor structures 
		\begin{eqn}
			G_{\mu \nu \rho \sigma}
				&= \sfrac{1}{2} \big( g_{\mu \rho} g_{\nu \sigma} 
					+ g_{\mu \sigma} g_{\nu \rho} \big),
		\hspace{.5cm}
			\tr_{\mu \nu \rho \sigma}
				=  g_{\mu \nu} g_{\rho \sigma},
		\hspace{.5cm}
			A_{\mu \nu \rho \sigma}
				= \frac{p_\mu p_\nu p_\rho p_\sigma}{p^4},
		\\
			B {}_{\mu \nu \rho \sigma}
				&= \frac{1}{2 p^2} \big( g_{\mu \nu} p_\rho p_\sigma + g_{\rho \sigma}p_\mu p_\nu \big),
		\hspace{.5cm}
			\tilde{C} {}_{\mu \nu \rho \sigma}
				= \frac{1}{4 p^2} \big( g_{\mu \rho} p_\nu p_\sigma + g_{\mu \sigma}p_\nu p_\rho + g_{\nu \rho} p_\mu p_\sigma + g_{\nu \sigma} p_\mu p_\rho\big).
		\end{eqn}
The $d$-dimensional graviton propagator in terms of these tensor structures is then
\begin{eqn}  \Delta_{\mu \nu \rho \sigma}(p) = \frac{\ii}{p^2}\Bigg[G_{\mu \nu \rho \sigma} + \frac{1}{2-d}{\rm tr}_{\mu \nu \rho \sigma} - \frac{(2\beta - d +2)\big[\alpha(d-2)(2\beta -3d +2)+2\beta +d^2 -3d +2\big]}{(d-2)(\beta-d+1)^2} A_{\mu \nu \rho \sigma}  \\ 
  +  \frac{2(2\beta -d +2)}{(d-2)(\beta-d+1)}B_{\mu \nu \rho \sigma} + (4\alpha-2)\tilde{C}_{\mu \nu \rho \sigma} \Bigg]. \ \ \ \ \ \ \ \ \ \ \ \ \ \ \ \ \ \ \ \ \ \ \ \ \ \ \ \ \ \ \ \ \ \ \ \ \ \ \ \ \ \ \ \ \ \ \ \ \ \ \ \ \ \ \ \ \ \ \ \ \ \ \ \ 
\end{eqn}
Specifying to 4 dimensions, the graviton propagator becomes
\bea\label{eq:graviton_propagator}  \Delta_{\mu \nu \rho \sigma}(p) = \frac{\ii}{p^2}\left[G_{\mu \nu \rho \sigma} 
-\frac{1}{2} {\rm tr}_{\mu \nu \rho \sigma} - \frac{2(\beta-1)\big[2\alpha(\beta-5)+\beta+3 \big]}{(\beta-3)^2}A_{\mu \nu \rho \sigma} + \frac{2(\beta-1)}{\beta-3}B_{\mu \nu \rho \sigma} +(4\alpha-2)\tilde{C}_{\mu \nu \rho \sigma} \right].
\eea
        
	We represent the graviton propagator diagrammatically as
		\begin{eqn}
    	\begin{tikzpicture}[baseline=(v1.base)]
    		\begin{feynman}
    			\vertex (v1) {$\mu \nu$};
				\vertex at ($(v1) + (2cm, 0)$) (v2) {$\rho \sigma$};
    			\diagram*{
    				(v1) --[boson, momentum = $p$] (v2) 
    			};
    		\end{feynman}
    	\end{tikzpicture}
			&=  \Delta_{\mu \nu \rho \sigma}(p).
		\end{eqn}

\subsubsection{The Fourier transform convention and derivative interactions}
	
	In each case above the given propagator relates to the corresponding free position-space two-point function via a Fourier transform in the usual way, e.g. for a free scalar field
		\begin{eqn}
		\label{eq:scalar-propagator-mom-pos}
			\ev{\phi(x) \phi(y)}_0
				&=  D(x,y)
				= \int \frac{\dd[d]{k}}{(2 \pi)^d} \frac{\ii}{k^2-m^2 +i\varepsilon} \ee^{\ii k(x-y)}.
		\end{eqn}
	Now, observe that in eq. (\ref{eq:scalar-propagator-mom-pos}) we are free to choose the sign of the momentum $k$ in the exponent. This choice relates to the sign of the momentum in the Feynman rule for a vertex as follows. Consider for example the tree-level contribution to the two-point function $\ev{\phi(x) \partial_\mu \phi(y)}$. Proceeding as above and using the Fourier expansion of the propagator yields
		\begin{eqn}
			\ev{\phi(x) \partial_\mu \phi(y)}
				&= \ii \pdv{y^\mu} 
					\int \frac{\dd[d]{k}}{(2 \pi)^d}
						\frac{\ee^{\ii k (x-y)}}{k^2 - m^2 + i\varepsilon}
				= \int \frac{\dd[d]{k}}{(2 \pi)^d} (- \ii k_\mu)
					\frac{\ii}{k^2 - m^2 + i\varepsilon} \ee^{\ii k(x-y)}.
		\end{eqn}
	This convention is related to the diagrammatic approach as follows. The single contributing diagram (in a free theory) is a line carrying the momentum $k$ from $x$ to $y$, corresponding to the momentum-space propagator $\ii / (k^2 - m^2)$. The field at $x$ has the trivial external vertex factor of $1$, while the derivative at $\phi$ yields a momentum factor whose sign convention must be chosen. We choose the convention that {\em all outgoing momenta are positive}, meaning that if the momentum $k$ points from $x$ to $y$ then the vertex factor at $y$ is $- \ii k_\mu$:
		
		\begin{eqn}	
%
			&= (- \ii k_\mu) \frac{\ii}{k^2 - m^2 + i \varepsilon}.
		\end{eqn}
	Comparing to the previous result we see that our chosen convention does indeed correspond to the exponential sign choice $\ee^{\ii k(x-y)}$, whereas the opposite choice $\ee^{- \ii k(x-y)}$ would correspond to writing the {\em incoming} momenta as positive.

\subsection{Volume external insertions}
\label{sec:volume-external-insertions}

When calculating a correlator that includes an invariantized quantity it follows that at $X$ we will have not only the ``standard'' external vertex factor but also an infinite series of external vertices arising from the invariantization. As in Sect.~\ref{sec:observ}, we call these latter {\em coordinate corrections}. For both the volume factor and the scalar curvature, it is possible to have standard insertions and coordinate corrections at the same order.

From Eq.~(\ref{eq:invar-volume-expansion}) we have the expansion of the invariantized volume factor,
		\begin{eqn}
			\sqrt{ - \det \mathcal{G}}
				&= 1 + \kappa \bigg( \sfrac{1}{2} h - \partial_\mu \mathsf{X}_1^\mu \bigg)
					+ \kappa^2 \bigg(
						\sfrac{1}{2} \partial_\mu \mathsf{X}_1^\mu \partial_\nu \mathsf{X}_1^\nu
						+ \mathsf{X}_1^\mu \partial_\mu \partial_\nu \mathsf{X}_1^\nu
						+ \sfrac{1}{2} \partial_\mu \mathsf{X}_{1\nu} \partial^\nu \mathsf{X}_1^\mu
						- \partial_\mu \mathsf{X}_2^\mu
		\\
				&\hspace{7cm}
						- \sfrac{1}{2} \mathsf{X}_1^\mu \partial_\mu h
						- \sfrac{1}{2} h \partial_\mu \mathsf{X}_1^\mu
						+ \sfrac{1}{8} h^2
						- \sfrac{1}{4} h_{\mu \nu} h^{\mu \nu}
					\bigg)
					+ O(\kappa^3).
		\end{eqn}
	Just as for the invariantized scalar field this expansion yields both standard external vertices and coordinate corrections, but unlike the invariantized scalar field there are both types at all nontrivial orders.

Before proceeding we find it useful to rewrite the above expressions for the $\mathsf{X}$'s more explicitly. Our first step is to repackage the information in $J_1$ as a constant tensor acting on the single object $\partial_\alpha h_{\mu \nu}$. For $J_1$ we have
		\begin{eqn}
		\label{eq:j1-tensor}
			J_1^\mu
				&= \mathcal{J}_1^{\mu \alpha \rho \sigma}
					\partial_\alpha h_{\rho \sigma},
		&
			\mathcal{J}_1^{\mu \alpha \rho \sigma}
				&= \eta^{\mu \rho} \eta^{\alpha \sigma} 
					- \sfrac{1}{2} \eta^{\mu \alpha} \eta^{\rho \sigma},
		\end{eqn}
%
%
Writing the Green function as
		\begin{eqn}
			G(x, x')
				&= \int \frac{\dd[d]{p}}{(2 \pi)^d} 
					\Bigg( - \frac{1}{p^2} \Bigg) \ee^{\ii p (x-x')},
		\end{eqn}
	it follows that we can write $\mathsf{X}_1$ as
		\begin{eqn}
		\label{eq:X1-repackaged}
			\mathsf{X}_1^\mu(x) = \mathcal{J}_1^{\mu \alpha \rho \sigma}
				\int_{x', p} \Bigg( - \frac{1}{p^2} \Bigg) \partial_\alpha h_{\rho \sigma}(x') 
					\ee^{\ii p(x - x')},
		\end{eqn}
	introducing the shorthand $\int_x = \int \dd[d]{x}$ and $\int_p = \int \dd[d]{p} / (2 \pi)^d$.

\subsubsection{One-point: standard}

The standard term at $O(\kappa)$, $\kappa h / 2$, yields a one-point external vertex, which we determine by considering the two-point function
		\begin{eqn}
			\ev{ \sfrac{1}{2} \kappa h(x) h_{\rho \sigma}(y)}
				&= \sfrac{1}{2} \kappa \eta^{\mu \nu} \ev{h_{\mu \nu}(x) h_{\rho \sigma}(y)}.
		\end{eqn}
	Thus the external vertex factor for this term is
		\begin{eqn}
			\begin{tikzpicture}[baseline=(b.base)]
    		\begin{feynman}
    			\vertex [dot] (v1) {};
				\vertex at ($(v1) + (0, -0.09cm)$) (b);
				\vertex at ($(v1) + (1cm, 0)$) (v2);
    			\diagram*{
    				(v1) --[boson, momentum=$p$] (v2) 
    			};
    		\end{feynman}
    	\end{tikzpicture}
			&= 	
			\sfrac{1}{2}\kappa \eta^{\mu \nu}.
		\end{eqn}

\subsubsection{One-point: coordinate corrections}

The coordinate correction term at $O(\kappa)$, $ - \kappa \partial_\mu \mathsf{X}_1^\mu$, also yields a one-point external vertex. Using Eq.~(\ref{eq:X1-repackaged}) we obtain this vertex also from a two-point function:
		\begin{eqn}
			\ev{\bigg( - \kappa \partial_\mu \mathsf{X}_1^\mu(x) \bigg) h_{\rho \sigma}(y)}
				&= \kappa \mathcal{J}_1^{\alpha \beta \mu \nu} \int_{x', p} \frac{1}{p^2} 
					\ev{\partial_\alpha \partial_\beta h_{\mu \nu}(x') h_{\rho \sigma}(y)} 
						\ee^{\ii p(x-x')}
		\\
				&= - \kappa \mathcal{J}_1^{\alpha \beta \mu \nu}
					\int_{x', p, p'} \frac{1}{p^2} p'_\alpha p'_\beta \Delta_{\mu \nu \rho \sigma}(p')
						\ee^{\ii p'(x'-y)} \ee^{\ii p(x-x')}.
		\end{eqn}
	The $x'$ integral sets $p = p'$,
		\begin{eqn}
			\ev{\bigg( - \kappa \partial_\mu \mathsf{X}_1^\mu(x) \bigg) h_{\rho \sigma}(y)}
				&= - \kappa \mathcal{J}_1^{\alpha \beta \mu \nu} \int_p \frac{1}{p^2} p_\alpha p_\beta
					\Delta_{\mu \nu \rho \sigma}(p) \ee^{\ii p(x-y)},
		\end{eqn}
	from which we read off the external vertex
		\begin{eqn}
			\begin{tikzpicture}[baseline=(b.base)]
    		\begin{feynman}
    			\vertex [crossed dot] (v1) {};
				\vertex at ($(v1) + (0, -0.09cm)$) (b);
				\vertex at ($(v1) + (1cm, 0)$) (v2);
    			\diagram*{
    				(v1) --[boson, momentum=$p$] (v2) 
    			};
    		\end{feynman}
    	\end{tikzpicture}
			=
				- \kappa \frac{1}{p^2} \mathcal{J}_1^{\alpha \beta \mu \nu} p_\alpha p_\beta
				= - \kappa \frac{1}{p^2} \bigg( p^\mu p^\nu - \sfrac{1}{2} p^2 \eta^{\mu \nu} \bigg).
		\end{eqn}

\subsection{Scalar curvature external insertions}
\label{sec:scalar-curvature-external-insertions}

\subsubsection{One-point: standard}

The linear term in the expansion of $R$ is
		\begin{eqn}\label{eq:R1}
			R_1
				&= \partial_\mu \partial_\nu h^{\mu \nu}
					- \partial^2 h,
		\end{eqn}
	yielding the external vertex factor
		\begin{eqn}
          \begin{tikzpicture}[baseline=(b.base)]
    		\begin{feynman}
    			\vertex [dot] (v1) {};
				\vertex at ($(v1) + (0, -0.09cm)$) (b);
				\vertex at ($(v1) + (1cm, 0)$) (v2);
    			\diagram*{
    				(v1) --[boson, momentum=$p$] (v2) 
    			};
    		\end{feynman}
    	\end{tikzpicture}
			&= E_h^{\mu \nu}(p)
			= \kappa \big( p^2 \eta^{\mu \nu} - p^\mu p^\nu \big).
		\end{eqn}

\subsubsection{Two-point: standard}

The quadratic term in the expansion of $R$ is
		\begin{eqn}
			R_2
				&= h^{\mu \nu} \partial_\mu \partial_\nu h
					- \sfrac{1}{4} \partial_\mu h \partial^\mu h
					- \partial_\mu h^{\mu \nu} \partial_\alpha h {}_\nu {}^\alpha
					+ \partial_\mu h \partial_\nu h^{\mu \nu}
		\\
				&\hspace{5cm}
					- 2 h^{\mu \nu} \partial_\nu \partial_\alpha h {}_\mu {}^\alpha
					+ h^{\mu \nu} \partial^2 h_{\mu \nu}
					- \sfrac{1}{2} \partial_\alpha h_{\mu \nu} \partial^\mu h^{\nu \alpha}
					+ \sfrac{3}{4} \partial_\alpha h_{\mu \nu} \partial^\alpha h^{\mu \nu}.
		\end{eqn}
	We denote the resulting external vertex by
		\begin{eqn}
    	\begin{tikzpicture}[baseline=(v.base)]
    		\begin{feynman}
    			\vertex[dot] (v) {};
				\vertex at ($(v) + (1cm, 1cm)$) (v1) {$\mu \nu$};
				\vertex at ($(v) + (1.4cm, 0)$) (v2) {$\rho \sigma$};
    			\diagram*{
    				(v) --[boson, momentum = $p_1$] (v1);
    				(v) --[boson, momentum' = $p_2$] (v2);
				};
    		\end{feynman}
    	\end{tikzpicture}
			&= E_{h^2}^{\mu \nu \rho \sigma}(p_1, p_2),
		\end{eqn}
	and it is given explicitly by
		\begin{eqn}
			E_{h^2}^{\mu \nu \rho \sigma}(p_1, p_2)
				&= \kappa^2 \bigg(
						- p_2^\mu p_2^\nu \eta^{\rho \sigma} 
							- p_1^\rho p_1^\sigma \eta^{\mu \nu}
						+ \sfrac{1}{2} (p_1 \cdot p_2) \eta^{\mu \nu} \eta^{\rho \sigma}
						+ 2 p_1^\mu p_2^\rho \eta^{\nu \sigma}
						- p_1^\rho p_2^\sigma \eta^{\mu \nu}
						- p_1^\mu p_2^\nu \eta^{\rho \sigma}
		\\
				&\hspace{1cm}
						+ 2 p_2^\mu p_2^\rho \eta^{\nu \sigma}
						+ 2 p_1^\mu p_1^\rho \eta^{\nu \sigma}
						- \eta^{\mu \rho} \eta^{\nu \sigma} \big(p_1^2 + p_2^2\big)
						+ p_1^\rho p_2^\mu \eta^{\nu \sigma}
						- \sfrac{3}{2} (p_1 \cdot p_2) \eta^{\mu \rho} \eta^{\nu \sigma} 
					\bigg).
		\end{eqn}

\subsubsection{Two-point: coordinate corrections}

The invariantized scalar curvature also receives a coordinate correction at $O(\kappa^2)$, given by (\ref{eq:invar-R-expanded-from-scalar}):
		\begin{eqn}
            \label{eq:scalar-curvature-coordinate-correction_position}
			- \kappa^2 \mathsf{X}_1^\alpha \partial_\alpha R_1
				&= - \kappa^2 
					\partial_\beta \big( \partial_\mu \partial_\nu h^{\mu \nu}(x)
					- \partial^2 h(x) \big)
					\int \dd[d]{x'} G(x,x') \bigg( 
					\partial_\alpha h^{\alpha \beta}(x') - \sfrac{1}{2} \partial^\beta h(x') \bigg).
		\end{eqn}
	This yields an external vertex
		\begin{eqn}
		\label{eq:scalar-curvature-coordinate-correction}
		    	\begin{tikzpicture}[baseline=(v.base)]
    		\begin{feynman}
    			\vertex[crossed dot] (v) {};
				\vertex at ($(v) + (1cm, 1cm)$) (v1) {$\mu \nu$};
				\vertex at ($(v) + (1.4cm, 0)$) (v2) {$\rho \sigma$};
    			\diagram*{
    				(v) --[boson, momentum = $p_1$] (v1);
    				(v) --[boson, momentum' = $p_2$] (v2);
				};
    		\end{feynman}
    	\end{tikzpicture}
				= \tilde{E}_{h^2}^{\mu \nu \rho \sigma}(p_1, p_2)
				&= \frac{\kappa^2}{2} \Bigg\{
						\frac{1}{p_1^2} \big( - \eta^{\mu \nu} (p_1 \cdot p_2) 
							+ p_1^\mu p_2^\nu + p_1^\nu p_2^\mu \big)
							\big( p_2^\rho p_2^\sigma - p_2^2 \eta^{\rho \sigma} \big)
		\\
				&\hspace{3cm}
						+ \frac{1}{p_2^2} \big( - \eta^{\rho \sigma} (p_1 \cdot p_2) 
							+ p_1^\rho p_2^\sigma + p_1^\sigma p_2^\rho \big)
							\big( p_1^\mu p_1^\nu - p_1^2 \eta^{\mu \nu} \big)
					\Bigg\}.
		\end{eqn}

\noindent Note that this two-point coordinate correction vertex in momentum space contains a scalar field propagator denominator arising from the integral over the Green's function in Eq.~(\ref{eq:scalar-curvature-coordinate-correction_position}).

\subsubsection{Three-point: coordinate corrections}

The three-point coordinate corrections are also needed for our calculation.  Only the third term in parentheses in Eq.~(\ref{eq:invariant-ricci-as-scalar-field}) leads to a non-zero contribution for the curvature correlation function to the order we consider.  In position space, this contribution is

            \begin{eqn}
            \label{eq:h3_scalar-curvature-coordinate-correction_position}
                - \kappa^3 \partial_\alpha R_1 \mathsf{X}_2^\alpha  &= - \kappa^3 \partial_\alpha\big( \partial_\mu \partial_\nu h^{\mu \nu}(x)
					- \partial^2 h(x) \big)
                    \\
				&  \hspace{2cm}
                  \times  \int \textrm{d}^dx' G(x,x')\left[J_2^\alpha(x')
                    + K_1(x') \int \textrm{d}^dx'' 
                    G(x',x'')\big(\partial_\lambda h^{\lambda\alpha}(x'') - \frac{1}{2}\partial^\alpha h(x'') \big) \right],
            \end{eqn}

\noindent where $J_2^\alpha$ and $K_1$ are given in Eq.~(\ref{eq:K1andJ2}).  The Fourier transform of this vertex is denoted schematically as

            \begin{eqn}
		\label{eq:h3-coordinate-correction} 	
			\begin{tikzpicture}[baseline=(v.base)]
    		\begin{feynman}
    			\vertex[crossed dot] (v) {};
				\vertex at ($(v) + (-1.5cm, 0)$) (v1) {$\eta \lambda$};
				\vertex at ($(v) + (1.06, 1.06cm)$) (v2) {$\sigma \xi$};
                    \vertex at ($(v) + (1.06cm, -1.06cm)$) (v3) {$\mu \nu$};
    			\diagram*{
    				(v) --[boson, momentum' = $p_1$] (v1);
    				(v) --[boson, momentum' = $p_2$] (v2);
                        (v) --[boson, momentum' = $p_3$] (v3);
				};
    		\end{feynman}
    	\end{tikzpicture}
          = \tilde{E}_{h^3}^{\eta \lambda \sigma \xi \mu \nu}(p_1, p_2, p_3).
            \end{eqn}

\noindent The resulting expression is quite lengthy, with the combinatorics leading to six different permutations of the momenta, so we do not display the full expression here.  Note, however, that the Fourier transform of the right side of Eq.~(\ref{eq:h3_scalar-curvature-coordinate-correction_position}) leads to terms that include a factor of $1/(p_1+p_2)^2$, arising from the outer integral over $x'$.  Other terms appear with a similar factor for all possible pairs of momentum combinations.  To the order we are considering, this vertex can only appear as the source of a tadpole loop.  Since only (massless) gravitons propagate in the loops, naively this diagram should vanish.  However, the scalar propagator denominator arising from the integral over the outer Green's function in Eq.~(\ref{eq:h3_scalar-curvature-coordinate-correction_position}) can contain a sum over two momenta that includes the tadpole loop momentum and the external momentum.  Thus, the vertex of the tadpole diagram in our calculation, when combined with the graviton propagator in the loop, acquires the propagator structure of a diagram with an external momentum flowing through a loop with two vertices and does not vanish.  All of the terms to ${\cal O}(\kappa^3)$ in Eq.~(\ref{eq:RwCoordinateCorrections}) contribute to the three-leg vertex of the tadpole diagram, but only the last term $- \kappa^3 \partial_\alpha R_1 \mathsf{X}_2^\alpha$ leads to a sum over different momenta in a propagator denominator associated with the vertex, leading to the only non-vanishing contribution from the tadpole diagram to our final result. 

\subsection{The graviton self-interactions}

Expanding the Einstein-Hilbert Lagrangian $(-2/\kappa^2) \sqrt{-g} R$ in $h_{\mu \nu}$ yields an infinite series of graviton self-interactions. For the purposes of this work we need only the three-graviton vertex, which is nevertheless quite lengthy. In this section we provide the corresponding terms in the Lagrangian, from which the resulting vertex may be obtained by the standard Wick-contraction algorithm.

\subsubsection{Cubic}

In terms of the expansion coefficients of $\sqrt{-g}$ and $R$ given in secs. \ref{sec:volume-factor-expansion} and \ref{sec:scalar-curvature-expansion} the $O(\kappa)$ terms in the Einstein-Hilbert Lagrangian are
		\begin{eqn}
			L_{h^3}
				&= - 2 \kappa \bigg( R_3 + \gamma_1 R_2 + \gamma_2 R_1 \bigg),
		\end{eqn}
	using the fact that $\gamma_0 = 1$ and $R_0 = 0$. In terms of $h_{\mu \nu}$ this is
		\begin{eqn} 
			L_{h^3}
				&= \kappa \Bigg\{
						\sfrac{3}{2} h^{\mu \nu} \partial_\mu h^{\alpha \beta}
							\partial_\nu h_{\alpha \beta}
						- \sfrac{1}{2} h^{\mu \nu} \partial_\mu h \partial_\nu h
						+ 2 h^{\mu \nu} \partial_\mu h \partial_\alpha h {}_\nu {}^\alpha
						+ 2 h^{\mu \nu} \partial_\mu h {}_\nu {}^\alpha
							\partial_\alpha h
						+ 2 h^{\mu \nu} h {}_\mu {}^\alpha
							\partial_\nu \partial_\alpha h
		\\
				&\hspace{1cm}
						- h h^{\mu \nu} \partial_\mu \partial_\nu h
						- h^{\mu \nu} \partial_\alpha h \partial^\alpha h_{\mu \nu}
						+ \sfrac{1}{4} h \partial_\mu h \partial^\mu h
						- 2 h^{\mu \nu} \partial_\alpha h {}_\mu {}^\alpha
							\partial_\beta h {}_\nu {}^\beta
						- 4 h^{\mu \nu} \partial_\mu h {}_\nu {}^\alpha
							\partial_\beta h {}_\alpha {}^\beta
		\\
				&\hspace{1cm}
						+ h \partial_\mu h^{\mu \nu} \partial_\alpha h {}_\nu {}^\alpha
						+ 2 h^{\mu \nu} \partial^\alpha h_{\mu \nu} \partial_\beta 
							h {}_\alpha {}^\beta
						- h \partial_\mu h \partial_\nu h^{\mu \nu}
						- 2 h^{\mu \nu} h^{\alpha \beta}
							\partial_\mu \partial_\alpha h_{\nu \beta}
						+ 2 h^{\mu \nu} h^{\alpha \beta} 
							\partial_\mu \partial_\nu h_{\alpha \beta}
		\\
				&\hspace{1cm}
						- 4 h^{\mu \nu} h {}_\mu {}^\alpha 
							\partial_\alpha \partial_\beta h {}_\nu {}^\beta
						+ 2 h h^{\mu \nu} \partial_\mu \partial_\alpha h {}_\nu {}^\alpha
						+ \sfrac{1}{2} h_{\mu \nu} h^{\mu \nu} \partial_\alpha \partial_\beta
							h^{\alpha \beta}
						- \sfrac{1}{4} h^2 \partial_\mu \partial_\nu h^{\mu \nu}
						+ 2 h^{\mu \nu} h {}_\mu {}^\alpha \partial^2 h_{\nu \alpha}
		\\
				&\hspace{1cm}
						- h h^{\mu \nu} \partial^2 h_{\mu \nu}
						- \sfrac{1}{2} h_{\mu \nu} h^{\mu \nu} \partial^2 h
						+ \sfrac{1}{4} h^2 \partial^2 h
						- 2 h^{\mu \nu} \partial_\mu h {}_\alpha {}^\beta
							\partial_\beta h {}_\nu {}^\alpha
						- h^{\mu \nu} \partial_\alpha h {}_\mu {}^\beta
							\partial_\beta h {}_\nu {}^\alpha
		\\
				&\hspace{3cm}
						+ 3 h^{\mu \nu} \partial_\alpha h_{\nu \beta}
							\partial^\alpha h {}_\mu {}^\beta
						+ \sfrac{1}{2} h \partial_\mu h {}_\nu {}^\alpha
							\partial_\alpha h^{\mu \nu}
						- \sfrac{3}{4} h \partial_\alpha h_{\mu \nu}
							\partial^\alpha h^{\mu \nu}
					\Bigg\}.
            \label{eq:Lhhh}
		\end{eqn}
        The corresponding momentum space vertex requires a permutation over all three momenta, leading to six momentum combinations for every term in Eq.~(\ref{eq:Lhhh}).  Given the length of the resulting expression, we do not present it here.  We denote the vertex as
		\begin{eqn}
    	\begin{tikzpicture}[baseline=(v.base)]
    		\begin{feynman}
    			\vertex (v);
				\vertex at ($(v) + (1cm, 1cm)$) (h1) {$\mu \nu$};
				\vertex at ($(v) + (1cm, -1cm)$) (h2) {$\rho \sigma$};
				\vertex at ($(v) + (-1.4cm, 0)$) (h3) {$\alpha \beta$}; 
    			\diagram*{
    				(v) --[boson, momentum = $p_1$] (h1);
    				(v) --[boson, momentum = $p_2$] (h2);
    				(v) --[boson, momentum = $p_3$] (h3);
    			};
    		\end{feynman}
    	\end{tikzpicture}
			&= V^{\mu \nu \rho \sigma \alpha \beta}_{h^3}(p_1, p_2, p_3).
		\end{eqn}

\subsubsection{Quartic}

	The four-graviton vertex is denoted diagrammatically as
		\begin{eqn}
    	\begin{tikzpicture}[baseline=(v.base)]
    		\begin{feynman}
    			\vertex (v);
				\vertex at ($(v) + (1cm, 1cm)$) (h1) {$\mu \nu$};
				\vertex at ($(v) + (1cm, -1cm)$) (h2) {$\rho \sigma$};
				\vertex at ($(v) + (-1cm, -1cm)$) (h3) {$\alpha \beta$}; 
				\vertex at ($(v) + (-1cm, 1cm)$) (h4) {$\gamma \delta$};    			
    			\diagram*{
    				(v) --[boson, momentum = $p_1$] (h1);
    				(v) --[boson, momentum = $p_2$] (h2);
    				(v) --[boson, momentum = $p_3$] (h3);
    				(v) --[boson, momentum = $p_4$] (h4);
    			};
    		\end{feynman}
    	\end{tikzpicture}
			&= V^{\mu \nu \rho \sigma \alpha \beta \gamma \delta}_{h^4}(p_1, p_2, p_3, p_4).
		\end{eqn}

      \noindent This vertex appears only in tadpole diagrams to the order we work here, and these diagrams vanish when the mass of the particle in the loop is zero.  Thus, we do not display the lengthy expression for this vertex explicitly.

\subsection{The ghost-graviton vertex}

The ghost-graviton interaction terms are given in Eq.~(\ref{eq:ghost_lagrangian}), leading to
		\begin{eqn}
			L_{c \bar{c} h}
				&= \kappa
					\Bigg\{
						\Bigg( 1 - \frac{2 (1 + \beta)}{d} \Bigg) 
							\bar{c}^\mu h_{\nu \rho} \partial^\rho \partial_\mu c^\nu
						- \frac{1 + \beta}{d} 
							\big( \bar{c}^\mu \partial_\mu c^\nu \partial_\nu h
								+ \bar{c}^\mu c^\nu \partial_\mu \partial_\nu h \big)
						- \frac{2(1 + \beta)}{d} 
							\bar{c}^\mu \partial_\mu h_{\nu \rho} \partial^\rho c^\nu
		\\
				&\hspace{2cm}
						+ \bar{c}^\mu \partial_\mu c^\nu \partial_\rho {h_\nu}^\rho
						+ \bar{c}^\mu c^\nu \partial_\nu \partial_\rho {h_\nu}^\rho
						+ \bar{c}^\mu h_{\mu \nu} \partial^2 c^\nu
						+ \bar{c}^\mu \partial_\nu h_{\mu \rho} \partial^\rho c^\nu
						+ \bar{c}^\mu \partial_\rho h_{\mu \nu} \partial^\rho c^\nu
					\Bigg\}.
		\end{eqn}
	This yields a three-point vertex, which we denote
		\begin{eqn}
    	\begin{tikzpicture}[baseline=(v.base)]
    		\begin{feynman}
    			\vertex (v);
				\vertex at ($(v) + (1cm, 1cm)$) (cbar) {$\mu$};
				\vertex at ($(v) + (1cm, -1cm)$) (c) {$\nu$};
				\vertex at ($(v) + (-1.4cm, 0)$) (grav) {$\rho \sigma$};    			
    			\diagram*{
    				(cbar) --[charged scalar] (v) 
						--[charged scalar, momentum' = $k$] (c),
					(v) --[boson, momentum' = $p$] (grav)
    			};
    		\end{feynman}
    	\end{tikzpicture}
			&= V^{\mu \nu \rho \sigma}_{\bar{c} c h}(k,p),
		\end{eqn}
	where we omit mention of the antighost momentum because it does not appear in the vertex (since no derivatives act on $\bar{c}$ in the Lagrangian). Explicitly, this vertex is
		\begin{eqn}
			V^{\mu \nu \rho \sigma}_{\bar{c} c h}(k,p)
				&= - \ii \kappa
					\Bigg\{
						\Bigg( 1 - \frac{2(1 + \beta)}{d} \Bigg) 
							k^\mu k^\rho \eta^{\nu \sigma}
						- \frac{1 + \beta}{d} 
							\big( k^\mu p^\nu \eta^{\rho \sigma}
								+ p^\mu p^\nu \eta^{\rho \sigma} \big)
						- \frac{2(1 + \beta)}{d} p^\mu k^\rho \eta^{\nu \sigma}
		\\
				&\hspace{2cm}
						+ k^\mu p^\rho \eta^{\nu \sigma}
						+ p^\nu p^\rho \eta^{\mu \sigma}
						+ k^2 \eta^{\mu \rho} \eta^{\nu \sigma}
						+ k^\rho p^\nu \eta^{\mu \sigma}
						+ (k \cdot p) \eta^{\mu \rho} \eta^{\nu \sigma}
					\Bigg\}.
		\end{eqn}

\subsection{\label{higher_deriv} Higher derivative operator insertions}

At one-loop order we also need the tree-level insertions of the ${\cal O}(R^2)$ operators.  We thus need $\sqrt{-g}R^2$ and $\sqrt{-g}C^2$ to quadratic order.

\subsubsection{Quadratic $\sqrt{-g}R^2$ vertex}

 The operator $\sqrt{-g}R^2$ expanded to quadratic order is
\begin{eqn}\label{eq:quadR2vertex}
			L_{R^2, h^2}
				&= c_1 \kappa^2
					\Bigg\{ 
                       \partial_\beta \partial_\alpha h^{\alpha \beta} \partial_\delta \partial_\gamma h^{\gamma \delta}
                       - 2 \partial_\beta \partial_\alpha h^{\alpha \beta} \partial^2 h
                       + (\partial^2 h)^2
\Bigg\}.
		\end{eqn}

\subsubsection{Quadratic $\sqrt{-g}C^2$ vertex}

The operator $\sqrt{-g}C^2$ expanded to quadratic order is
\begin{eqn}
      L_{C^2, h^2}
				&= c_2 \kappa^2
					\Bigg\{ -\sfrac{1}{2}\partial_\rho \partial_\sigma h \partial^\rho \partial^\sigma h
        - \partial_\alpha \partial^\rho h^{\alpha \beta} \partial_\sigma \partial_\beta h_\rho^\sigma
        + \partial^\rho \partial^\beta h \partial_\sigma \partial_\beta h_\rho^\sigma
        -\partial_\alpha \partial^\rho h^{\alpha \beta} \partial_\sigma \partial_\rho h_\beta^\sigma
    \\
				&\hspace{1cm}
            + \partial^\rho \partial^\beta h \partial_\sigma \partial_\rho h_\beta^\sigma
        +\sfrac{1}{3}\partial_\beta \partial_\alpha h^{\alpha \beta} \partial_\sigma \partial_\rho h^{\sigma \rho}
             -\sfrac{2}{3}\partial^2 h \partial_\sigma \partial_\rho h^{\rho \sigma}
             -\sfrac{1}{2} \partial^2 h^{\alpha \beta} \partial^2 h_{\alpha \beta}
             \\
				&\hspace{1cm}
            +2 \partial_\alpha \partial^\rho h^{\alpha \beta} \partial^2 h_{\beta \rho}
            -\partial^\rho \partial^\beta h \partial^2 h_{\beta \rho}
            +\sfrac{1}{3} \partial^2 h \partial^2 h
            +\partial_\beta \partial_\alpha h_{\rho \sigma} \partial^\sigma \partial^\rho h^{\alpha \beta}
            \\
				&\hspace{1cm}
            - \partial_\beta \partial_\rho h_{\alpha \sigma} \partial^\sigma \partial^\rho h^{\alpha \beta}
            - \partial_\sigma \partial_\beta h_{\alpha \rho} \partial^\sigma \partial^\rho h^{\alpha \beta}
            + \partial_\sigma \partial_\rho h_{\alpha \beta} \partial^\sigma \partial^\rho h^{\alpha \beta}       
     \Bigg\}.
		\end{eqn}

We denote these vertices by a solid square box,

        \bea \begin{tikzpicture}[baseline=(b.base)]
    		\begin{feynman}
    			\vertex [] (v1) {$\mu \nu$};
				\vertex [square dot] at ($(v1) + (1cm, 0)$) (b) {};
				\vertex [] at ($(v1) + (2cm, 0)$) (v2) {$\rho \sigma$};
    			\diagram*{
    				(v1) --[boson, momentum = $p_1$] (b)
                             --[boson, momentum = $p_2$] (v2)
    			};
    		\end{feynman}
    	\end{tikzpicture}	
        = V_{h^2, (R^2, C^2)}^{\mu \nu \rho \sigma}  \left(p_1,p_2\right).
\eea

\section{\label{sec:corr}Euclidean correlation functions}

We use the Feynman rules for gravity and the coordinate corrections discussed in previous sections to compute Euclidean correlation functions, with the hope of comparing them to results from numerical lattice gravity calculations.  One of the main results of this work is the calculation of the curvature-curvature correlation function, $\ev{\mathcal{R}(x) \mathcal{R}(y)}$, which we compute through next-to-leading (one-loop) order.  At leading order, the tree-level diagram gives a result that is analytic in the external momentum, and its Fourier transform to position space leads to (derivatives of) a delta function.  Thus, the effective theory at leading order vanishes apart from a contact term and is not the leading behavior that would be seen in a numerical lattice calculation, if the lattice theory were indeed to reproduce the effective theory in the low energy limit.  This tree-level result was obtained already decades ago \cite{Modanese:1992ir}.  At next-to-leading order, the correlation function $\ev{\mathcal{R}(x) \mathcal{R}(y)}$ receives one-loop, non-analytic contributions that lead to a power-law fall off with the source-sink separation in position space.  We calculate this expression and show that, once we account for the coordinate corrections implicit in the construction of the invariantized scalar curvature, our final result is completely independent of the gauge parameters $\alpha$ and $\beta$ of the generalized de Donder gauge.  This cancellation of the gauge dependence is highly non-trivial, and it is a strong cross-check that our result is correct.  
The final expression for the correlation function is fully predicted by the low energy effective theory, as it involves only Newton's constant as an input parameter.

We also consider the volume-volume correlation function $\ev{\sqrt{-\mathcal{G}(x)} \sqrt{-\mathcal{G}(y)}}$.  The lower dimensionality of this correlation function compared to the curvature-curvature function implies a leading order, tree-level contribution proportional to $1/q^2$, with $q$ the external momentum.  Thus, the momentum-space expression is non-analytic even at leading order, and its position-space Fourier transform leads to a power-law fall off in the source-sink separation.  We calculate this tree-level contribution and show that, as long as one includes the coordinate corrections, it is also independent of the gauge parameters $\alpha$ and $\beta$, providing a further cross-check of the machinery.  Thus, the leading order contribution to the volume-volume correlator does not vanish, and could potentially be compared to lattice gravity simulations, since again, it is fully predicted by the effective theory and depends only on Newton's constant.  

We are able to extend our result for the volume-volume correlation function to next-to-leading order without performing an explicit one-loop calculation by taking advantage of the small number of local counter-terms available to renormalize the one-loop divergences.  The same linear combination of ${\cal O}(R^2)$ operators appears in both correlation functions, where the low energy constants that multiply the ${\cal O}(R^2)$ operators absorb the logarithmic divergences of the one-loop insertions of the Einstein-Hilbert action.  Using dimensional regularization, and specializing to the $\overline{MS}$ scheme, the low energy constants multiplying the $\sqrt{-g}R^2$ and $\sqrt{-g}C^2$ terms in Eq.~(\ref{eq:R2Lagrangian}) take the form
\bea  c_i = c_i^r + \frac{1}{16\pi^2}\left[\frac{1}{\varepsilon} - \gamma_{\rm E} + \ln{(4 \pi)} \right]\Gamma_i,
\eea
where the $1/\varepsilon$ pole is an ultra-violet divergence, and the $\Gamma_i$ are the coefficients required for the cancellation of the divergences appearing in the one-loop contributions to the correlation functions.  The renormalized low energy constants $c_i^r$ are finite, but they are undetermined within the effective theory.  They are scale dependent, and this scale dependence cancels against the logarithmic scale dependence that accompanies the regulated logarithmic divergences coming from loop diagrams.  Note that this appears to contradict the conventional wisdom concerning the one-loop renormalizablity of pure gravity \cite{tHooft:1974toh}, but it is necessary, as we reiterate.  Although the contributions of ${\cal O}(R^2)$ operators vanish in on-shell S-matrix elements of tree-level graviton scattering, they do appear in the correlation functions considered in this work, as we show by explicit calculation.  The field-redefinition that can be applied to remove them from the Lagrangian \cite{AccettulliHuber:2019jqo} reintroduces them to the external operators of the correlation functions, leading to the same result as that of the untransformed Lagrangian.  Their next-to-leading order (tree-level) contributions are gauge-invariant, as are the one-loop divergences their counter terms must cancel, so that the $\Gamma_i$ can be determined unambiguously.  Given that this must be true of all correlation functions for the effective theory to be consistent, the NLO one-loop result for $\ev{\mathcal{R}(x) \mathcal{R}(y)}$ uniquely fixes the NLO one-loop contribution to $\ev{\sqrt{-\mathcal{G}(x)} \sqrt{-\mathcal{G}(y)}}$ as well.  This assumes that the logarithmic divergence that typically accompanies the logarithmic dependence of an energy scale appears with the same coefficient.  In the correlation functions considered here, the only energy scale appearing in the logarithms is the external momentum scale, so we might expect this relation to hold.  The relation can be modified in gravity, however, by the appearance of the Gauss-Bonnet term, an evanescent operator whose coefficient is proportional to a $1/\varepsilon$ pole in dimensional regularization. Since the Gauss-Bonnet term appears without an accompanying logarithmic dependence on an energy scale \cite{Bern:2015xsa}, it can modify the relation between the logarithmic divergence and the logarithmic dependence on the energy scale.  However, the Gauss-Bonnet term does not contribute to our correlation functions to the order we are working, so we expect that the relation between the pole structure and the logarithmic dependence on the external momentum does hold.

We work in momentum space and Lorentz signature for the perturbative calculations, applying the Euclidean continuation and the Fourier transform to position space at the end.  We argue that the Euclidean continuation must be applied with some care for the volume correlation function because of a subtlety associated with the conformal mode \cite{Gibbons:1978ac, Mazur:1989by, tHooft:2002umw}.  We apply the perturbative form of the Gibbons-Hawking prescription \cite{Gibbons:1978ac}, and we show that this is required for our correlation functions to remain positive definite.

\subsection{\label{curvature_treelevel} Curvature correlation function at tree level}

The curvature-curvature correlation function $\langle \mathcal{R}(x) \mathcal{R}(y) \rangle$ was first computed at tree-level in \cite{Modanese:1992ir}.  This correlation function does not receive coordinate corrections to this order, and our result agrees with that of Ref.~\cite{Modanese:1992ir}. 
 This result is given by the insertion of $R_1$ from Eq.~\ref{eq:R1} for the external operators, and it is independent of $\alpha$ and $\beta$ when calculated in the generalized de Donder gauge.  We find  
\begin{eqn}
		\label{eq:curvature-tree-0}
			\begin{tikzpicture}[baseline=(b.base)]
    		\begin{feynman}
    			\vertex [dot] (v1) {};
				\vertex at ($(v1) + (0, -0.09cm)$) (b);
				\vertex [dot] at ($(v1) + (2cm, 0)$) (v2) {};
    			\diagram*{
    				(v1) --[boson, momentum = $q$] (v2) 
    			};
    		\end{feynman}
    	\end{tikzpicture}			
				= \ii A_{{\cal R}}
				&= - \frac{3}{2} \ii \kappa^2 q^2,
		\end{eqn}
for the result in momentum space.  

\subsection{The volume-volume two-point function at tree-level}
\label{sec:volume-2pt-function}

We next consider the two-point function of the volume factor $\sqrt{-g}$ at leading order. We recall from Sec. \ref{sec:volume-factor-expansion} that the standard expansion of the volume factor is
		\begin{eqn}
			\sqrt{-g}
				&= 1 + \sfrac{1}{2} \kappa h + O(\kappa^2),
		\end{eqn}
	which is augmented in the invariantized volume factor by a coordinate correction term,
		\begin{eqn}
			\sqrt{ - \det \mathcal{G}}
				&= 1 + \kappa \bigg( \sfrac{1}{2} h - \partial_\mu \mathsf{X}_1^\mu \bigg).
		\end{eqn}
	The two-point function of the invariantized volume factor therefore receives three contributions at tree level, corresponding to both external vertices being standard; one being a coordinate correction and the other standard; and both being a coordinate correction:
		\begin{eqn}
			\ev{\sqrt{-\det \mathcal{G}(x)} \sqrt{-\det \mathcal{G}(y)}}
				&= 1 + \kappa^2 \Bigg\{
						\ev{ \bigg( \sfrac{1}{2} h(x) \bigg) \bigg( \sfrac{1}{2} h(y) \bigg)}
						- 2 \ev{ \bigg( \partial_\mu \mathsf{X}_1^\mu (x) \bigg) 
							\bigg( \sfrac{1}{2} h(y) \bigg)}
						+ \ev{ \bigg( \partial_\mu \mathsf{X}_1^\mu (x) \bigg) 
							\bigg( \partial_\mu \mathsf{X}_1^\mu (y) \bigg)}
					\Bigg\}.
		\end{eqn}
	Note that by Lorentz invariance the two cross-terms must be equal. In momentum space we therefore find three diagrams at this order. The standard contribution is
		\begin{eqn}
		\label{eq:volume-tree-0-setup}
			\begin{tikzpicture}[baseline=(b.base)]
    		\begin{feynman}
    			\vertex [dot] (v1) {};
				\vertex at ($(v1) + (0, -0.09cm)$) (b);
				\vertex [dot] at ($(v1) + (2cm, 0)$) (v2) {};
    			\diagram*{
    				(v1) --[boson, momentum = $q$] (v2) 
    			};
    		\end{feynman}
    	\end{tikzpicture}			
				= \ii A_0
				&= \sfrac{1}{4} \kappa^2 \eta^{\mu \nu} \eta^{\rho \sigma} 
					\Delta_{\mu \nu \rho \sigma}(q).
		\end{eqn}
	With one standard vertex, and including the factor of $2$ to account for the coordinate correction being on either end, we have
		\begin{eqn}
		\label{eq:volume-tree-1-setup}
			\begin{tikzpicture}[baseline=(b.base)]
    		\begin{feynman}
    			\vertex [crossed dot] (v1) {};
				\vertex at ($(v1) + (0, -0.09cm)$) (b);
				\vertex [dot] at ($(v1) + (2cm, 0)$) (v2) {};
    			\diagram*{
    				(v1) --[boson, momentum = $q$] (v2) 
    			};
    		\end{feynman}
    	\end{tikzpicture}	
			= \ii A_1
				&= - \kappa^2 \sfrac{1}{q^2} \mathcal{J}_1^{\alpha \beta \mu \nu} 
					p_\alpha p_\beta \eta^{\rho \sigma} \Delta_{\mu \nu \rho \sigma}(p).
		\end{eqn}
	Finally, with coordinate corrections on both ends,
		\begin{eqn}
		\label{eq:volume-tree-2-setup}
			\begin{tikzpicture}[baseline=(b.base)]
    		\begin{feynman}
    			\vertex [crossed dot] (v1) {};
				\vertex at ($(v1) + (0, -0.09cm)$) (b);
				\vertex [crossed dot] at ($(v1) + (2cm, 0)$) (v2) {};
    			\diagram*{
    				(v1) --[boson, momentum = $q$] (v2) 
    			};
    		\end{feynman}
    	\end{tikzpicture}	
				= \ii A_2
				&= \kappa^2 \sfrac{1}{q^4} \mathcal{J}_1^{\alpha \beta \mu \nu}
					\mathcal{J}_1^{\gamma \delta \rho \sigma}
					q_\alpha q_\beta q_\gamma q_\delta \Delta_{\mu \nu \rho \sigma}(q).
		\end{eqn}
	Note that without accounting for coordinate corrections we would only have the diagram $A_0$, which we show below is not sufficient to obtain a gauge-invariant result. 

\subsubsection{In a simple gauge}

Before calculating the above diagrams for general values of the gauge parameters $(\alpha, \beta)$ we evaluate them in harmonic gauge, with $\alpha = 1/2$, $\beta = (d/2) - 1$. Further, since we are only working at tree level we can safely set $d = 4$. In this gauge the graviton propagator becomes
		\begin{eqn}
			\Delta_{\mu \nu \rho \sigma}(q)
				&= \frac{\ii}{2q^2} \big( \eta_{\mu \rho} \eta_{\nu \sigma} 
					+ \eta_{\mu \sigma} \eta_{\nu \rho} 
					- \eta_{\mu \nu} \eta_{\rho \sigma} \big).
		\end{eqn}
	The standard diagram then becomes
		\begin{eqn}
			\ii A_0
				&= \frac{1}{8 q^2} \ii \kappa^2 \eta^{\mu \nu} \eta^{\rho \sigma}
					\big( \eta_{\mu \rho} \eta_{\nu \sigma} 
					+ \eta_{\mu \sigma} \eta_{\nu \rho} 
					- \eta_{\mu \nu} \eta_{\rho \sigma} \big)
				= - \frac{\ii \kappa^2}{q^2}.
		\end{eqn}
	We happen to find the same value for the first coordinate correction diagram,
		\begin{eqn}
			\ii A_1
				&= - \ii \kappa^2 \frac{1}{2q^4} 
					\big( \eta^{\alpha \mu} \eta^{\beta \nu} 
						- \sfrac{1}{2} \eta^{\alpha \beta} \eta^{\mu \nu} \big)
					q_\alpha q_\beta \eta^{\rho \sigma}
						\big( \eta_{\mu \rho} \eta_{\nu \sigma} + \eta_{\mu \sigma} \eta_{\nu \rho}
							- \eta_{\mu \nu} \eta_{\rho \sigma} \big)
				= - \frac{\ii \kappa^2}{q^2},
		\end{eqn}
	while for the second we have
		\begin{eqn}
			\ii A_2
				&= \ii \kappa^2 \frac{1}{2 q^6}
					\big( \eta^{\alpha \mu} \eta^{\beta \nu} 
						- \sfrac{1}{2} \eta^{\alpha \beta} \eta^{\mu \nu} \big)
					\big( \eta^{\gamma \rho} \eta^{\delta \sigma}
						- \sfrac{1}{2} \eta^{\gamma \delta} \eta^{\rho \sigma} \big)
					\big( \eta_{\mu \rho} \eta_{\nu \sigma} + \eta_{\mu \sigma} \eta_{\nu \rho}
						- \eta_{\mu \nu} \eta_{\rho \sigma} \big)
				= \frac{\ii \kappa^2}{2q^2}.
		\end{eqn}
	We thus find that the tree-level two-point function of the invariantized volume factor is
		\begin{eqn}
			\ii A_0 + \ii A_1 + \ii A_2
				&= - \frac{3 \ii \kappa^2}{2 q^2}.
		\end{eqn}

\subsubsection{In a general gauge}

We now subject our machinery to its first real test: if we leave $\alpha$ and $\beta$ arbitrary, do we find a gauge-invariant result for the two-point function of the invariantized volume factor?

We reevaluate the expressions (\ref{eq:volume-tree-0-setup}), (\ref{eq:volume-tree-1-setup}), and (\ref{eq:volume-tree-2-setup}) for the three tree-level diagrams, now using the more complicated form (\ref{eq:graviton_propagator}) for the graviton propagator.  
	Such a complicated propagator makes even the simplest of calculations quite tedious.  We performed the calculations using specialized packages written for {\sc Mathematica}, with one of us using {\sc xAct} and the other using {\sc FeynCalc} \cite{Mertig:1990an, Shtabovenko:2016sxi, Shtabovenko:2020gxv}.

For the first diagram we find
		\begin{eqn}
			\ii A_0
				= \sfrac{1}{4} \kappa^2 \eta^{\mu \nu} \eta^{\rho \sigma} \Delta_{\mu \nu \rho \sigma}
				= \frac{2 (1 + 4(\alpha - 1) - 2 \alpha)}{(\beta - 3)^2} \frac{\ii \kappa^2}{q^2}.
		\end{eqn}
	Note that this diagram is {\em gauge-dependent!} Firstly, this is what we expect -- this is the only diagram that occurs at tree level in $\ev{\sqrt{-g(x)} \sqrt{-g(y)}}$ without coordinate corrections, and $\sqrt{-g}$ is not gauge-invariant, so its two-point function should not be either. Secondly, this means that, if $\ev{\sqrt{-\det\mathcal{G}(x)} \sqrt{-\det\mathcal{G}(y)}}$ is to be gauge-invariant, the gauge parameters must cancel in a nontrivial manner.

Indeed, this is exactly what we find. For the first coordinate correction diagram we have
		\begin{eqn}
			\ii A_1
				&= - \frac{16(1 + \alpha) + 2(1 + \beta) - 4 (3 + 2 \alpha + 2 \beta)}{(\beta - 3)^2}
					\frac{\ii \kappa^2}{q^2},
		\end{eqn}
	and for the second
		\begin{eqn}
			\ii A_2
				&= \frac{64(\alpha - 1) + 4(\beta + 1)^2 + 16(5-2 \alpha + 4 \beta) 
						- 16(2 + 3 \beta + \beta^2)}{8(\beta-3)^2}
					\frac{\ii \kappa^2}{q^2}.
		\end{eqn}
	Since all three prefactors share a common denominator we can focus on the sum of the numerators, which simplifies nicely:
		\begin{eqn}
			\bigg[ 2 (1 + 4(\alpha - 1) - 2 \alpha) \bigg]
				&- \bigg[ 16(1 + \alpha) + 2(1 + \beta) - 4 (3 + 2 \alpha + 2 \beta) \bigg]
		\\
				&+ \sfrac{1}{8}\bigg[ 64(\alpha - 1) + 4(\beta + 1)^2 + 16(5-2 \alpha + 4 \beta) 
						- 16(2 + 3 \beta + \beta^2) \bigg]
		\\
				&\hspace{-2cm} = - \sfrac{3}{2} (\beta - 3)^2.
		\end{eqn}
	This then cancels the shared denominator, as it must, leaving precisely our prior result:
		\begin{eqn}\label{eq:volume-tree-0-final}
			\ii A_{{\cal G}} \equiv A_0 + \ii A_1 + \ii A_2
				&= - \frac{3 \ii \kappa^2}{2 q^2}.
		\end{eqn}
	We see that even though the expression for each of the three diagrams is not separately gauge-invariant, their sum {\em is} -- indicating that our construction of a gauge-invariant volume factor has been successful.

\subsection{\label{analytic_continuation} Analytic continuation and position space correlation functions}

Since the focus of this work is a comparison of the results of the low energy effective theory of gravity with that of Euclidean lattice calculations, we need our correlation functions in Euclidean position-space.  The naive continuation of gravity to Euclidean space is unbounded from below, so that there is no guarantee of a positive energy spectrum, and no guarantee of reflection positivity.  In order to have a consistent Euclidean theory, one may choose the integration contour of the conformal mode to be opposite that of the usual convention.  This well-known prescription is often invoked to define a non-perturbative gravitational path integral \cite{Gibbons:1978ac}.  Although defining the admissible metrics in the full, non-perturbative Euclidean gravitational path integral remains an important open problem \cite{Witten:2021nzp}, the Gibbons-Hawking prescription ought to be sufficient for perturbation theory about flat space.  We find that this standard prescription does in fact ensure that the Euclidean correlation functions calculated here are positive definite.  This is reassuring, since these correlation functions are universal long-distance predictions of gravity, making them independent of the ultra-violet completion.  The underlying theory is presumably unitary, thus implying the reflection positivity of a properly defined Euclidean theory.

We show that this modified continuation is needed for the volume correlation function by first demonstrating that at tree-level the conformal mode is the only thing that propagates between the external operators.  The York decomposition of the metric \cite{York:1973ia} leads to
\bea\label{eq:YorkDecomp}  h_{\mu \nu}= h_{\mu \nu}^\perp + \partial_\mu a_\nu^\perp + \partial_\nu a_\mu^\perp + \partial_\mu \partial_\nu a +\frac{1}{4}\eta_{\mu \nu} \Phi,
\eea
where $h_{\mu \nu}^\perp$ is a transverse and traceless tensor, i.e. $g^{\mu \nu} h_{\mu \nu}^\perp=\partial^\mu h_{\mu \nu}^\perp=0$, the vector mode $a_\mu$ is decomposed into a traceless mode $\partial_\nu a_\nu^\perp=0$ and a scalar mode $a$, and $\Phi$ is the gauge-invariant conformal mode.  From Eq.~(\ref{eq:invar-volume-expansion}) the expansion of the external operator $\sqrt{(-{\rm det}{\cal G})}$ involving only a single graviton leg is $1/2 h - \partial^\mu X_{1,\mu}$.  Inserting the decomposition Eq.~(\ref{eq:YorkDecomp}) into this, we find
\bea\label{eq:volconformal}  \frac{1}{2} h - \partial^\mu X_{1,\mu} = -\frac{3}{4}\Phi
\eea
where everything except for the conformal mode vanishes.  Thus, the tree-level diagram of the volume correlation function propagates only the conformal mode.  (Note that without the coordinate correction term in Eq.~(\ref{eq:volconformal}), the operator would also include the scalar $a$ mode.)  

The momentum space result for the volume correlator at leading order, Eq.~(\ref{eq:volume-tree-0-final}), is just an overall constant times the momentum space propagator of a scalar particle, and as such its Fourier transform appears to be a standard field theory exercise.  However, because the propagating degree of freedom is the conformal mode, the Euclidean continuation is not that of an ordinary real scalar field, but is defined by a rotation of contour that is the opposite of the usual procedure \cite{Gibbons:1978ac, Mazur:1989by}.  This choice for the integration contour removes the divergences associated with the unboundedness of the conformal mode in the Euclidean theory. It is also required to make our correlation functions positive definite, as we demonstrate below.

The standard Fourier transform of a scalar propagator to position space is
\bea \int \frac{d^4q}{(2\pi)^4} \frac{i}{q^2 + i \varepsilon} e^{-i q \cdot x} = -\frac{1}{4\pi^2 x^2},
\eea
and upon Euclidean continuation,
\bea  -\frac{1}{4\pi^2 x^2} \to \frac{1}{4\pi^2 x_E^2},
\eea
with $x_E$ the Euclidean distance interval.  This propagator corresponds to the Euclidean two-point correlation function of scalar fields $\langle 0| \phi(x_E) \phi(0) |0 \rangle$, and it is positive, as required by reflection positivity. 

Since the result for the tree-level volume correlator comes from the propagation of the conformal mode, when we regulate the expression for the purpose of taking the Fourier transform, we adopt the opposite $i\varepsilon$ prescription from that of a standard propagator,
\bea  i A_{\cal G} = -\frac{3 i \kappa^2}{2(q^2 - i\varepsilon)}.
\eea
The Fourier transform of this expression picks up an extra minus sign compared to the standard Feynman propagator due to the opposite contour rotation implied by the change in sign of the $i\varepsilon$ term.  This leads to
\bea  \int \frac{d^4 q}{(2\pi)^4} i A_{\cal G} e^{-i q \cdot x} = \frac{-3 \kappa^2}{2}\left(\frac{1}{4\pi^2 x^2}\right)
\to \frac{3 \kappa^2}{8 \pi^2 x^2_E},
\eea
where the Euclidean continuation of this correlation function is positive as expected if we are to have a well-defined Euclidean theory.

We also consider the Fourier transform of the tree-level curvature correlator $\langle \mathcal{R}(x) \mathcal{R}(y) \rangle$.  
%
%
To evaluate it, we first write the Fourier transform of both sides of Eq.~(\ref{eq:curvature-tree-0}) as
\bea\label{eq:treelevelR_FT} \int \frac{d^4 q}{(2\pi)^4} i A_{\cal R} e^{-i q \cdot x} = -\frac{3\kappa^2}{2} \partial^{4} \int \frac{d^4 q}{(2\pi)^4}\frac{i}{q^2+i\varepsilon} e^{-i q \cdot x}.
\eea
Here, since the amplitude $i A_{\cal R}$ is everywhere analytic, its contour deformation should not matter, despite the fact that the curvature correlator at tree-level also only propagates the conformal mode.  We choose to follow the standard continuation of the contour integral when evaluating the Fourier transform.  The Fourier transform appearing in Eq.~(\ref{eq:treelevelR_FT}) is just the massless Feynman scalar propagator in position space
\bea \int \frac{d^4q}{(2\pi)^4} \frac{i}{q^2 + i \varepsilon} e^{-i q \cdot x} = -\frac{1}{4\pi^2 x^2}.
\eea
We use this and the fact that the covariant Laplacian of the position-space propagator is a four-dimensional delta function to write,
\bea \int \frac{d^4 q}{(2\pi)^4} i A_{\cal R} e^{-i q \cdot x}= -\frac{3\kappa^2}{2}\partial^2\left[\partial^2\left(-\frac{1}{4\pi^2 x^2} \right) \right] =  \frac{3\kappa^2}{2}\partial^2 \left[-i\delta^4(x)\right]
\eea
Under Euclidean continuation we have $\partial^2 \to -\partial^2_E$ and $-i \delta^4(x) \to -\delta^4(x_E)$, where the delta function is now over $R^4$ instead of $R^{1,3}$.  This leads to 
\bea \int \frac{d^4 q}{(2\pi)^4} i A_{\cal R} e^{-i q \cdot x} \to \frac{3\kappa^2}{2}\partial_E^2 \delta^4(x_E),
\eea
which vanishes away from the origin.   


\subsection{\label{curvature_1loop} Curvature correlation function at one-loop}

In order to present the calculation of the curvature correlation function to one-loop order, we need to consider the effective Lagrangian at $O(R^2)$, Eq.~(\ref{eq:R2Lagrangian}).  This is a convenient basis for the $O(R^2)$ operators, since the insertion of the Weyl squared tensor in tree-level diagrams vanishes for the correlation functions we consider.  This is expected because the external operators propagate only the conformal mode (at tree-level), and the Weyl tensor is conformally invariant.  

The $\sqrt{-g}R^2$ operator, expanded to quadratic order in $h_{\mu \nu}$, is given by $L_{R^2, h^2}$ in Eq.~(\ref{eq:quadR2vertex}).  When inserted into the tree-level diagram, it yields
\bea \begin{tikzpicture}[baseline=(b.base)]
    		\begin{feynman}
    			\vertex [dot] (v1) {};
				\vertex [square dot] at ($(v1) + (1cm, 0)$) (b) {};
				\vertex [dot] at ($(v1) + (2cm, 0)$) (v2) {};
    			\diagram*{
    				(v1) --[boson, momentum = $q$] (b)
                             --[boson, momentum = $q$] (v2)
    			};
    		\end{feynman}
    	\end{tikzpicture}	
        =\ii A_{{\cal R}}^{p^4, {\rm tree-level}} = - \frac{9}{2} \ii c_1\kappa^4 q^4.
\eea
This result is independent of $\alpha$ and $\beta$ when calculated in the generalized de Donder gauge.  This is as it must be, since the coefficient $c_1$ does not appear in any other diagrams to this order, so no additional terms could cancel a residual gauge-dependence.

The full next-to-leading order expression for the curvature correlation function can be written as
\begin{eqn}
		\label{eq:curvature-tree-0-setup}
				 \ii A_{{\cal R}}^{p^4}
				&= i{A_{\cal R}^{\rm one-loop}} - \frac{9}{2} \ii c_1\kappa^4 q^4,
		\end{eqn}
where the term $iA_{\cal R}^{\rm one-loop}$ represents the one-loop insertions of the Einstein-Hilbert action in the curvature correlation function.  At one-loop, the coordinate corrections must be taken into account for this quantity.  The expression for $iA_{\cal R}^{\rm one-loop}$ is equal to the sum over the following eight Feynman diagrams:

\bea\label{eq:RR_diagram1}  \begin{tikzpicture}[baseline=(b.base)]
    		\begin{feynman}
    			\vertex [dot] (v1) {};
				\vertex at ($(v1) + (0, -0.09cm)$) (b);
				\vertex [dot] at ($(v1) + (2cm, 0)$) (v2) {};
    			\diagram*{
    				(v2) --[boson, momentum = $\ell$] (v1) 
                             --[boson, half left, momentum = $\ell + q$] (v2) 
    			};
    		\end{feynman}
    	\end{tikzpicture}			
        = \frac{1}{2}\int \frac{d^d \ell}{(2\pi)^d} \Bigg[E_{h^2}^{\mu \nu \eta \lambda}(\ell+q,-\ell)
           \Delta_{\mu \nu \rho \sigma}(\ell+q)\Delta_{\eta \lambda \alpha \beta}(\ell)
           E_{h^2}^{\alpha \beta \rho \sigma}(\ell,-\ell-q) \Bigg],
\eea

\bea \label{eq:RR_diagram2} \begin{tikzpicture}[baseline=(b.base)]
    		\begin{feynman}
    			\vertex [dot] (v1) {};
				\vertex at ($(v1) + (1.5cm, 0)$) (v3);
				\vertex [dot] at ($(v1) + (2.5cm, 0)$) (v2) {};
    			\diagram*{
    				(v3) --[boson, momentum = $\ell$] (v1) 
                             --[boson, half left, momentum = $\ell + q$] (v3) 
                             --[boson, momentum' = $q$] (v2) 
    			};
    		\end{feynman}
    	\end{tikzpicture}			
          = \int \frac{d^d \ell}{(2\pi)^d} \Bigg[E_{h^2}^{\mu \nu \eta \lambda}(\ell+q,-\ell)
            \Delta_{\mu \nu \rho \sigma}(\ell+q)\Delta_{\eta \lambda \alpha \beta}(\ell)V_{h^3}^{\alpha \beta \rho \sigma \delta \xi}(\ell,-\ell-q,q) \Bigg]\Delta_{\delta \xi \varepsilon \gamma}(q)E_h^{\varepsilon \gamma}(q),
\eea

\bea \label{eq:RR_diagram3}  \begin{tikzpicture}[baseline=(b.base)]
    		\begin{feynman}
    			\vertex [dot] (v1) {};
				\vertex at ($(v1) + (1cm, 0)$) (v3);
                    \vertex at ($(v1) + (2cm, 0)$) (v4);
				\vertex [dot] at ($(v1) + (3cm, 0)$) (v2) {};
    			\diagram*{
                        (v1) --[boson, momentum = $q$] (v3),
    				     (v4)  --[boson, half left, momentum = $\ell$] (v3)
                              --[boson, half left, momentum = $\ell + q$] (v4) 
                             --[boson, momentum = $q$] (v2) 
    			};
    		\end{feynman}
    	\end{tikzpicture}	 
          = \frac{1}{2} E_{h}^{\mu \nu}(q) \Delta_{\mu \nu \eta \lambda}(q)\int \frac{d^d \ell}{(2\pi)^d}\Bigg[
                V_{h^3}^{\eta \lambda \rho \sigma \alpha \beta}(-q,q+\ell,-\ell)
                \Delta_{\rho \sigma \delta \xi}(\ell+q)
               \nonumber \\ 
                \times \Delta_{\alpha \beta \varepsilon \gamma}(\ell)
                V_{h^3}^{\delta \xi \varepsilon \gamma \zeta \psi}(-\ell-q,\ell,q)\Bigg]
                 \Delta_{\zeta \psi \phi \chi}(q) E_{h}^{\phi \chi}(q), \ \ \ \ \ \ \ \ \ \ \ \ \ \ 
\eea

\bea \label{eq:RR_diagram4}  \begin{tikzpicture}[baseline=(b.base)]
    		\begin{feynman}
    			\vertex [dot] (v1) {};
				\vertex at ($(v1) + (1cm, 0)$) (v3);
                    \vertex at ($(v1) + (2cm, 0)$) (v4);
				\vertex [dot] at ($(v1) + (3cm, 0)$) (v2) {};
    			\diagram*{
                        (v1) --[boson, momentum = $q$] (v3),
    				     (v4)  --[charged scalar, half left, momentum = $\ell$] (v3)
                              --[charged scalar, half left, momentum = $\ell + q$] (v4) 
                             --[boson, momentum = $q$] (v2) 
    			};
    		\end{feynman}
    	\end{tikzpicture}	  = -E_h^{\mu \nu}(q)\Delta_{\mu \nu \eta \lambda}(q) \int \frac{d^d \ell}{(2\pi)^d} \Bigg[
                V_{\bar{c}ch}^{\rho \sigma \eta \lambda}(\ell+q,-q)S^{\sigma \alpha}(\ell+q)
                S^{\rho \beta}(\ell) V_{\bar{c}ch}^{\alpha \beta \delta \xi}(\ell,q) \Bigg]
              \nonumber   \\
                \times\Delta_{\delta \xi \varepsilon \gamma}(q) E_h^{\varepsilon \gamma}(q), \ \ \ \ \ \ \ \ \ 
\eea

\bea \label{eq:RR_diagram5}  \begin{tikzpicture}[baseline=(b.base)]
    		\begin{feynman}
    			\vertex [crossed dot] (v1) {};
				\vertex at ($(v1) + (0, -0.09cm)$) (b);
				\vertex [crossed dot] at ($(v1) + (2cm, 0)$) (v2) {};
    			\diagram*{
    				(v2) --[boson, momentum = $\ell$] (v1) 
                             --[boson, half left, momentum = $\ell + q$] (v2) 
    			};
    		\end{feynman}
    	\end{tikzpicture} = \frac{1}{2}\int \frac{d^d \ell}{(2\pi)^d} \Bigg[\tilde{E}_{h^2}^{\mu \nu \eta \lambda}(\ell+q,-\ell)
           \Delta_{\mu \nu \rho \sigma}(\ell+q)\Delta_{\eta \lambda \alpha \beta}(\ell)
           \tilde{E}_{h^2}^{\alpha \beta \rho \sigma}(\ell,-\ell-q) \Bigg],
\eea

\bea \label{eq:RR_diagram6}  \begin{tikzpicture}[baseline=(b.base)]
    		\begin{feynman}
    			\vertex [crossed dot] (v1) {};
				\vertex at ($(v1) + (1.5cm, 0)$) (v3);
				\vertex [dot] at ($(v1) + (2.5cm, 0)$) (v2) {};
    			\diagram*{
    				(v3) --[boson, momentum = $\ell$] (v1) 
                             --[boson, half left, momentum = $\ell + q$] (v3) 
                             --[boson, momentum' = $q$] (v2) 
    			};
    		\end{feynman}
    	\end{tikzpicture}		
        = \int \frac{d^d \ell}{(2\pi)^d} \Bigg[\tilde{E}_{h^2}^{\mu \nu \eta \lambda}(\ell+q,-\ell)
            \Delta_{\mu \nu \rho \sigma}(\ell+q)\Delta_{\eta \lambda \alpha \beta}(\ell)V_{h^3}^{\alpha \beta \rho \sigma \delta \xi}(\ell,-\ell-q,q) \Bigg]\Delta_{\delta \xi \varepsilon \gamma}(q)E_h^{\varepsilon \gamma}(q),
\eea

\bea \label{eq:RR_diagram7}  \begin{tikzpicture}[baseline=(b.base)]
    		\begin{feynman}
    			\vertex [crossed dot] (v1) {};
				\vertex at ($(v1) + (0, -0.09cm)$) (b);
				\vertex [dot] at ($(v1) + (2cm, 0)$) (v2) {};
    			\diagram*{
    				(v2) --[boson, momentum = $\ell$] (v1) 
                             --[boson, half left, momentum = $\ell + q$] (v2) 
    			};
    		\end{feynman}
    	\end{tikzpicture}		
        = \int \frac{d^d \ell}{(2\pi)^d} \Bigg[\tilde{E}_{h^2}^{\mu \nu \eta \lambda}(\ell+q,-\ell)
           \Delta_{\mu \nu \rho \sigma}(\ell+q)\Delta_{\eta \lambda \alpha \beta}(\ell)
           E_{h^2}^{\alpha \beta \rho \sigma}(\ell,-\ell-q) \Bigg],
\eea

\bea \label{eq:RR_diagram8} \begin{tikzpicture}[baseline=(b.base)]
    		\begin{feynman}
    			\vertex [crossed dot] (v1) {};
				\vertex [dot] at ($(v1) + (2cm, 0)$) (v2) {};
    			\diagram*{
    				v1 --[boson, out=55, in=125, loop, min distance=2.5cm, momentum'={[arrow shorten=0.15mm,arrow distance=3mm]\(\ell\)}] v1 
                             --[boson, momentum' = $q$] (v2) 
    			};
    		\end{feynman}
    	\end{tikzpicture}		
          = \int \frac{d^d \ell}{(2\pi)^d}\Bigg[\tilde{E}_{h^3}^{\mu \nu \eta \lambda \rho \sigma}(\ell,-\ell,q)
                \Delta_{\mu \nu \eta \lambda}(\ell) \Bigg] \Delta_{\rho \sigma \alpha \beta}(q) E_h^{\alpha \beta}(q),
\eea
where the vertices and propagators in the above equations are given in Sec.~\ref{sec:Feyn}, $q$ is the external momentum, and $\ell$ is the loop momentum.  The diagrams given by Eqs.~(\ref{eq:RR_diagram1}), (\ref{eq:RR_diagram2}), (\ref{eq:RR_diagram3}), and (\ref{eq:RR_diagram4}) are the standard field theory diagrams, with Eq.~(\ref{eq:RR_diagram4}) containing a ghost/anti-ghost loop as required by the usual Fadeev-Popov quantization prescription.
Given the enormous complexity of the graviton propagator in the general de Donder gauge, as well as the very lengthy three-graviton vertex, the diagram with the internal graviton loop has tens of thousands of terms at an intermediate stage of the calculation, requiring the use of a computer algebra package, in our case the {\sc FeynCalc} package \cite{Mertig:1990an, Shtabovenko:2016sxi, Shtabovenko:2020gxv} for {\sc Mathematica}.  {\sc FeynCalc} performs a tensor decomposition to reduce the expression of a one-loop Feynman diagram to a set of scalar basis integrals, the Passarino-Veltman integrals \cite{Passarino:1978jh}; we review these integrals in Appendix \ref{App:PV}.  The scalar Passarino-Veltman integrals are labeled by $A_0$, $B_0$, $C_0$ etc., where $A_0$ results from a tadpole diagram, $B_0$ from a loop with two vertices, $C_0$ from a loop with three vertices, etc.  Although the loops in the above diagrams have at most two vertices, and thus two propagators, in the generalized de Donder gauge there are terms in the graviton propagator with powers of momentum in the denominator up to $p^6$, leading to contributions from Passarino-Veltman integrals up to and including $F_0$ from these diagrams.  We omit additional tadpole diagrams, which vanish identically when only massless particles propagate in the loops.  We are careful to include factors of two associated with the symmetry factors of the diagrams, as well as for the mirror factors of the diagrams with an additional reflected counterpart that is not pictured.

The result for these standard field theory diagrams is
\bea  i A_{{\cal R}, \rm stand}^{\rm one-loop} = \frac{\kappa^4 q^4 \beta}{4(\beta-3)^3}
              \Bigg[\big(2\beta^2 + 5\beta - 9\big)B_0\big(q^2,0,0\big) - 3\big(\beta-1\big)q^2 C_0\big(0,q^2,q^2,0,0,0\big) \Bigg],
\eea
where we see that all Passarino-Veltman integrals beyond $C_0$ cancel in this subset of diagrams.  Also, the $\alpha$ dependence cancels in a non-trivial way among the diagrams.  Some dependence on $\beta$ remains, however, showing that the standard field theory diagrams are not sufficient to yield a gauge-invariant result.  The coordinate correction diagrams are given by Eqs.~(\ref{eq:RR_diagram5}), (\ref{eq:RR_diagram6}), (\ref{eq:RR_diagram7}), and (\ref{eq:RR_diagram8}).  Note that the coordinate correction vertex, being non-local, contains one or more scalar propagators, depending on the order of the coordinate corrections.  At second order the scalar propagators can introduce a non-trivial momentum flow, such that the tadpole diagram of Eq.~(\ref{eq:RR_diagram8}) does not vanish, contrary to the usual intuition.  Summing these four diagrams, we find
\bea  i A_{{\cal R}, \rm cc}^{\rm one-loop} = \frac{\kappa^4 q^4}{16(\beta-3)^3}
              \Bigg[\big(-9\beta^3 -11\beta^2 +9\beta +27\big)B_0\big(q^2,0,0\big) + 12\beta\big(\beta-1\big) q^2 C_0\big(0,q^2,q^2,0,0,0\big) \Bigg],
\eea
where again the Passarino-Veltman integrals beyond $C_0$ cancel, the $\alpha$ dependence cancels, but the $\beta$ dependence does not.  Adding all eight diagrams together yields
\bea  i A_{{\cal R}}^{\rm one-loop} = -\frac{\kappa^4}{16}q^4 B_0\big(q^2,0,0\big),
\eea
which is now independent of $\beta$, as expected for a gauge-invariant correlation function.  Taking the result for $B_0(q^2,0,0)$ from Appendix \ref{App:PV}, we find that the complete NLO result for the curvature correlator is
\bea \label{eq:ARp4}  i A_{{\cal R}}^{p^4} = \frac{i\kappa^4}{256\pi^2}q^4 \ln(-q^2) + {\rm local \ terms},
\eea
where the local terms are analytic terms proportional to $q^4$.  Upon Fourier transforming to position space, the local terms lead to (derivatives of) delta functions and can be ignored for our purposes.  The Fourier transform to position space of the first term in Eq.~(\ref{eq:ARp4}) can be obtained from the result of Appendix~\ref{App:Fourier}, and yields
\bea  {\cal F}(iA_{\cal R}^{p^4}) = \frac{3\kappa^4}{4\pi^4 x^8},
\eea
where $x$ is the source-sink separation, and we have ignored contact terms.  A straightforward Euclidean continuation of this result leads to
\bea {\cal F}(iA_{\cal R}^{p^4}) \rightarrow \frac{3\kappa^4}{4\pi^4 x_E^8} = \frac{768 G^2}{\pi^2 x_E^8},
\eea
where we also rewrite the expression in terms of Newton's constant.

\subsection{\label{volume_1loop} Volume correlation function at one-loop}

The volume correlation function to one-loop can be calculated by evaluating the same diagrams, except using the external operator $\sqrt{-\det({\cal G})}$ in place of ${\cal R}$.  This calculation involves the third order coordinate corrections, requiring a lengthy expansion that we do not undertake.  Instead, we take a short-cut.  We exploit the fact that the next-to-leading order contribution from the tree-level insertion of the $O(R^2)$ operators leads to the same linear combination of local counterterms for the curvature and volume correlators.  Having established that evanescent contributions do not appear in our correlators to this order, we see that a determination of the coefficient of the $1/\varepsilon$ pole in the tree-level counterterm canceling the one-loop ultra-violet divergence also fixes the coefficient of the $\ln{(-q^2)}$ term in the one-loop contribution. 

The tree-level insertion of the $O(R^2)$ operators in our chosen basis leads to
\bea \begin{tikzpicture}[baseline=(b.base)]
    		\begin{feynman}
    			\vertex [dot] (v1) {};
				\vertex [square dot] at ($(v1) + (1cm, 0)$) (b) {};
				\vertex [dot] at ($(v1) + (2cm, 0)$) (v2) {};
    			\diagram*{
    				(v1) --[boson, momentum = $q$] (b)
                             --[boson, momentum = $q$] (v2)
    			};
    		\end{feynman}
    	\end{tikzpicture}	
        = \ii A_{{\cal G}}^{p^4, {\rm tree-level}} = - \frac{9}{2} \ii c_1\kappa^4,
\eea
where again the $\alpha$ and $\beta$ dependence cancels, leading to a gauge-invariant result for this contribution.  The coefficient $c_1$ is the same low energy constant that appears in $\ii A_{{\cal R}}^{p^4, {\rm tree-level}}$.  Given that the low energy constant always appears with the same counterterm coefficient for the $1/\varepsilon$ pole, which in turn accompanies the one-loop logarithm in a fixed linear combination, we are thus able to infer the full next-to-leading order contribution to the volume correlator from the result for the curvature correlator.  The result is 
\bea  i A_{{\cal G}}^{p^4} = \frac{i\kappa^4}{256\pi^2} \ln(-q^2) + {\rm local \ terms},
\eea
where the local terms are not given explicitly, since they Fourier transform to a four-dimensional delta function in position space.
Making use of the Fourier transform of Appendix~\ref{App:Fourier}, we find
\bea  {\cal F}\Bigg(iA_{\cal G}^{p^4}\Bigg) = \frac{\kappa^4}{256\pi^4 x^4}.
\eea
A straightforward Euclidean continuation of this result yields
\bea {\cal F}\Bigg(iA_{\cal G}^{p^4}\Bigg) \rightarrow \frac{\kappa^4}{256\pi^4 x_E^4} = \frac{4 G^2}{\pi^2 x_E^4},
\eea
where we have also expressed the answer in terms of Newton's constant $G$.

In summary, our result for the Euclidean volume correlator in position space through next-to-leading order is
\bea  {\cal F}\Bigg(iA_{\cal G}^{p^2}+iA_{\cal G}^{p^4}\Bigg) \rightarrow \frac{12G}{\pi x_E^2} + \frac{4 G^2}{\pi^2 x_E^4}.
\eea

\subsection{\label{sec:summary_final} Summary of results and making contact with the lattice}

To summarize our main results, we find for the curvature-curvature correlator in Euclidean position space, 
\bea {\cal F}(iA_{\cal R})=\frac{768 G^2}{\pi^2 x_E^8} + O(G^3),
\eea
where we have discarded the contact terms that vanish away from the origin.  The only low energy constant appearing is Newton's constant $G$, since higher order low energy constants appear only in contact terms to this order. This result is thus a universal prediction of the effective theory that could in principle be compared to lattice results.  Similarly we find for the volume-volume correlator in Euclidean position space,
\bea {\cal F}(iA_{\cal G})=\frac{12 G}{\pi x_E^2} + \frac{4 G^2}{\pi^2 x_E^4} + O(G^3),
\eea
where this expression also omits contact terms.  It differs from the curvature correlator in that its leading-order (tree-level) contribution is non-vanishing away from the origin.  The expression includes the next-to-leading order contribution also calculated in this work.  

Since these results are gauge-invariant and universal, any lattice formulation that makes contact with the correct low energy theory should reproduce them.  As an example, we have in mind the lattice gravity formulation known as Euclidean dynamical triangulations (EDT) \cite{Agishtein:1991cv, Ambjorn:1991pq, Catterall:1994pg, Laiho:2016nlp, Dai:2024vjc}.  This approach does not introduce a metric, but rather sums over the space of invariant discrete geometries, which are constructed by gluing together equilateral 4-simplices.  Correlation functions are constructed by taking the geodesic distance between source and sink, defined by the minimal number of hops between adjacent simplices needed to traverse the distance.  Because the geometry fluctuates, there is no sense in which the source and sink locations are fixed, and the geodesic distance between them is necessarily measured against the fluctuating geometry, as in the relational approach discussed at length in this work.  By formulating the correlation functions in EDT only in terms of the invariant (discrete) geometry, it seems likely that the gauge-invariant correlation functions will be recovered in the continuum limit, or perhaps in the low energy limit at scales well below the lattice cut-off, if no continuum limit exists.

\section{\label{sec:conclude}Conclusion}

This work gives a detailed review of the relational approach, which is needed to construct diffeomorphism-invariant observables in quantum gravity.  The relational approach provides a systematic method for computing the coordinate corrections that result from the fluctuation of spacetime and must be accounted for.  The perturbative expansion of the coordinate corrections leads to new Feynman diagrams that are to be evaluated for a given correlation function, and only once these corrections are included does the theory produce gauge-invariant results.  We verify this explicitly for the curvature and volume correlation functions, which we compute.  The cancellation of gauge-dependence in our expressions provides further evidence for the consistency of the relational formalism, as well as a strong cross-check of our results.  The fact that our expressions for the correlation functions are positive is reassuring, since this is a consequence of reflection positivity, which we might expect to hold for the universal, long-distance predictions of a more fundamental (unitary) underlying theory.  In summary, the expressions for the curvature and volume correlation functions are the main result of this work, and we hope that they will provide a useful point of comparison to numerical lattice gravity calculations.  

\section*{Acknowledgments}
The authors thank Simon Catterall, Markus Fr\"ob, Jay Hubisz, Alex Maloney, Marc Schiffer, Judah Unmuth-Yockey, and Scott Watson for valuable discussions.  JL and KR were supported
by the U.S. Department of Energy (DOE), Office of Science, Office of High Energy Physics under Award Number DE-SC0009998.

\appendix

\section{\label{App:PV}Passarino-Veltman Integrals}

Passarino and Veltman showed that all one-loop integrals with various tensor structures can be reduced to scalar integrals \cite{Passarino:1978jh}.  The simplest of the scalar integrals arises from tadpole diagrams, taking the form,
\bea  A_0\big(m^2\big) = \int \frac{d^d k}{(2\pi)^d} \frac{1}{k^2-m^2}.
\eea
This integral leads to a result proportional to $m^2$, so that $A_0(0)=0$.  Since we consider only massless particles propagating in loops, this integral vanishes for all the processes considered in this work.

Feynman diagrams with two legs within a loop lead to integrals of the form,
\bea  B_0\big(q^2,m_1^2,m_2^2\big) = \int \frac{d^d k}{(2\pi)^d} \frac{1}{\big(k^2-m_1^2\big)\big((k+q)^2-m_2^2\big)},
\eea
which, assuming $d=4-2\varepsilon$, can be simplified to
\bea B_0\big(q^2,m^2,m^2\big) = \frac{i}{16\pi^2}\Bigg(\frac{1}{\varepsilon} -\gamma +\ln(4\pi)
                        -\int_0^1 dx \ln\big[m^2 - q^2 x(1-x)\big] \Bigg).
\eea
In our case, we are only interested in loops containing massless propagators, leading to the simple form,
\bea  B_0\big(q^2,0,0\big) = \frac{i}{16\pi^2}\Bigg(\frac{1}{\varepsilon} -\gamma +\ln(4\pi)
                            +2 -\ln\big(-q^2\big) \Bigg).
\eea

Feynman diagrams with three legs in the loop introduce the integral,
\bea  C_0\big(q_1^2,q_2^2,m_1^2,m_2^2,m_3^2\big) = \int \frac{d^d k}{(2\pi)^d}\frac{1}{\big(k^2-m_1^2\big)\big((k+q_1)^2-m_2^2\big)\big((k+q_2)^2-m_3^2\big)},
\eea
and integrals with additional propagators in the loop continue this pattern, with labels $D_0$, $E_0$, etc.  Diagrams with up to six propagators in the loop appear at intermediate stages of our calculation, though everything beyond $B_0$ ends up canceling in the final result.

\section{\label{App:Fourier} Fourier transform}

We wish to find the 4-dimensional Fourier transform of $\ln(-q^2)$, which we denote by $T(x)$, i.e. $T(x)\equiv{\cal F}\big(\ln(-q^2)\big)$.  We add a term $m^2-i\varepsilon$ to the argument of the log in order to regulate the Fourier transform integral.  Though we will end up taking the limit as $m$ and $\varepsilon$ go to zero, it is convenient to introduce them in this form in order to express our result in terms of the scalar Feynman propagator at an intermediate stage of the calculation. We thus consider the Fourier transform,
\bea
{\cal F}\big(T(x)\big)= \ln(-q^2+m^2-i\varepsilon),
\eea
which we invert using the following trick.  First, we differentiate both sides with respect to momentum, 
\bea
\frac{\partial^2}{\partial q^2}{\cal F}\big(T(x)\big)= \frac{\partial^2}{\partial q^2}\ln(-q^2+m^2-i\varepsilon).
\eea
The left-hand side of this expression becomes
\bea \frac{\partial^2}{\partial q^2}\int d^4 x  T(x) e^{i q\cdot x} = \int d^4 x \big(-x^2 T(x)\big)  e^{i q \cdot x},
\eea
implying that
\bea -x^2 T(x) = \int \frac{d^4 q}{(2\pi)^4}\Bigg[ \frac{\partial^2}{\partial q^2}\ln(-q^2+m^2-i\varepsilon)
             \Bigg]e^{-i q\cdot x}.
\eea

Taking the momentum derivatives, this can be rewritten as
\bea\label{eq:Fourier_long} -x^2 T(x) = -i\Bigg(8 + 2 \frac{\Box}{m}\frac{\partial}{\partial m} \Bigg)\int \frac{d^4 q}{(2\pi)^4}
             \frac{i}{q^2-m^2+i\varepsilon} e^{-i q\cdot x},
\eea
where we have written the right-hand side in terms of the Feynman propagator in position space,
\bea D(x) = \int \frac{d^4 q}{(2\pi)^4}\frac{i}{q^2-m^2+i\varepsilon} e^{-i q\cdot x}.
\eea
The position-space Feynman propagator has the well-known massless limit,
\bea\label{eq:Feynprop}  \lim_{m\to 0}D(x)=-\frac{1}{4\pi^2 x^2}.
\eea
The position space propagator also leads to the following relation in the massless limit,

\bea  \lim_{m\to 0} \frac{1}{m}\frac{\partial}{\partial m} D(x) = \frac{1}{4\pi^2}
                  \ln\Bigg(\frac{\sqrt{x^2-i\varepsilon}}{2} \Bigg),
\eea
which implies
\bea \label{eq:Feynprop_deriv} \lim_{m\to 0} 2 \frac{\Box}{m}\frac{\partial}{\partial m}D(x) = \frac{1}{\pi^2 x^2}.
\eea
Substituting Eqs.~(\ref{eq:Feynprop}) and (\ref{eq:Feynprop_deriv}) into Eq.~(\ref{eq:Fourier_long}), we find in the massless limit that
\bea  -x^2 T(x) = \frac{i}{\pi^2 x^2},
\eea
which leads to the final result for the Fourier transform,
\bea  T(x) \equiv {\cal F}\big(\ln(-q^2)\big)  = -\frac{i}{\pi^2 x^4}.
\eea

We also want the Fourier transform  ${\cal F} \left(q^4 \ln(-q^2)\right)$, which can be obtained using
\bea \int \frac{d^4 q}{(2\pi)^4} q^4 \ln\big(-q^2 \big) e^{-i q\cdot x} = 
               \Box^2\int \frac{d^4 q}{(2\pi)^4} \ln\big(-q^2\big) e^{-i q\cdot x}= \Box^2 \Bigg( -\frac{i}{\pi^2 x^4}\Bigg).
\eea
Evaluating the covariant Laplacian on the rightmost side of this equation yields
\bea  {\cal F}\big(q^4\ln(-q^2)\big) = -\frac{192i}{\pi^2 x^8}.
\eea
%

\bibliographystyle{apsrev4-1}
\bibliography{refs.bib}

\end{document}